\documentclass[longauth]{aa} 

\usepackage{graphicx}
\usepackage{txfonts}
\usepackage[colorlinks=true,citecolor=blue,linkcolor=blue]{hyperref}
\usepackage{natbib}
\bibpunct{(}{)}{;}{a}{}{,} 
\usepackage{enumitem}
\usepackage{bm}
\usepackage{upgreek}
\usepackage{textcomp}
\usepackage[dvipsnames]{xcolor}
\usepackage{fontawesome}
\usepackage{scrextend}

\newcommand{\ho}{$H_0$}
\newcommand{\lensname}{WGD 2038$-$4008}
\newcommand{\lcdm}{$\Lambda$CDM}
\newcommand{\hunit}{km s$^{-1}$ Mpc$^{-1}$}
\newcommand{\Dd}{D_{\rm d}}
\newcommand{\Ds}{D_{\rm s}}
\newcommand{\Dds}{D_{\rm ds}}
\newcommand{\Ddt}{D_{\Delta t}}
\newcommand{\balpha}{\bm{\alpha}}
\newcommand{\bbeta}{\bm{\beta}}
\newcommand{\btheta}{\bm{\theta}}
\newcommand{\rmd}{{\rm d}}

\newcommand{\ajs}[1]{{{#1}}}
\newcommand{\ajsii}[1]{{{#1}}}
\newcommand{\ajsiii}[1]{{{#1}}}
\newcommand{\ajsiv}[1]{{{#1}}}
\newcommand{\ajsv}[1]{{{#1}}}
\newcommand{\kw}[1]{{{#1}}}
\newcommand{\suyu}[1]{{{#1}}}

\newcommand{\reply}[1]{{{#1}}}
\newcommand{\lreply}[1]{{{#1}}}

\begin{document} 

   \title{TDCOSMO}

   \subtitle{IX. Systematic comparison between lens modelling \ajs{software programs}: Time-delay prediction for \lensname}
        
   \titlerunning{Systematic comparison between lens modelling software programs}
   \authorrunning{Shajib, Wong, Birrer, Suyu, Treu et al.}

   \author{A.~J.~Shajib\inst{\ref{inst1}, \ref{inst2}, \ref{inst3}}\fnmsep\thanks{NHFP Einstein Fellow}
          \and K.~C.~Wong\inst{\ref{inst4}, \ref{inst5}}
          \and S.~Birrer\inst{\ref{inst6}, \ref{inst7}}
          \and S.~H.~Suyu\inst{\ref{inst8}, \ref{inst9}, \ref{inst10}}
          \and T.~Treu\inst{\ref{inst3}}\fnmsep\thanks{Packard Fellow}
          \and E.~J.~Buckley-Geer\inst{\ref{inst11}, \ref{inst1}}
          \and H.~Lin\inst{\ref{inst11}}
          \and C.~E.~Rusu\inst{\ref{inst4}}
          \and J.~Poh\inst{\ref{inst1}, \ref{inst2}}
          \and A.~Palmese\inst{\ref{inst12},\ref{inst11}}\fnmsep\thanks{NHFP Einstein Fellow}
          \and A.~Agnello\inst{\ref{inst13}}
          \and M.~W.~Auger-Williams\inst{\ref{inst14}, \ref{inst15}}
          \and A.~Galan\inst{\ref{inst16}}
          \and S.~Schuldt\inst{\ref{inst8}, \ref{inst9}}
          \and D.~Sluse\inst{\ref{inst17}}
          \and F.~Courbin\inst{\ref{inst16}}
          \and J.~Frieman\inst{\ref{inst1}, \ref{inst2}, \ref{inst11}}
          \and M.~Millon\inst{\ref{inst16}}
          }

   \institute{
                         Department  of  Astronomy  \&  Astrophysics,  University  of Chicago, Chicago, IL 60637, USA \\
                     \email{ajshajib@uchicago.edu}\label{inst1}
                  \and
                     Kavli Institute for Cosmological Physics, University of Chicago, Chicago, IL 60637, USA\label{inst2}
                 \and
                         Department of Physics and Astronomy, University of California, Los Angeles, CA 90095, USA\label{inst3}
                 \and
                    National Astronomical Observatory of Japan (NAOJ), National Institutes of Natural Sciences, 2-21 Osawa, Mitaka, Tokyo 181-8588, Japan\label{inst4}
                 \and
                    Kavli IPMU (WPI), UTIAS, The University of Tokyo, Kashiwa, Chiba 277-8583, Japan\label{inst5}
                 \and
                        Kavli Institute for Particle Astrophysics and Cosmology and Department of Physics, Stanford University, Stanford, CA 94305, USA\label{inst6}
                 \and
                    SLAC National Accelerator Laboratory, Menlo Park, CA 94025, USA\label{inst7}
                 \and
                        Max Planck Institute for Astrophysics, Karl-Schwarzschild-Str. 1, 85748 Garching, Germany\label{inst8}
                 \and
                        Technische Universit\"at M\"unchen, Physik-Department, James-Franck-Str. 1, 85748 Garching, Germany\label{inst9}
                 \and
                        Institute of Astronomy and Astrophysics, Academia Sinica, 11F of ASMAB, No.1, Section 4, Roosevelt Road, Taipei 10617, Taiwan\label{inst10}
                \and
                        Fermi National Accelerator Laboratory, P. O. Box 500, Batavia, IL 60510, USA\label{inst11}
                \and
                Department of Astronomy, University of California, Berkeley, 501 Campbell Hall, Berkeley, CA 94720, USA\label{inst12}
                \and
                    DARK, Niels Bohr Institute, Jagtvej 128, 2200 Copenhagen, Denmark\label{inst13}
                \and
                        Institute of Astronomy, Madingley Road, Cambridge CB3 0HA, UK\label{inst14}
                \and
                    Kavli Institute for Cosmology, University of Cambridge, Madingley Road, Cambridge CB3 0HA, UK\label{inst15}
                \and
                        Institute of Physics, Laboratoire d'Astrophysique, Ecole Polytechnique F\'ed\'erale de Lausanne (EPFL), Observatoire de Sauverny, CH-1290 Versoix, Switzerland\label{inst16}
        \and
            STAR Institute, Quartier Agora - All\'ee du six Ao\^ut, 19c B-4000 Liege, Belgium\label{inst17}
        }
        
   \date{Received May XX, XXXX; accepted XX XX, XXXX}

 
  \abstract
  {The importance of alternative methods for measuring the Hubble constant, such as time-delay cosmography, is highlighted by the recent Hubble tension. It is paramount to thoroughly investigate and rule out systematic biases in all measurement methods before we can accept new physics as the source of this tension. In this study, we perform a check for systematic biases in the lens modelling procedure of time-delay cosmography by comparing independent and blind time-delay predictions of the system WGD 2038$-$4008 from two teams using two different software programs: \textsc{glee} and \textsc{lenstronomy}. The predicted time delays from the two teams incorporate the stellar kinematics of the deflector and the external convergence from line-of-sight structures. The un-blinded time-delay predictions from the two teams agree within $1.2\sigma$, implying that once the time delay is measured the inferred Hubble constant will also be mutually consistent. However, there is a $\sim$4$\sigma$ discrepancy between the power-law model slope and external shear, which is a significant discrepancy at the level of lens models before the stellar kinematics and the external convergence are incorporated. We identify the difference in the reconstructed point spread function (PSF) to be the source of this discrepancy. When the same reconstructed PSF was used by both teams, we achieved excellent agreement, within $\sim$0.6$\sigma$, indicating that potential systematics stemming from source reconstruction algorithms and investigator choices are well under control. We recommend that future studies supersample the PSF as needed and marginalize over multiple algorithms or realizations for the PSF reconstruction to mitigate the systematics associated with the PSF. A future study will measure the time delays of the system WGD 2038$-$4008 and infer the Hubble constant based on our mass models.
  }



   \maketitle
%
\section{Introduction} \label{sec:introduction}

The Hubble constant, \ho, is a central cosmological parameter as it sets the expansion rate of the Universe. Consequently, precise knowledge of its value is crucial for our understanding of the Cosmos, and it also has important implications in extragalactic astrophysics. However, different methods have measured the Hubble constant with \ajsiv{discrepant} values, producing the so-called Hubble tension \citep[e.g.][]{Freedman21}. Mapping of the temperature fluctuations of the cosmic microwave background allows one to measure the Hubble parameter $H(z\approx1100)$ at the last scattering surface, and then the Hubble constant, \ho, at the current epoch is extrapolated using $\Lambda$ cold dark matter (\lcdm) cosmology. This early-Universe probe resulted in constraints of $H_0 = 67.4 \pm 0.5$ \hunit\ \citep{PlanckCollaboration18} and $H_0 = 67.6 \pm 1.1$ \hunit\ \citep{Aiola20}. In the local Universe, \ho\ is typically measured by building a cosmic distance ladder up to type Ia supernovae (SNe) in the Hubble flow by calibrating their absolute magnitudes with intermediate distance probes. The Supernova \ho for the Equation of State of dark energy (SH0ES) team used Cepheids and parallax distances to calibrate the cosmic distance ladder, measuring $H_0 = 73.04 \pm 1.04$ \hunit\ \citep{Riess22}, which is in 5$\sigma$ tension with the \textit{Planck} measurement. The Carnegie--Chicago Hubble Project used the tip of the red giant branch to calibrate the distance ladder, measuring $H_0 = 69.6 \pm 1.9$ \hunit\ \citep{Freedman19, Freedman20}, which, interestingly, is statistically consistent with both the SH0ES and \textit{Planck} measurements. Several other local probes strengthened the Hubble tension, for example the Megamaser Cosmology Project measured $H_0 = 73.9 \pm 3.0$ \hunit\ \citep{Pesce20}, the Tully--Fisher method calibrated with Cepheids measured $H_0 = 75.1 \pm 0.2 \pm 3.0$ \hunit\ \citep{Kourkchi20}, and the surface brightness fluctuation method measured $H_0 = 73.7 \pm 0.7 \pm 2.4$ \hunit\ \citep{Blakeslee21}. If systematics in these measurements can be ruled out as the source of this Hubble tension, new physics beyond the standard \lcdm\ cosmology will be required to resolve the tension \citep[e.g.][]{Poulin19, Knox20, Efstathiou21}. Therefore, thoroughly investigating the potential systematics that are as yet unknown in each of the probes is paramount.

Strong-lensing time delays provide an independent probe of the Hubble constant \citep{Refsdal64}. The \ajs{delays} between the arrival times of photons corresponding to different images of the background source depend on the cosmological distances involved in the strong-lensing system, and thus these delays allow us to measure a combination of these distances, called the `time-delay distance' \citep{Suyu10}.  The time-delay distance is inversely proportional to \ho\ and weakly dependent on other cosmological parameters. Although early implementations of this method in the 1990s and the early 2000s suffered from limitations in data quality and analysis techniques, both of these aspects have improved by a large margin over the past decade \citep[for a review with a historical perspective, see][]{Treu16b}. Inferring \ho\ from the time delays \ajs{requires}: (i) measuring the time delays, \ajs{(ii) measuring the redshifts of the deflector and the background quasar,} (iii) modelling the mass distribution in the central deflector to compute the Fermat potential differences between the image positions, and (iv) estimating the extra lensing contribution from the line-of-sight (LOS) mass distribution between the background source and the observer. Thanks to breakthroughs in all of these factors, the \ho\ Lenses In the COSMOGRAIL's Wellspring (H0LiCOW) and the Strong-lensing   High Angular Resolution Programme (SHARP) collaborations measured $H_0 = 73.3_{-1.8}^{+1.7}$ \hunit\ from a sample of six strongly lensed quasar systems \citep{Suyu17, Bonvin17, Birrer19b, Chen19, Rusu20, Wong20}. The STRong-lensing Insights into the Dark Energy Survey (STRIDES) collaboration analysed a seventh lens system to measure $H_0 = 74.2_{-3.0}^{+2.7}$ \hunit\ \citep{Shajib20}. It is noteworthy that six out of these seven analyses were performed blindly, with only the first one being a non-blind analysis. The H0LiCOW, STRIDES, Cosmological Monitoring of Gravitational Lenses (COSMOGRAIL), and SHARP collaborations have united under the umbrella of the Time-Delay COSMOgraphy (TDCOSMO) collaboration.

The TDCOSMO collaboration has already performed a number of tests to search for previously unknown systematics. \citet[][TDCOSMO-I]{Millon20} checked for systematics arising from the current treatments of the stellar kinematics, LOS mass distribution, and the choice of lens model families, finding no evidence for unaccounted errors. \citet[][TDCOSMO-III]{Gilman20} find that dark sub-halos -- which are ignored in lens modelling through the assumption of smooth mass profiles -- also do not systematically bias the \ho\ inference, adding negligible random uncertainty. \citet[][TDCOSMO-IV]{Birrer20} relaxed the assumption of the power-law mass distribution in the deflector galaxies \ajs{to allow maximal degeneracy in the mass distribution} under the mass-sheet transformation \citep[MST;][]{Falco85}. By constraining the mass distribution from the stellar kinematics only, these authors inferred $H_0 = 74.5_{-6.1}^{+5.6}$ \hunit, that is, relaxing the power-law assumption leads to an increase in \ho\ uncertainty from 2.2\% to 7.9\% for the sample of the seven analysed systems. To regain the lost precision, TDCOSMO-IV combined an external sample of galaxy--galaxy strong lenses from the Sloan Lens ACS\footnote{\lreply{Advanced Camera for Surveys}} (SLACS) survey to add more information on the galaxy mass distribution, {under the assumption that the SLACS lenses and the TDCOSMO lenses belong to the same galaxy population}. Adding a sample of 33 SLACS lenses improved the precision to 5.4\%. Although the point estimate of \ho\ shifted to $H_0 = 67.4_{-3.2}^{+4.1}$ \hunit\ with the addition of the SLACS lenses, this value is still consistent with all the previous TDCOSMO measurements within 1$\sigma$. \citet[][TDCOSMO-V]{Birrer21c} forecasted that a future sample of 40 time-delay lenses with spatially resolved stellar kinematics and an external lens sample of 200 non-time-delay lenses will be able to infer \ho\ with 1.2--1.3\% precision, which is necessary to independently settle the Hubble tension at the $\sim$5$\sigma$ confidence level. \ajsiv{\citet[][TDCOSMO-VII]{VandeVyvere22} find that the systematic bias in the measured \ho\ arising from the boxy-ness or discy-ness of the deflector galaxy is <1\% and thus insignificant.}  Blind data challenges are also important tests for the presence of systematics. The Time-Delay Challenge validated the robustness of the methods currently used  to measure time delays from quasar light curves \citep{Dobler15, Liao15}. The Time-Delay Lens Modelling Challenge similarly validated the modelling techniques currently used to recover the ground truth when the shapes of the underlying galaxy mass profiles are known \citep{Ding21b}.

In this paper we present the results of an experiment to search for potential systematics in the lens modelling -- within specific assumed mass profile families -- that may arise from different modelling \ajs{software programs} used by different investigators. In this experiment, two teams using different \ajs{software programs} independently modelled the strongly lensed quasar system \lensname\ to the level required for cosmographic application (i.e. to the noise level; \citealt{Agnello18c}). The two modelling \ajs{software programs} being compared are \textsc{glee}\footnote{\textsc{glee} is developed by A.~Halkola and S.~H.~Suyu \citep{Suyu10b,Suyu12b}.} and \textsc{lenstronomy}\footnote{The lead developer of \textsc{lenstronomy} is S. Birrer. \textsc{Lenstronomy} also received numerous contributions from the community. The full list of contributors is provided at: \url{https://github.com/lenstronomy/lenstronomy/blob/main/AUTHORS.rst}.}. The core members of the \textsc{glee} team are K.~C.~Wong and S.~H.~Suyu; the core members of the \textsc{lenstronomy} team are A.~J.~Shajib, S.~Birrer, and T.~Treu. Both of the \ajs{software programs} have previously been used for lens modelling in cosmographic analyses by the TDCOSMO collaboration  -- five systems with \textsc{glee} and two systems with \textsc{lenstronomy}. Although \citet{Birrer16} performed a cosmographic analysis outside the TDCOSMO umbrella using \textsc{lenstronomy} for the system RXJ1131$-$1231, which was previously analysed by the H0LiCOW collaboration using \textsc{glee}, a systematic blind comparison between the two \ajs{software programs} on the same lens system has not been \ajs{done previously}. Both \ajs{software programs} perform parametric modelling of the deflector mass distribution, but they differ in the method used for source reconstruction. Whereas \textsc{glee} uses a pixel-based source reconstruction with regularization conditions \citep{Suyu06}, \textsc{lenstronomy} uses a basis set of parameterized profiles for source reconstruction \citep{Birrer15, Birrer18, Birrer21b}.

In addition to the software architectures, differences in the lens models may arise from 
modelling choices made by an investigator in such modelling processes. Our experiment also encompasses this human aspect of the modelling process by having the two teams work independently and blindly. However, to facilitate a fair comparison between the model predictions, we established a baseline model setup with minimal specifications that was agreed upon by the two teams before performing their own analyses. After each team separately completed their internal systematic checks and \ajs{went} through an internal review by the TDCOSMO collaboration, the lens models were frozen and the model predictions were un-blinded to make comparisons between the two teams. As the time delay for this system has not yet been measured with sufficient precision for an \ho\ measurement, we leave the \ho\ inference from our models to be done in the future. However, we predict the time delays for this system as a function of \ho\ after marginalizing over the inferences from the two modelling \ajs{software programs}. As a result, our `preemptive' lens models enforce an additional layer of blindness for the future \ho\ measurement from this system. 

The baseline models for comparison have two different lens model setups: (i) a power-law mass model and (ii) a two-component mass model that individually accounts for the dark and baryonic components. 
It is well known that conventional parametric models such as the power-law model impose assumptions that break the mass-sheet degeneracy (MSD; e.g. \citealt{Birrer20,Kochanek20}). However, a lens model is still useful for extracting the relevant lensing information (i.e. the Fermat potential difference) from the data, which can then be processed to allow the additional freedom along the MSD following TDCOSMO-IV. Although techniques to extract lensing information without relying on parametric models have recently~been proposed \citep[e.g.][]{Birrer21}, they have not yet been applied to real systems for rigorous lens modelling similar to the TDCOSMO analyses.
Furthermore, no evidence has so far demonstrated that the simply parametrized models are \ajs{not an adequate description}, \ajs{and the necessity or physical reality of a mass component that acts as a physical mass sheet has not been demonstrated}. For all these reasons, until new evidence is gathered to inform new choices, simply parametrized lens models are going to be the baseline in TDCOSMO analyses. Therefore, it is important to compare the modelling methods based on these \ajs{software programs} to check for systematic differences as performed in this paper. 

\ajs{In this paper we only predict the time delays for \lensname\ based on our lens models, as the actual time delays for this system are yet to be measured and thus the \ho\ cannot be inferred. Measuring \ho\ based on the lens models presented in this paper is left for a future paper.}

This paper is organized as follows. In Sect. \ref{sec:theory_framework} we provide a brief review of the strong lensing formalism to establish the notations and describe the Bayesian inference framework of our model predictions. The observables in our analysis are described in Sect. \ref{sec:data}. We present the baseline models that are common to both teams in Sect. \ref{sec:baseline_models}. The modelling procedures and results are presented by the \textsc{glee} and \textsc{lenstronomy} teams in Sects. \ref{sec:glee_modelling} and \ref{sec:lenstronomy_modelling}, respectively. We compare and discuss the results from the two teams  in Sect. \ref{sec:software_comparison} and conclude the paper in Sect. \ref{sec:conclusion}. \lreply{Sects.} \ref{sec:introduction}--\ref{sec:lenstronomy_modelling} \ajsiv{were} written prior to the un-blinding. After un-blinding on October 22, 2021, Sects. \ref{sec:software_comparison} and \ref{sec:conclusion} \ajsiv{were} written and no major edits \ajsiv{were} done to Sects. \ref{sec:introduction}--\ref{sec:lenstronomy_modelling}, except for minor fixes for typos and grammatical errors.


\section{Framework of the lens modelling} \label{sec:theory_framework}

In this section we describe the theoretical framework for our analysis. We give a brief overview of the strong lensing formalism in Sect.~\ref{sec:lensing_formalism}, discuss the MSD in Sect.~\ref{sec:mst}, explain our modelling of the stellar kinematics in Sect.~\ref{sec:kinematics_theory}, and present the Bayesian inference framework for our analysis in Sect.~\ref{sec:bayesian_inference}.

\subsection{Strong lensing formalism} \label{sec:lensing_formalism}

The goal of this section is to provide the necessary definitions in strong lensing and establish the notation. This formalism was developed in multiple previous studies (see e.g. \citealt{Schneider92, Blandford92}) and has been implemented in numerous previous TDCOSMO analyses \citep[e.g.][]{Suyu10, Birrer19b, Shajib20}.

The delay $\Delta t_{\rm XY}$ between arrival times of photons corresponding to images labelled as X and Y is given by
\begin{equation} \label{eq:time_delay_full}
        \begin{split}
        \Delta t_{\rm XY} &= \frac{1+z_{\rm d}}{c} \frac{\Dd \Ds }{\Dds} \left[ \frac{(\btheta_{\rm X} - \bbeta)^2}{2} - \frac{(\btheta_{\rm Y} - \bbeta)^2}{2} - \psi(\btheta_{\rm X}) + \psi(\btheta_{\rm Y})  \right].
        \end{split}
\end{equation}
Here, $\Dd$ is the angular diameter distance to the deflector, $\Ds$ is that to the source, and $\Dds$ is that between the deflector and the source, $z_{\rm d}$ is the deflector redshift, $c$ is the speed of light, $\btheta$ is the image position, $\bbeta$ is the un-lensed source position, and $\psi$ is the deflection potential that is related to the deflection angle as $\nabla \psi \equiv \balpha$ and the convergence as $\nabla^2 \psi = 2 \kappa$. The convergence is the surface mass density scaled by the critical density as $\kappa \equiv \Sigma/\Sigma_{\rm crit}$ with 
\begin{equation}
        \Sigma_{\rm crit} = \frac{c^2 \Ds}{4 \uppi G \Dds \Dd}.
\end{equation}
 The Fermat potential $\phi$ is defined by combining the geometric delay term with the deflection potential as
\begin{equation}
        \phi(\btheta) \equiv \frac{(\btheta - \bbeta)^2}{2} - \psi(\btheta).
\end{equation}
The so-called time-delay distance is defined as
\begin{equation}
        \Ddt \equiv (1+z_{\rm d}) \frac{\Dd \Ds }{\Dds}.
\end{equation}
Each distance term contains a factor of $H_{0}^{-1}$, which cancel out such that $\Ddt \propto H_{0}^{-1}$. \lreply{Eq.} \ref{eq:time_delay_full} can be written in short form as
\begin{equation}
\label{eq:time_delay_shortform}
        \Delta t_{\rm XY} = \frac{\Ddt}{c} \left[ \phi(\btheta_{\rm X})  - \phi(\btheta_{\rm Y}) \right] \equiv \frac{\Ddt}{c} \Delta \phi_{\rm XY} .
\end{equation}

\subsection{Mass-sheet degeneracy} \label{sec:mst}

The imaging observables of the lensing phenomenon -- the image positions and the flux ratios -- remain invariant under the transformation
\begin{equation}
        \begin{split}
        &\kappa(\btheta) \to \kappa_{\lambda} (\btheta) = \lambda \kappa (\btheta) + 1 - \lambda, \\
        &\bbeta \to \bbeta^\prime = \lambda \bbeta,
        \end{split}
\end{equation}
which is referred \ajs{to} as the MST \citep[][]{Falco85}. The invariance of the observables under this transformation gives rise to the MSD. We note that the magnifications are not invariant under the MST (although magnification ratios are), and thus strongly lensed standard candles can break the MSD \citep{Bertin06}.

We can separate all of the mass contributing to lensing of the background source into two components as
\begin{equation}
        \kappa_{\rm true} = \kappa_{\rm cen} + \kappa_{\rm ext},
\end{equation}
where $\kappa_{\rm cen}$ is the convergence 
from the central deflector and $\kappa_{\rm ext}$ is the convergence from all the LOS mass distribution -- except the central deflector -- projected onto the plane of the central deflector (i.e. the image plane). \ajs{In some cases, the central deflector may have nearby companions or satellites, or nearby LOS perturbing galaxies that are explicitly accounted for in the lens model, for example RXJ1131$-$1231, HE 0435$-$1223, and ES J0408$-$5354 \citep{Suyu13, Wong17, Shajib20}. We consider these additional mass components to be included in $\kappa_{\rm cen}$.} As the mass distribution of the central deflector goes to zero at very large radius, we have 
\begin{equation} \label{eq:true_kappa}
        \lim_{\theta \to \infty} \kappa_{\rm true} (\theta) = \kappa_{\rm ext}.   
\end{equation}
Therefore, $\kappa_{\rm ext}$ can be interpreted as lensing mass in the 3D space far from or `external' to the central deflector. Let $\kappa_{\rm model}^{\prime}$ be the model convergence that can reproduce the imaging observables. However, due to the MSD, $\kappa_{\rm model}^{\prime}$ is not a unique solution and we cannot ascertain that $\kappa_{\rm true} = \kappa_{\rm model}^{\prime}$. If we impose the condition $\lim_{\theta \to \infty} \kappa_{\rm model}^{\prime} = 0$, then $\kappa_{\rm model}^{\prime}$ is a mass-sheet transform of $\kappa_{\rm true}$ with the rescaling factor $\lambda = 1/(1 - \kappa_{\rm ext})$ as
\begin{equation} \label{eq:external_mst}
        \kappa_{\rm true} \to \kappa'_{\rm model} = \frac{1}{1-\kappa_{\rm ext}}(\kappa_{\rm cen} + \kappa_{\rm ext}) - \frac{\kappa_{\rm ext}}{1 - \kappa_{\rm ext}} = \frac{\kappa_{\rm cen}}{1 - \kappa_{\rm ext}}.
\end{equation}
If the external convergence $\kappa_{\rm ext}$ can be independently estimated by studying the lens environment, then the true lensing convergence $\kappa_{\rm true}$ can be recovered from $\kappa_{\rm model}^{\prime}$ through the corresponding inverse MST. However, the lens model $\kappa_{\rm model}$ that we actually constrain can be an internal MST of $\kappa_{\rm model}^{\prime}$ as
\begin{equation} \label{eq:internal_mst}
        \kappa_{\rm model}' = \lambda_{\rm int} \kappa_{\rm model} + 1 - \lambda_{\rm int}.
\end{equation}
Interestingly, both $\kappa_{\rm model}$ and $\kappa_{\rm model}^{\prime}$ can go to zero at $\theta \to \infty$ by construction. In such a case, $\lambda_{\rm int}$ is not a constant and it satisfies $\lim_{\theta \to \infty} = 1$ \citep{Schneider14}. We can combine Eqs. \ref{eq:true_kappa}, \ref{eq:external_mst}, and \ref{eq:internal_mst} to write the relation between the true mass distribution $\kappa_{\rm true}$ and the modelled mass distribution $\kappa_{\rm model}$ as
\begin{equation}
        \kappa_{\rm true} = (1 - \kappa_{\rm ext}) \left[\lambda_{\rm int}\kappa_{\rm model} + 1 -\lambda_{\rm int} \right] + \kappa_{\rm ext}.
\end{equation}
To constrain $\lambda_{\rm int}$, we require observables that rescale with the MST, for example the stellar kinematics. Although such observables rescale with $\lambda_{\rm int}(1- \kappa_{\rm ext})$, the external convergence $\kappa_{\rm ext}$ is independently estimated from the LOS properties leaving only $\lambda_{\rm int}$ to be constrained from those observables. The LOS velocity dispersion rescales with the MST as
\begin{equation}
                \sigma_{\rm los} \to \sigma_{\rm los}^{\prime} = \sqrt{\lambda}\sigma_{\rm los}.
\end{equation}
%
This rescaling is only valid for a pure MST, \ajs{such as} the external MST, and is approximately valid for an internal MST with single aperture kinematics. However, this is not valid for internal MST with spatially resolved kinematics \citep[][]{Chen21, Yildirim21}. The time delay rescales with the MST as
\begin{equation}
        \Delta t \to \Delta t^{\prime} = \lambda \Delta t.
\end{equation}
As a result, we need to correct the time delays $\Delta t_{\rm model}$ predicted by the model $\kappa_{\rm model}$ as 
\begin{equation}
\label{eq:time_delay_true}
        \Delta t_{\rm true} = (1 - \kappa_{\rm ext}) \lambda_{\rm int} \Delta t_{\rm model}.
\end{equation}
In the next section, we describe our framework for \ajs{the} kinematics analysis.

\subsection{Kinematics analysis} \label{sec:kinematics_theory}

The stellar velocity dispersion probes the 3D mass distribution of the deflector galaxy that is deprojected from $\kappa_{\rm cen}$. We adopt the spherical Jeans equation that connects the velocity dispersion with the gravitational potential $\Phi(r)$ as
\begin{equation}
        \frac{\rmd \left( l(r)\ \sigma_{\rm r}(r)^2 \right)}{\rmd r} + \frac{2 \beta_{\rm ani}(r)\ l(r) \  \sigma_{\rm r}(r)^2}{r} = - l(r) \ \frac{\rmd \Phi(r)}{\rmd r}.
\end{equation}
Here, $l(r)$ is the 3D luminosity density, $\sigma_{\rm r} (r)$ is the radial velocity dispersion, and $\beta_{\rm ani} (r)$ is the anisotropy parameter that relates $\sigma_{\rm r}$ to the tangential velocity dispersion $\sigma_{\rm t}$ as
\begin{equation}
        \beta_{\rm ani}(r) \equiv 1 - \frac{\sigma_{\rm t}^2(r)}{\sigma_{\rm r}^2(r)}.
\end{equation}
The observable quantity is the luminosity-weighted LOS velocity dispersion, which we can obtain by solving the Jeans equation as
\begin{equation} \label{eq:jeans_solution}
        \sigma_{\rm los}^2(R) = \frac{2G}{I(R)} \int_R^{\infty} \mathcal{K}_{\beta} \left(\frac{r}{R} \right) \frac{l(r)\ M(r)}{r} \ \rmd r,
\end{equation}
where $G$ is the gravitational constant, $I(R)$ is the surface brightness, and $M(r)$ is the 3D enclosed mass within radius $r$ \citep[Eqs. A15--A16 of][]{Mamon05}. The function $\mathcal{K}_{\beta}(r/R)$ depends on the parameterization of $\beta_{\rm ani}(r)$. We adopt the Osipkov--Merritt parameterization given by
\begin{equation}
        \beta_{\rm ani}(r) = \frac{r^2}{r^2 + r_{\rm ani}^2},
\end{equation}
where $r_{\rm ani}$ is a scaling radius \citep{Osipkov79, Merritt85, Merritt85b}. For this parameterization, the form of $\mathcal{K}_{\beta}(r/R)$ is given by
\begin{equation}
\begin{split}
                \mathcal{K}_{\beta} \left(u\equiv\frac{r}{R} \right) &= \frac{u_{\rm ani}^2 + 1/2}{(u_{\rm ani}+1)^{3/2}}    \left( \frac{u^2+u_{\rm ani}^2}{u} \right) \tan^{-1} \left( \sqrt{\frac{u^2-1}{u^2_{\rm ani}+1}} \right)  \\
                &\qquad\qquad\qquad - \frac{1/2}{u_{\rm ani}^2 + 1} \sqrt{1 - \frac{1}{u^2}},
\end{split}     
\end{equation}
with $u_{\rm ani} \equiv r_{\rm ani}/R$ \citep{Mamon05}. The observed aperture-averaged velocity dispersion is
\begin{equation} \label{eq:jeans_solution_convolved}
        \sigma_{\rm ap}^2 = \frac{\int_{\rm ap} \left[ I(R) \sigma_{\rm los}^2(R) \right] * \mathcal{S}\ \rmd x \rmd y}{\int_{\rm ap} I(R) * \mathcal{S} \ \rmd x \rmd y},
\end{equation}
where $\int_{\rm ap}$ denotes integration over the aperture and $* \mathcal{S}$ denotes convolution with the seeing. Thus, the lens-model-predicted LOS velocity dispersion can be written in the form
\begin{equation} \label{eq:vdisp_ap}
        \sigma_{\rm ap,\ model}^2 = \frac{\Ds}{\Dds} c^2 J(\xi_{\rm lens}, \xi_{\rm light}, \beta_{\rm ani}),
\end{equation}
where $\xi_{\rm lens}$ is the set of mass model parameters \ajsiv{and} $\xi_{\rm light}$ is the set of light distribution parameters. \ajs{The internal and external MST parameters modify the lens-model-predicted velocity dispersion as}  
\begin{equation} \label{eq:vdisp_ap}
        \sigma_{\rm ap,\ true}^2 = (1 - \kappa_{\rm ext}) \lambda_{\rm int} \ \sigma_{\rm ap,\ model}^2.
\end{equation}
The dependence of $\sigma_{\rm ap}$ on the cosmology is fully captured in the \ajs{$\Ds/\Dds$} term. The function $J$ is independent of cosmology as all of its arguments are expressed in angular units, \ajs{but \lreply{it should be noted} that $J$ is directly connected to the model convergence $\kappa_{\rm model}$ through the parameters $\xi_{\rm lens}$} \citep{Birrer16}.

\subsection{Bayesian inference} \label{sec:bayesian_inference}

We denote the set of all the observables as ${O} \equiv \{{O}_{\rm img}, {O}_{\rm kin} \}$, where ${O}_{\rm img}$ is the imaging data of the lens system and ${O}_{\rm kin}$ is the measured stellar velocity dispersion. Although data from spectroscopic and photometric \ajs{surveys} of the lens environment are necessary to estimate the external convergence, we fold in the estimated external convergence as the prior $p(\kappa_{\rm ext})$ in our inference. To predict the time delay for a given cosmology, we want to infer the Fermat potential difference $\Delta \phi$ between the corresponding image pairs. The Fermat potential difference $\Delta \phi(\xi, \kappa_{\rm ext}, \lambda_{\rm int})$ is a function of the set of model parameter $\xi \equiv \{\xi_{\rm lens}, \xi_{\rm light}, r_{\rm ani} \}$ in a model family $M$, external convergence $\kappa_{\rm ext}$, and internal MST parameter $\lambda_{\rm int}$. Thus, to obtain $p(\Delta \phi \mid O)$, we first aim to infer $p(\xi, \kappa_{\rm ext}, \lambda_{\rm int} \mid O)$. Applying Bayes' theorem, we can write
\begin{equation} \label{eq:master_bayes}
\begin{split}
        &p(\xi, \kappa_{\rm ext}, \lambda_{\rm int} \mid O) \propto p (O \mid \xi, \kappa_{\rm ext}, \lambda_{\rm int}) \ p(\xi, \kappa_{\rm ext}, \lambda_{\rm int}) \\
    & \qquad = p(O \mid \xi, \kappa_{\rm ext}, \lambda_{\rm int})\ p(\xi, \kappa_{\rm ext}) \ p(\lambda_{\rm int}) \\
    & \qquad = \int p(O \mid \xi, M, S, D_{\rm s/ds}, \kappa_{\rm ext}, \lambda_{\rm int})\ p(\xi, \kappa_{\rm ext} \mid M, S) \\
    & \qquad \qquad \qquad \times p(\lambda_{\rm int}) \ \rmd S \ \rmd D_{\rm s/ds} \ \rmd M.
\end{split}
\end{equation}
Here, $S$ is the set of lens model hyper-parameters that is only relevant for $O_{\rm img}$, and $D_{\rm s/ds}$ is a short notation for the distance ratio $D_{\rm s/ds} \equiv \Ds / \Dds$. We explicitly separate the hyper-parameters $S$ -- that need to be fixed during optimizing a lens model, for example the set of pixels for computing the image likelihood, resolution of the source reconstruction -- from the choice of lens model family $M$. The prior $p(\kappa_{\rm ext} \mid M)$ depends on the model family $M$, since the model-constrained shear is used to estimate $\kappa_{\rm ext}$ corresponding to $M$. Since $O_{\rm img}$ and $O_{\rm kin}$ are independent data, the likelihood term $p(O \mid \xi, M, S, D_{\rm s/ds}, \kappa_{\rm ext})$ can be decomposed as
\begin{equation} \label{eq:likelihood_decomposition}
\begin{split}
        p(O \mid \xi, M, S, D_{\rm s/ds}, \kappa_{\rm ext, \lambda_{\rm int}}) &= p(O_{\rm img} \mid \xi, M, S) \\
         & \times p(O_{\rm kin} \mid \xi, M, D_{\rm s/ds}, \kappa_{\rm ext}, \lambda_{\rm int}). 
\end{split}
\end{equation}
Then, we can first perform the following sub-integral within the right-hand side of Eq. \ref{eq:master_bayes}:
\begin{equation} \label{eq:imaging_evidence}
\begin{split}
        &\int p(O_{\rm img} \mid \xi, M, S)\ p(\xi \mid M, S)\ p(S)\ \rmd S \\ 
        & = \int p(\xi \mid O_{\rm img}, M, S) \ p(O_{\rm img} \mid M, S) \ p(S) \ \rmd S.
\end{split}
\end{equation}
Here, $p(O_{\rm img} \mid M, S)$ is the model evidence. We perform this integral in the form of the right-hand side of Eq. \ref{eq:imaging_evidence} for numerical convenience, as it allows us to first obtain the posterior $p(\xi \mid O_{\rm img}, M, S)$ using Monte Carlo sampling, and then combine the posteriors weighted by the model evidence to perform the integration in Eq. \ref{eq:imaging_evidence}. We use the Bayesian information criterion (BIC) as a proxy for the model evidence in our analysis \citep{Schwarz78}. The BIC is defined as
\begin{equation} \label{eq:bic_definition}
        \mathrm{BIC} = k \ln N_{\rm data} - 2 \ln \hat{\mathcal{L}},
\end{equation}
where $k$ is the number of free parameters, $N_{\rm data}$ is the number of data points, and $\hat{\mathcal{L}}$ is the \ajs{maximum} of the likelihood function $\mathcal{L}$. Both the BIC and directly computed model evidence were used in previous analyses for Bayesian model averaging \citep[BMA; e.g.][]{Madigan94, Hoeting99} in the context of lens modelling for cosmographic analysis (BIC: \citealt{Birrer19b, Chen19, Rusu20}; model evidence: \citealt{Shajib20}).

Specific implementations of the Bayesian inference framework presented in this section through sampling by each team are described in Sects. \ref{sec:glee_modelling} and \ref{sec:lenstronomy_modelling}.

\section{Imaging data and ancillary measurements} \label{sec:data}

The system \lensname\ was discovered from a combined search in the {Wide-field Infrared Survey Explorer} and \textit{Gaia} data over the Dark Energy Survey (DES) footprint \citep{Agnello18c}. The deflector redshift is $z_{\rm d} = 0.230 \pm 0.002$ and the source redshift is $z_{\rm s} = 0.777 \pm 0.001$ \citep{Agnello18c}. In this section we describe the imaging data and spectroscopic measurements used in our analysis.
 
\subsection{HST imaging} \label{sec:hst_imaging}

\begin{figure*}
        \centering
    \includegraphics[width=0.85\textwidth]{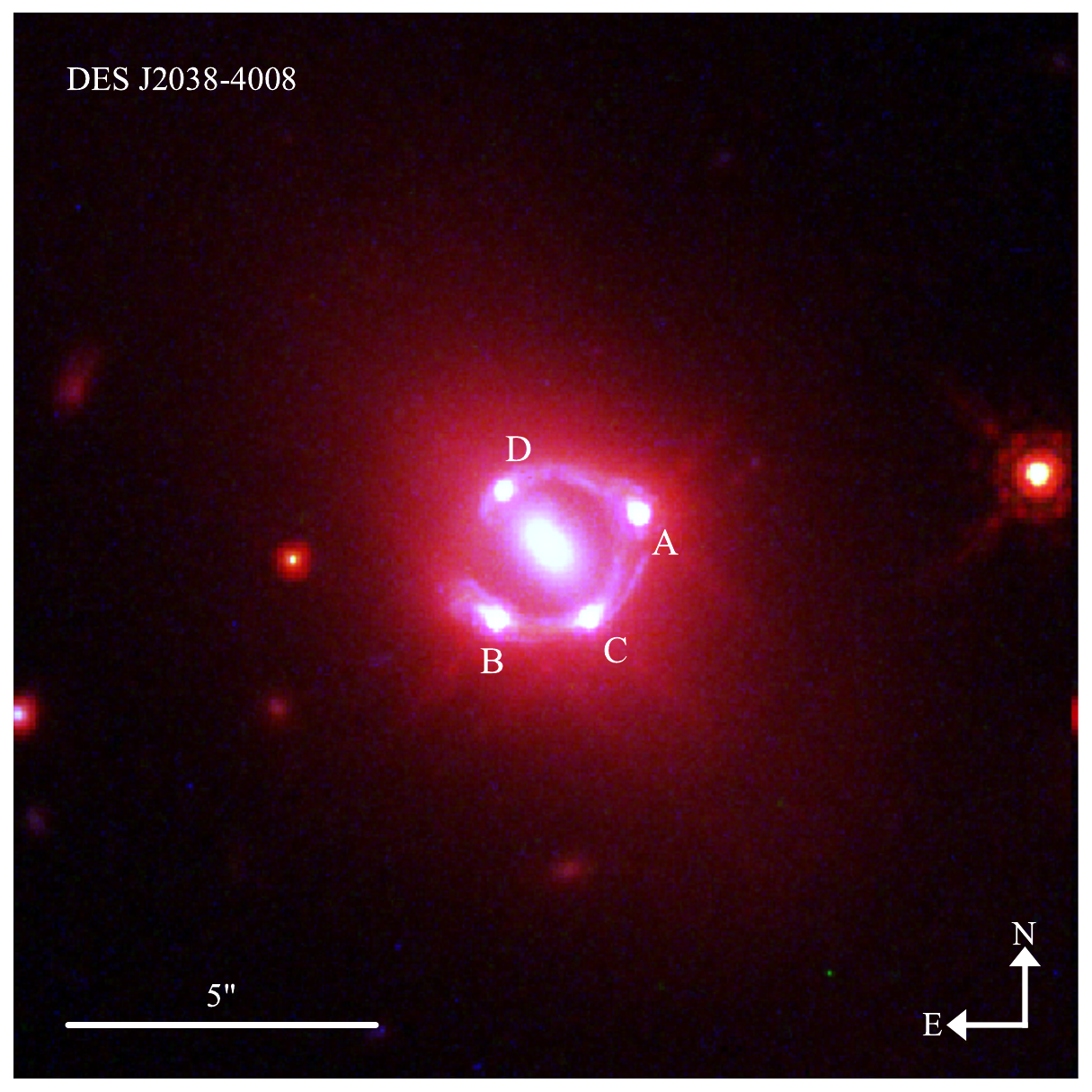}
    \caption{
    False-colour image of the lens systems \lensname. This RGB image is created from the F160W (red), F814W (green), and F475X (blue) filters of the HST WFC3. We adjusted the relative amplitudes between the three filters to achieve a higher contrast for better visualization. The four lensed quasar images are marked as A, B, C, and D.
    }
    \label{fig:rgb_cutout}
\end{figure*}

We obtained \textit{Hubble} Space Telescope ({HST)} imaging of the system \citep[GO-15320, PI: Treu;][]{Shajib19} using the Wide-Field Camera 3 (WFC3). The imaging was taken in three filters: F160W in the infrared (IR) channel, and F814W and F475X in the ultraviolet-visual (UVIS) channel. Four exposures were taken in each filter to cover the large dynamic range in surface brightness of the brighter quasar images and the fainter extended host galaxy. For the IR band, we adopted a four-point dither pattern and STEP100 readout sequence for the MULTIACCUM mode. The total exposure times are 2196.9 s, 1428.0 s, and 1158.0 s, respectively, in the three filters. We show a false-colour red-green-blue (RGB) image of the system created from the HST imaging in Fig. \ref{fig:rgb_cutout}.

\ajs{The point spread function (PSF) corresponding to each filter is estimated from stacking \ajsiii{4--6} stars that are within \ajsiv{each} corresponding HST image. These PSFs are only used as \reply{an} initial estimate \ajsiii{by both teams} and they are refined to more accurately match the PSF at the quasar image positions by iterative reconstruction during the lens model optimization (see Sects. \ref{sec:glee_modelling} and \ref{sec:lenstronomy_modelling} for more details on the iterative reconstruction).}

\subsection{Stellar velocity dispersion} \label{sec:velocity_dispersion_measurement}

\citet{Buckley-Geer20} measure the stellar velocity dispersion of the deflector from spectroscopic observation using the Gemini Multi-Object Spectrograph (GMOS-S) on the Gemini South Telescope. The measured velocity dispersion is $\sigma_{\rm los} = 296 \pm 19$ km s$^{-1}$ from a $0\farcs75\times1\arcsec$ rectangular \ajsiv{aperture}\ajs{, which is in agreement with a more recent measurement from the XShooter instrument on the Very Large Telescope \citep[VLT;][]{Melo21}}. \reply{We used the measurement from \citet{Buckley-Geer20} instead of the more precise measurement from \citet{Melo21} because the latter was published after the un-blinding, when the lens models were frozen and utilized the previous measurement.} The seeing full width at half maximum (FWHM) is 0\farcs9, and the exponent parameter of the Moffat PSF is $\beta=1.74$. 

\subsection{LOS environment} \label{sec:kappa_ext_measurement}

The LOS environment of the system \lensname\ was studied by \citet{Buckley-Geer20}. These authors estimated the external convergence based on the weighted galaxy number counts approach \citep{Greene13, Rusu17, Birrer19b, Rusu20}. The weighted number counts were obtained in two separate apertures with radii 45$\arcsec$ and 120$\arcsec$ centred on the lens system \ajs{from the DES multi-band imaging}. The magnitude limit of counted galaxies is $I = 22.5$ mag. The counts are weighted based on simple physical quantities, such as the inverse of the distance to the lens. \ajs{The spectroscopic redshifts were obtained from Gemini South GMOS-S and the photometric redshifts are based on DES multi-band photometry.} Analogous number counts are also obtained within a large number of different apertures with the same sizes along random LOSs in the DES footprint. By comparing the weighted number counts for the LOS around \lensname\ \ajsiv{with those} for random LOSs, the over- or under-density is estimated in terms of a weighted number count ratio. The external convergence is then estimated by comparing the weighted number count ratio with that from statistically similar LOSs from the Millennium simulation with computed external convergence \citep{Springel05, Hilbert09}. \ajs{If no external shear is considered, then the system \lensname\ was found to be along a LOS with approximately no overdensity within $\sim$1\% uncertainty. We provide the $\kappa_{\rm ext}$ \ajsiv{re-weighted based on the best-fit external shear magnitudes from our lens models} in Sects. \ref{sec:glee_modelling} and \ref{sec:lenstronomy_modelling}.} \citet{Buckley-Geer20} also find that no nearby LOS perturbers are significant enough that they need to be included explicitly in the lens mass modelling.

\section{Setup of baseline models} \label{sec:baseline_models}

In this section we describe the baseline models that were initially agreed upon by the two teams before performing separate and independent lens modelling. \lreply{In our baseline models, we use two families of mass models for the central deflector: (i) a power-law profile, and (ii) a composite profile with an elliptical NFW potential for the dark component and a \ajs{superposition of three Chameleon profiles (hereafter, triple Chameleon profile) in convergence} for the luminous component. We also add external shear to both types of mass model. For the light profile of the central deflector, we adopt a triple S\'ersic profile in all three bands in the models with the power-law mass profile. In the models with the composite mass profile, however, we adopt a triple Chameleon light profile in the F160W band linked with the triple Chameleon mass profile and a triple S\'ersic profile in the UVIS bands.}

%

\ajs{We adopted three Chameleon profiles to sufficiently account for the complexity in the light profile of the deflector. Moreover, we adopted the triple Chameleon light profile only for the F160W profile, since this is the only band that is connected to the luminous component of the convergence profile.}

\ajs{Although both teams adopted these baseline models, individual teams were allowed to make their own choices -- which may not necessarily be identical -- pertaining to other model specifications, for example parameter priors and fixing parameter values.}

In the next subsections we provide the definitions of the mass and light profiles in the baseline models.

\subsection{Mass profiles}

The two baseline lens model families we adopt are the power-law mass profile and the composite mass profile.

\subsubsection{Power-law mass profile}
We adopted the power-law elliptical mass distribution \citep[PEMD;][]{Barkana98} defined as
\begin{equation}
        \kappa_{\rm PL}(\theta_1, \theta_2) \equiv \frac{3 - \gamma}{2} \left[ \frac{\theta_{\rm E}}{\sqrt{q_{\rm m} \theta_1^2 + \theta_2^2/q_{\rm m}}} \right]^{\gamma - 1},
\end{equation}
where $\gamma$ is the logarithmic slope, $\theta_{\rm E}$ is the Einstein radius, and $q_{\rm m}$ is the axis ratio. The coordinates $(\theta_1,\ \theta_2)$ are in the coordinate frame that is aligned with the major and minor axes. The position angle of this frame is $\varphi_{\rm m}$ with respect to the RA--Dec frame.

\subsubsection{Composite mass profile}
The composite mass profile consists of two individual mass profiles for the baryonic and the dark components of the mass distribution. 

For the dark matter distribution, we adopt a Navarro--Frenk--White (NFW) profile with ellipticity defined in the potential. The 3D NFW profile in the spherical case is given by
\begin{equation}
        \rho_{\rm NFW}(r) \equiv \frac{\rho_{\rm s}}{\left(r/r_{\rm s}\right) \left(1 + r/r_{\rm s}\right)^2},
\end{equation}
where $\rho_{\rm s}$ is the density normalization, and $r_{\rm s}$ is the scale radius \citep{Navarro97}. We refer to \citet{Golse02} for the expressions of the lens potential and deflection angles associated with the elliptical NFW profile.

For the baryonic mass distribution, we adopt the Chameleon convergence profile. The Chameleon profile matches with the S\'ersic profile within a few per cent at 0.5--3$\theta_{\rm eff}$, where $\theta_{\rm eff}$ is the half-light or effective radius of the S\'ersic profile \citep{Dutton11}. The Chameleon profile is defined as the difference between two non-singular isothermal ellipsoids:
\begin{equation} \label{eq:chameleon}
        \begin{split}
        \kappa_{\rm Chm} (\theta_1, \theta_2) &\equiv \frac{a_0}{1 + q_{\rm m}} \left[ \frac{1}{\sqrt{\theta_1^2+\theta_2^2/q_{\rm m}^2 + 4w_{\rm c}^2/(1+q_{\rm m}^2)}} \right. \\
         & \quad \quad \quad \quad \left. - \frac{1}{\sqrt{\theta_1^2+\theta_2^2/q_{\rm m}^2 + 4w_{\rm t}^2/(1+q_{\rm m}^2)}} \right],
\end{split}
\end{equation}
where \ajs{$a_0$ is the normalization and $w_c$ and $w_t$ are the core sizes for the individual non-singular isothermal components in the Chameleon profile} \citep{Dutton11, Suyu14}. This profile is numerically convenient for computing lensing quantities using closed-form expressions unlike the S\'ersic profile.  


\subsection{Light profiles of the deflector}

\subsubsection{S\'ersic profile}

The S\'ersic profile is defined as
\begin{equation}
        I_{\text{S\'ersic}} (\theta_1, \theta_2) \equiv I_{\rm eff} \exp \left[-b_n \left\{\left(\frac{\sqrt{\theta_1^2 + \theta_2^2/q_{\rm L}^2}}{\theta_{\rm eff}/\sqrt{q_{\rm L}}} \right)^{1/n_{\rm s}} - 1 \right\} \right],
\end{equation}
where $I_{\rm eff}$ is the amplitude, $\theta_{\rm eff}$ is the effective radius \ajs{along the intermediate axis}, and $n_{\rm s}$ is the S\'ersic index \citep{Sersic68}. The factor $b_n$ is a normalizing factor so that $\theta_{\rm eff}$ is the half-light radius.

\subsubsection{Chameleon light profile}
In the composite baseline model, we use the same Chameleon profile from Eq. \ref{eq:chameleon} for the light profile of the deflector, but replacing the convergence amplitude $\kappa_0$ with the flux amplitude $I_0$.

\section{\textsc{glee} modelling} \label{sec:glee_modelling}

In this section we describe the \textsc{glee} modelling procedure.  \textsc{glee} is a software package developed by S. H. Suyu and A. Halkola \citep{Suyu10b,Suyu12c}.  The lensing mass distribution is described by a parameterized profile.  The lensed quasar images are modelled as point sources on the image plane convolved with the PSF.  The extended host galaxy of the \ajsiii{lensed quasar} is modelled on a $50\times50$ pixel grid with curvature regularization \citep{Suyu06}, spanning the range of source coordinates corresponding to the pixels within \ajsiii{a region containing the lensed arcs (hereafter, referred to as the `arcmask')}.  The quasar image amplitudes are independent of the extended host galaxy light distribution to allow for deviations due to microlensing, time delays, and substructure. The lens galaxy light distribution is represented as the sum of three S\'{e}rsic (or three Chameleon) profiles with a common centroid.

The lens model is constrained by the positions of the lensed quasar images and the surface brightness of the pixels of the lensed Einstein ring of the quasar host galaxy in the three HST bands that are fit simultaneously. The quasar positions are fixed to the positions of the point sources on the image plane (after they have stabilized) and are given a fixed Gaussian uncertainty of width 0\farcs004~to account for offsets due to substructure in the lens or LOS.  This uncertainty is small enough to satisfy astrometric requirements for cosmography \citep{Birrer19}. The quasar flux ratios are not used as constraints, as they can be affected by microlensing, which has been detected in this system \citep{Melo21}.  \ajsiii{We use} the initial PSF estimate in each band \ajsiii{that was} created from $\sim$4--6 bright stars within the HST image (Sect. \ref{sec:hst_imaging}).  We first model the lens separately in each band to iteratively update the respective PSFs using the lensed active galactic nucleus (AGN) images \citep{Chen16,Wong17,Rusu20}.  We then keep the `corrected' PSFs fixed and use them in our final models that simultaneously use the surface brightness distribution in all three bands as constraints.  We use the positions of the quasar images to align the cutouts in the three HST bands.  We do not enforce any similarity of pixel values at the same spatial position across different bands (i.e. the model flux at any position in one band is independent of the model flux in other bands).   In our Markov chain Monte Carlo (MCMC) sampling, we vary the light parameters of the lens galaxy and quasar images, the mass parameters of the lens galaxy, and the external shear.  The source position is also sampled in the modelling.  The quasar image positions are linked across all bands, but the other light parameters are allowed to vary independently.

\begin{figure*}
        \centering
    \includegraphics[width=\textwidth]{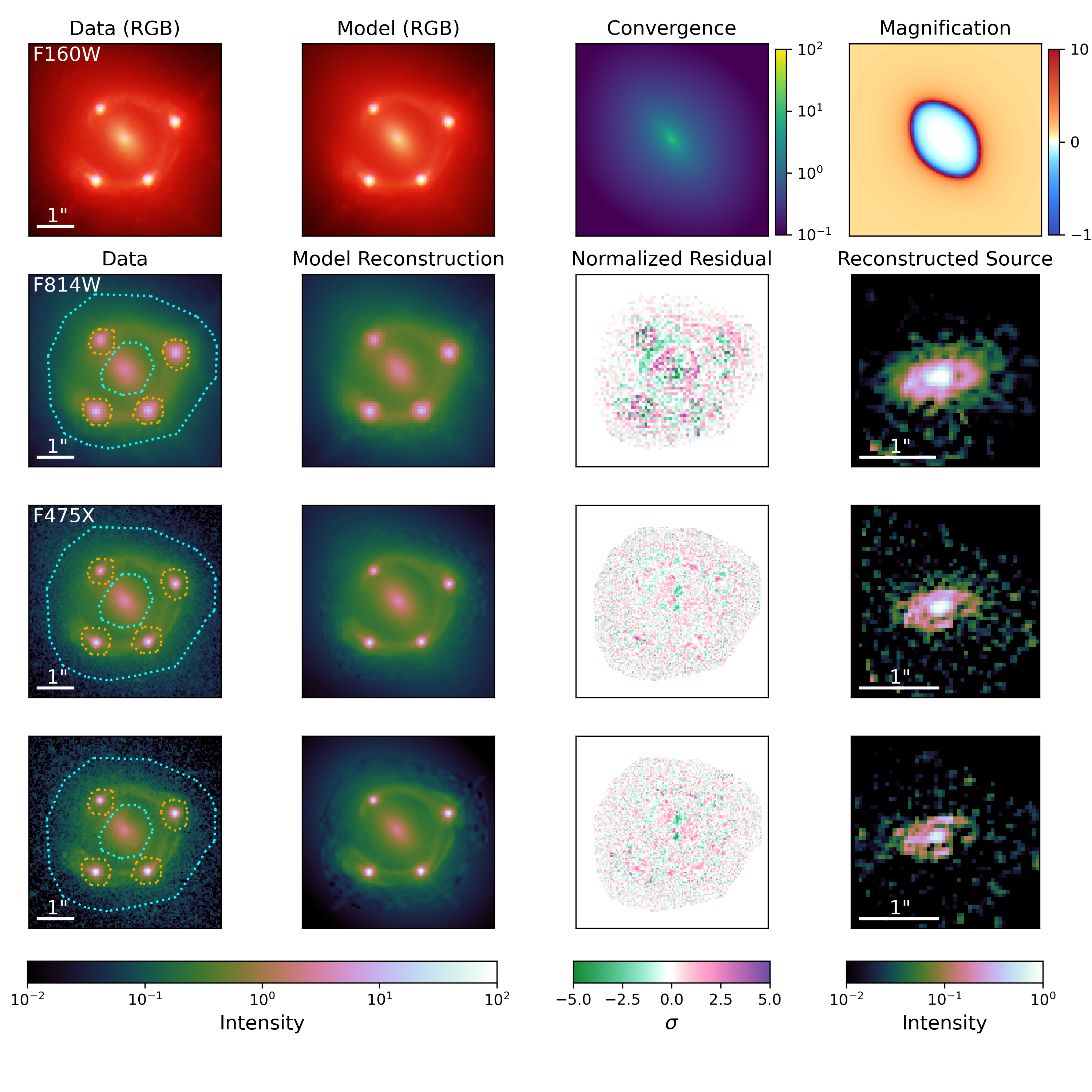}
    \caption{
    Fiducial power-law model results for IR/F160W (top row), UVIS/F814W (middle row), and UVIS/F475X (bottom row) \ajsiii{from \textsc{glee}}.  The maximum-likelihood model in the MCMC chain is shown.  Shown are the observed image (first column), the reconstructed image predicted by the model (second column), the normalized residual within and interior to the arcmask region (defined as the difference between the data and model, normalized by the estimated uncertainty of each pixel; third column), and the reconstructed source (right column).  In the first column, the dotted cyan lines indicate the arcmask (donut-shaped) region used for fitting the extended source, the dotted orange lines indicate the AGN mask region where the power-law weighting is applied, and the region outside the dotted cyan arcmask is used to further constrain the foreground lens light and (partly) the AGN light (but not the AGN host galaxy light since its corresponding lensed arcs are below the noise level in this outer region). The colour bars show the scale in the respective panels. The results shown here are for the fiducial power-law model, but the results for the other systematics tests (Sect.~\ref{subsec:glee_sys_test}) are qualitatively similar.
    }
    \label{fig:glee_rgb_pl}
\end{figure*}

We create cutouts of the HST images with dimensions of $5\farcs6 \times 5\farcs6$, which corresponds to a $140\times140$ pixel cutout for the UVIS/F475X and UVIS/F814W bands and a $70\times70$ pixel cutout for the IR/F160W band.  This conservative cutout size is chosen to include the entire region containing the lensed host galaxy arc light.  We define \ajs{the} arcmask around the \ajsiii{deflector galaxy} in each of the three bands, which encloses the region where we reconstruct the lensed arc from the extended quasar host galaxy.  The arcmask is used to calculate the likelihood involving the reconstructed lensed arc light, but the whole cutout is used for calculating the likelihood associated with the lens light.  The construction of the weight images and bad pixel masking for each cutout are analogous to the procedure in \citet{Wong17} and \citet{Rusu20}.  In order to avoid biasing the modelling due to large residuals from a PSF mismatch near the AGN image \ajsiv{positions}, we rescale the weights in those regions by a power-law model such that a pixel originally given an estimated 1$\sigma$ noise value of $\sigma_{\mathrm{img},i}$ is rescaled to a noise value of $A \times \sigma_{\mathrm{img},i}^{b}$.  The constants $A$ and $b$ are chosen for each band such that the normalized residuals (the residual flux of each pixel normalized by its 1$\sigma$ uncertainty) in the AGN image regions are approximately consistent with the normalized residuals in the rest of the arc region. \reply{We do not rescale the weights outside of the AGN image regions.} The arcmask region and the regions around the AGN with rescaled weights are shown in the first column of Fig.~\ref{fig:glee_rgb_pl}.

\subsection{Power-law model} \label{subsec:glee_pl_model}
Our fiducial power-law mass model uses the triple S\'{e}rsic parameterization for the lens galaxy light and has the additional free parameters: (i) position $(\theta_{1}, \theta_{2})$ of the mass centroid (allowed to vary independently from the centroid of the light distribution),
(ii) Einstein radius $\theta_{\mathrm{E}}$, (iii) minor-to-major axial ratio, \ajsiii{$q_{\rm m}$}, and associated position angle \ajsiii{${\varphi}_{\rm m}$} (measured east of north),
(iv) 3D slope of the power-law mass distribution $\gamma$, and (v) external shear $\gamma_{\mathrm{ext}}$ and associated position angle \ajsiii{$\varphi_{\rm ext}$} (measured east of north). 

We assume uniform priors on the model parameters over a wide physical range.  \lreply{Fig.}~\ref{fig:glee_rgb_pl} shows the data and the lens model results in all three bands for our fiducial power-law model, as well as the reconstructed sources.  Our model simultaneously reproduces the surface brightness structure of the lensed AGN and host galaxy in all bands.  The normalized residual in the third column shows the area within the arcmask, as well as the region interior to the arcmask.  In the IR/F160W band, there is an excess residual at the inner boundary of the arcmask (as well as outside of the arcmask, not shown in this figure) arising from the technical details of the PSF not being corrected outside of the arcmask.  We run a test where the pixels showing excess residual outside of the arcmask are downweighted and find no significant change in the model parameters.

\subsection{Composite model} \label{subsec:glee_comp_model}
Our composite model consists of a baryonic component linked to the light profile of the lens galaxy, plus a dark matter component.  The composite model assumes the triple Chameleon light profile for the lens galaxy in the IR/F160W band scaled by an overall mass-to-light (M/L) ratio.  The Chameleon light profiles link to parameters describing the light distribution to those of the mass distribution in a straightforward way, as they are fundamentally just a combination of isothermal profiles.  We keep the triple S\'{e}rsic model for the lens galaxy light in the UVIS bands to maintain consistent parameterization with the power-law models.  The dark matter component is modelled as an elliptical NFW \citep{Navarro96} halo with the centroid linked to the light centroid in the F160W band.

The fiducial composite model has the following free parameters in addition to the lens light parameters:
(i) \ajsiii{mass-to-light ratio ($M/L$)} for the baryonic component,
(ii)  NFW halo scale radius, $r_{\mathrm{s}}$, (iii) NFW halo normalization, $\kappa_{0,\mathrm{h}}$ \citep[defined as $\kappa_{0,\mathrm{h}} \equiv 4\kappa_{\mathrm{s}} \equiv 4 \rho_{\rm s} r_{\rm s} \Sigma_{\mathrm{crit}}^{-1}$;][]{Golse02},
(iv) NFW halo minor-to-major axial ratio, \ajsiii{$q_{\rm NFW}$}, and associated position angle, \ajsiii{$\varphi_{\rm NFW}$}, and (v) external shear, $\gamma_{\mathrm{ext}}$, and associated position angle, \ajsiii{$\varphi_{\rm ext}$}.

\begin{figure*}
        \centering
    \includegraphics[width=\textwidth]{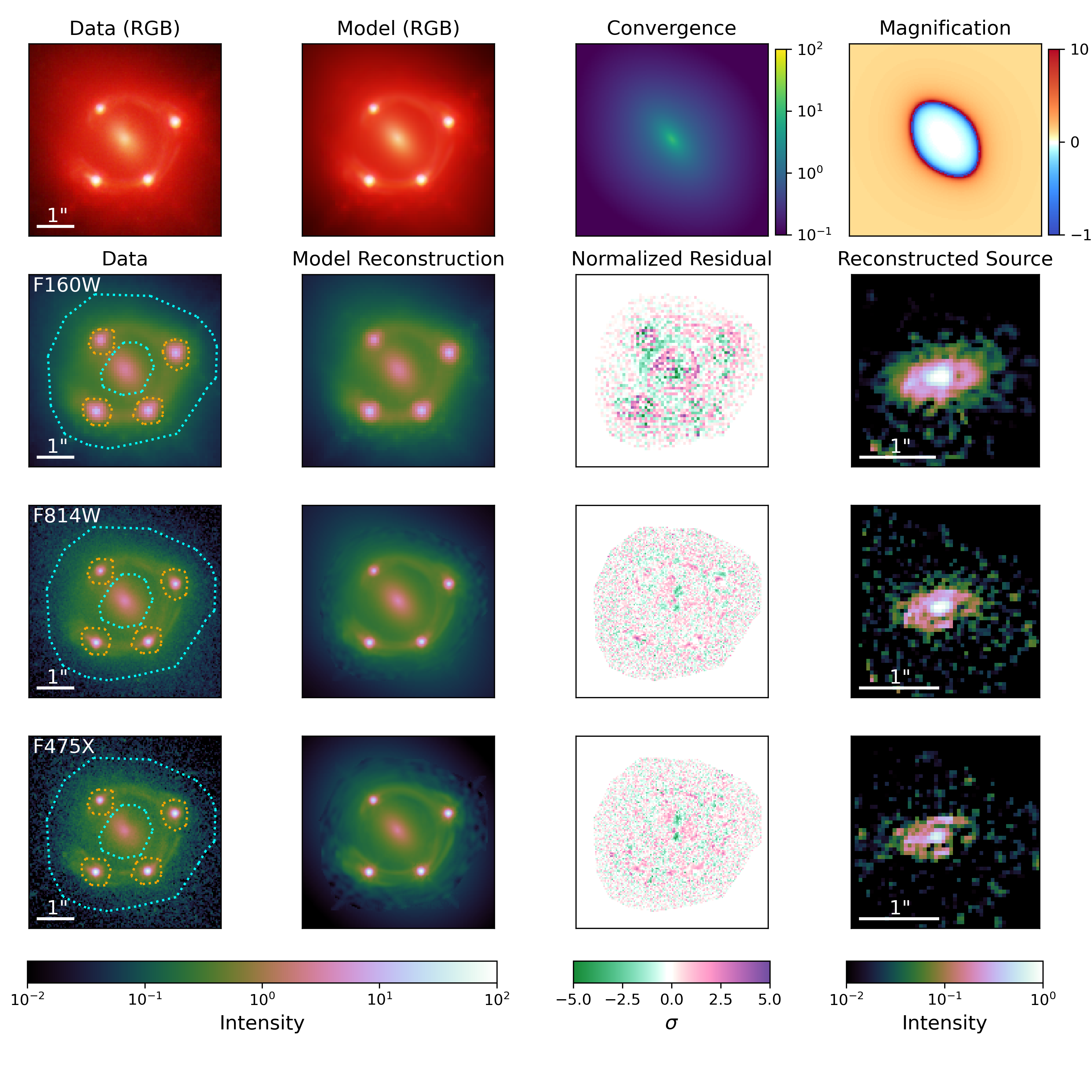}
    \caption{
    Same as Fig.~\ref{fig:glee_rgb_pl}, but for the fiducial composite model \ajsiii{from \textsc{glee}}.
    }
    \label{fig:glee_rgb_comp}
\end{figure*}

A Gaussian prior for the \ajsiii{$M/L$} of the baryonic component is employed, using the stellar mass constraint from \citet{Agnello18c} of $\log_{10} (M_{\star}/M_{\odot}) = 11.40^{+0.01}_{-0.08}$ for a Salpeter initial mass function (IMF).  Although this value is lower than our estimate derived from the photometry of our models of the lens light profile (see Sect.~\ref{sec:lenstronomy_modelling}), this prior has little influence on the result, as the model prefers an almost maximal M/L with little dark matter contribution (see Sect.~\ref{subsec:glee_results}).  We set a Gaussian prior of $r_{\mathrm{s}} = 22\farcs6 \pm 3\farcs1$ based on the results of \citet{Gavazzi07} for a sample of lenses in the SLACS survey \citep[][]{Bolton06}.  These lenses span a redshift and velocity dispersion range that includes \lensname, with a mean virial mass of $\langle M_{\mathrm{vir}} \rangle = 1.4_{-0.5}^{+0.6} \times 10^{13}~h^{-1}~M_{\odot}$.  All other parameters are given uniform priors.  The relative amplitudes of the three Chameleon profiles representing the stellar light distribution of the lens galaxy can vary within an MCMC chain. However, their relative amplitudes in the mass model initialization are necessarily fixed (due to the way that the \textsc{glee} user interface is set up), even though they share the same global \ajsiii{$M/L$} parameter. To account for this, we iteratively run a series of MCMC chains for the fiducial composite model and update the relative amplitudes of the three mass components to match that of the light components after each chain.  After several iterations, the predicted Fermat potential stabilizes, and we stop iterating.  We subsequently ran a test fiducial model using an updated version of \textsc{glee} in which the amplitudes of the mass components are directly linked to the light components and found that the results were unchanged.  \lreply{Fig.}~\ref{fig:glee_rgb_comp} shows the data and the lens model results in all three bands for our fiducial composite model, as well as the source reconstructions.

\subsection{Systematics tests} \label{subsec:glee_sys_test}
In this section we describe a variety of tests of the effects of various systematics in our modelling arising from different assumptions in the way we constructed the model that might affect the posterior.  In addition to the basic fiducial models described above, we perform inferences for both the power-law and composite models given the following sets of assumptions:
(i) a model where the regions near the AGN images are given zero weight rather than being scaled by a power-law weighting; (ii) a model where the region near the AGN images scaled by the power-law weighting is increased by one pixel around the outer edge; (iii) a model where the reconstructed source plane resolution in all bands is reduced to $40\times40$ pixels; (iv) a model where the reconstructed source plane resolution in all bands is increased to $60\times60$ pixels; and (v) a model with the arcmask region increased by one pixel on both the inner and outer edges.
We combined the MCMC chains from all of these tests, weighted by the BIC \citep[similar to][see Sect.~\ref{subsec:glee_bic}]{Rusu20}.

\subsection{Kinematics and external convergence} \label{subsec:glee_kin_kext}
We used the kinematics and external convergence constraints from \citet{Buckley-Geer20}.  We combined both LOS velocity dispersion measurements to constrain the lens models.  \citet{Buckley-Geer20} constrain the external convergence for different external shear amplitudes in steps of 0.01.  For each model, we use the distribution corresponding to the external shear that is closest to the median amplitude for that model.  We use importance sampling \citep[e.g.][]{Lewis02} to simultaneously combine the velocity dispersion and external convergence distributions in a manner similar to \citet{Wong17} and \citet{Rusu20}. For each set of lens parameters ${\nu}$ from our lens model chain, we draw a $\kappa_{\mathrm{ext}}$ sample from the distributions in \citet{Buckley-Geer20} and a sample of $r_{\mathrm{ani}}$ from the uniform distribution \ajsiii{$[0.5,5]\theta_{\mathrm{eff}}$ ($\theta_{\mathrm{eff}}$} is calculated from the lens light distribution in the IR/F160W band from the power-law model). From these together with the $\Dds/\Dd$ ratio (that is fixed given the fixed $\Omega_{\rm m}$ value of 0.3 in flat $\Lambda$CDM), we can compute the kinematics likelihood for the joint sample $\{\nu, \Omega_{\rm m}, \kappa_{\mathrm{ext}}, r_{\mathrm{ani}}\}$ via Eq. \ref{eq:vdisp_ap} and use this to weight the joint sample.  We can then combine the Fermat potential computed from our lens model parameters $\nu$ with values of $\kappa_{\mathrm{ext}}$ and $\Ddt$ to predict the time delays as a function of $H_0$ (via Eqs. \ref{eq:time_delay_shortform} and \ref{eq:time_delay_true}).

\subsection{BIC weighting} \label{subsec:glee_bic}

We weight our models using the BIC, defined in Eq.~(\ref{eq:bic_definition}).  We take $N_{\mathrm{data}}$ (the number of data points) to be the number of pixels in the image region across all three bands that are outside the fiducial AGN mask (so that we are comparing equal areas), plus eight (for the four AGN image positions), plus one (for the velocity dispersion).  $k$ (the number of free parameters) is taken to be the number of parameters in the model that are given uniform priors, plus two (for the source position), plus one (for the anisotropy radius to predict the velocity dispersion).  $\hat{L}$ (the maximum likelihood of the model from the MCMC sampling) is the product of the AGN position likelihood, the pixellated image plane likelihood, and the kinematic likelihood. The image plane likelihood is the Bayesian evidence of the pixelated source intensity reconstruction using the imaging data within the arcmask \citep[which marginalizes over the source surface brightness pixel parameters and is thus the likelihood of the lens parameters excluding the source pixel parameters; see Eqs. 12 and 13 in][]{Suyu10b} multiplied by the likelihood of the lens model parameters within the image plane region that excludes the arcmask.  We evaluate the BIC using the fiducial weight image and arcmask, as the majority of the models were optimized with these.

We estimate the variance in the BIC, $\sigma_{\mathrm{BIC}}^{2}$, by sampling the fiducial model with source resolutions of [47, 48, 49, 50, 51, 52, 53, 54, 56, 58, 60] pixels on a side (the $50\times50$ pixel case is just the original fiducial model), keeping the arcmask the same. Changing the source resolution in this way shifts the predicted time delays stochastically, but there is no overall trend with resolution, and the degree of the shifts are smaller than the scatter among the different models in the systematics tests we run.  We calculate the BIC for each of these models with different source resolutions and take the variance of this set of models as $\sigma_{\mathrm{BIC}}^{2}$.  We find $\sigma_{\mathrm{BIC}}^{2} \sim 36$ for the power-law models and $\sigma_{\mathrm{BIC}}^{2} \sim 34$ for the composite models.

To avoid biases due to our choice of lens model parameterization, we split the samples into the power-law and composite models and calculate the relative BIC and weighting for each set separately, similar to \citet{Birrer19b} and \citet{Rusu20}. Specifically, we weight a model with a given BIC of value $x$ by a function $f_{\mathrm{BIC}}(x)$, defined as the convolution
\begin{equation} \label{eq:f_bic}
f_{\mathrm{BIC}}(x) = h(x,\sigma_{\mathrm{BIC}}) \ast \mathrm{exp} \left( -\frac{x-\mathrm{BIC_{min}}}{2} \right),
\end{equation}
where $\mathrm{BIC_{min}}$ is the smallest BIC value within a set of models (power-law or composite), and $h$ is a Gaussian centred on $x$ with a variance of $\sigma_{\mathrm{BIC}}^{2}$.  \kw{The exponential term is a proxy to the evidence ratio.}  We follow the calculation of \citet{Yildirim20} in evaluating the convolution integral in Eq.~(\ref{eq:f_bic}). Once we weighted time delay distributions for the power-law and composite models, we combined these two with equal weight in the final inference.

\subsection{Modelling results with $\lambda_{\mathrm{int} = 1}$} \label{subsec:glee_results}
\begin{figure*}
        \centering
    \includegraphics[width=\textwidth]{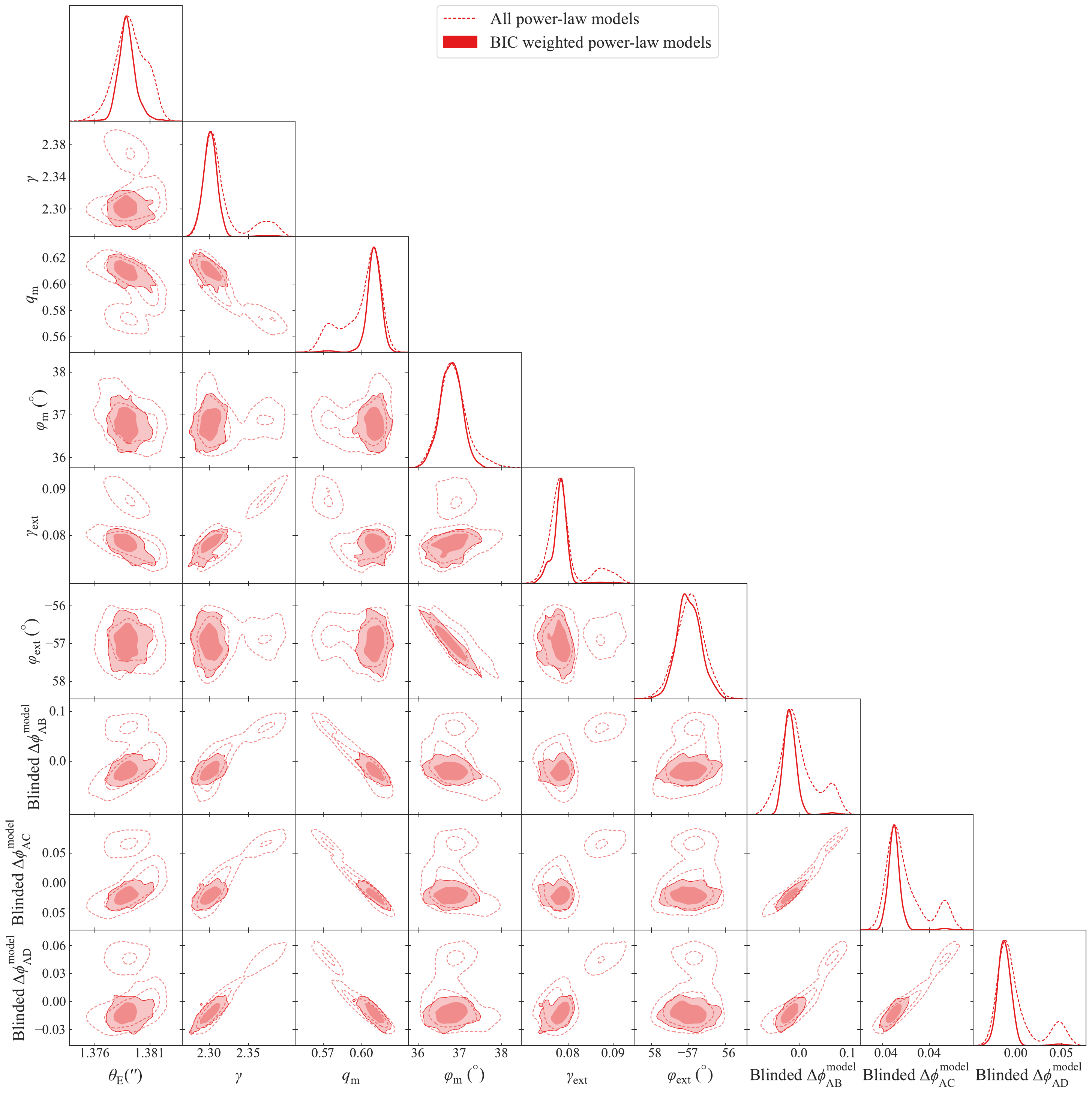}
    \caption{
    Marginalized parameter distributions from our power-law lens model results \ajsiii{from \textsc{glee}}. We show the combined results from our systematics tests (dashed red contours) with each model weighted equally, as well as the BIC-weighted model results (shaded red contours). The contours represent the 68.3 per cent and 95.4 per cent quantiles. 
    }
    \label{fig:glee_corner_spemd}
\end{figure*}

\begin{figure*}
        \centering
    \includegraphics[width=\textwidth]{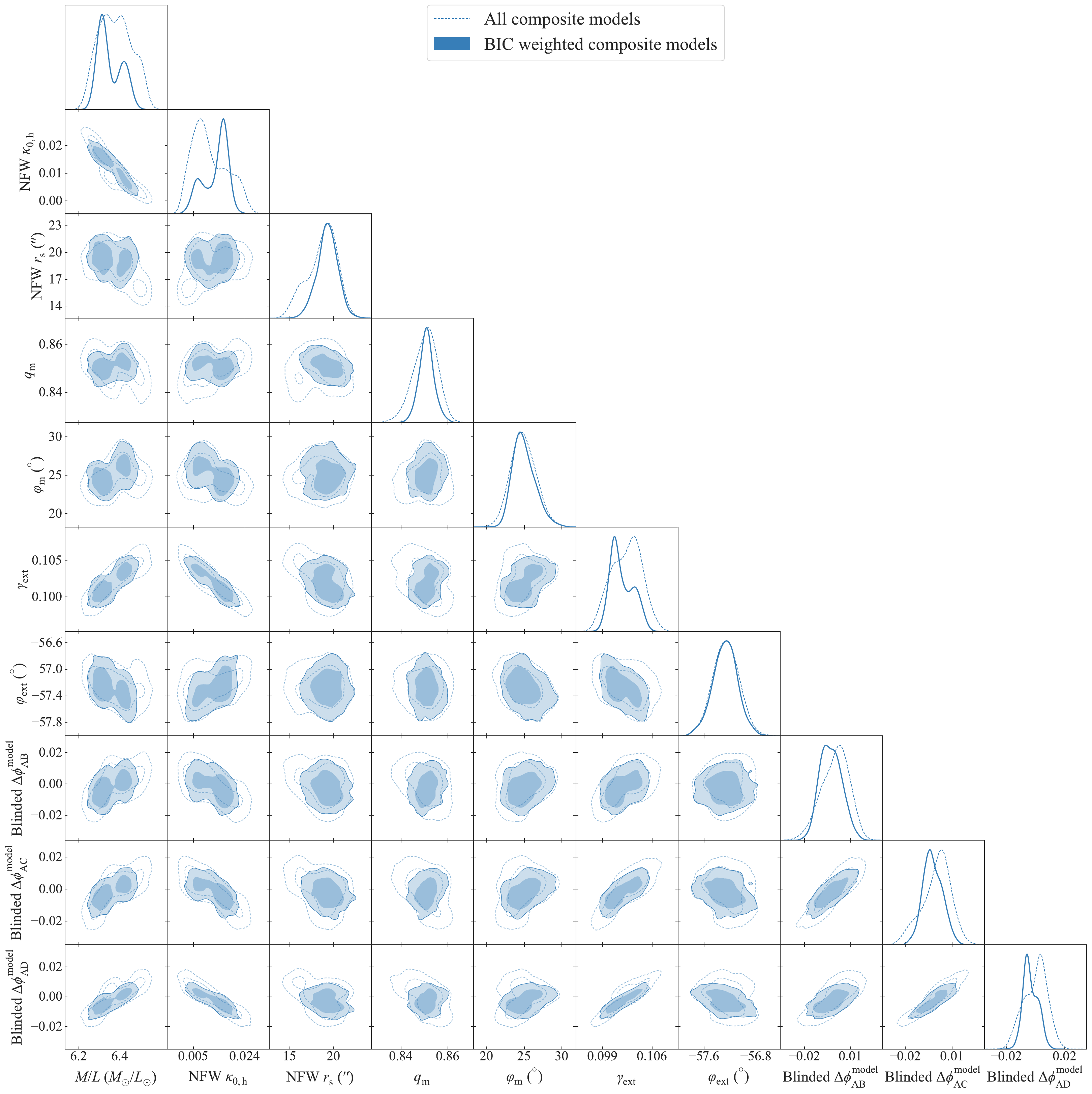}
    \caption{
    Marginalized parameter distributions from our composite lens model results \ajsiii{from \textsc{glee}}. We show the BIC-weighted model (shaded blue contours) and the combined results from our systematics tests (dashed blue contours). The contours represent the 68.3 per cent and 95.4 per cent quantiles. 
    }
    \label{fig:glee_corner_comp}
\end{figure*}

The marginalized parameter distributions of the power-law model are shown in Fig.~\ref{fig:glee_corner_spemd}.  We show the combined distributions of all power-law models where each model is given equal weight, as well as the BIC-weighted distribution.  \lreply{Fig.}~\ref{fig:glee_corner_comp} shows the similar parameter distribution for the composite models. \reply{The point estimates for the mass model parameters from the \textsc{glee} models are presented and compared with those from the \textsc{lenstronomy} models later in Sect. \ref{sec:model_param_compare}.} 
The reconstructed sources of each model are qualitatively very similar, which is an important consistency check of the two models.

The power-law model has a steep mass profile slope of \ajsiii{$\gamma = 2.30 \pm 0.01$}, but the parameters are consistent with the previous model of \citet{Shajib19}.  The various systematics tests do not show substantial variation.  The `island'-like feature in Fig. ~\ref{fig:glee_corner_spemd} comes from the model with a lower source plane resolution, but this model is downweighted by the BIC, so it does not affect our result.  The centroid of the mass and light profiles are consistent to within $\sim 0\farcs003$, and the model is able to fit the quasar positions to an rms of $\sim0\farcs005$.

\begin{figure}
        \centering
    \includegraphics[width=0.45\textwidth]{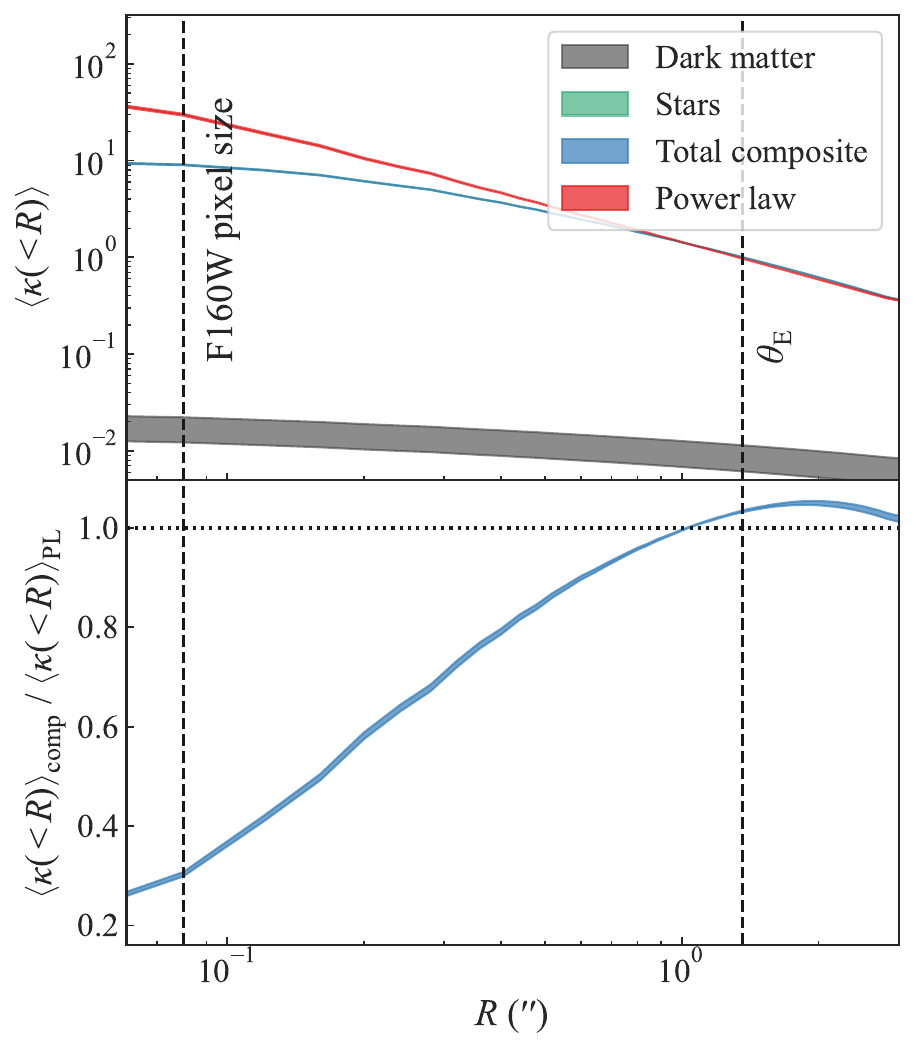}
    \caption{
    \lreply{Radial mass profiles of the central deflector constrained by the \textsc{glee} models.} {\bf Top:} Circularly averaged convergence $\langle \kappa(<R) \rangle$ as a function of radius for the \textsc{glee} power-law model (\ajsiv{red}) and composite model (\ajsiv{blue}). \reply{The shaded regions represent the 1$\sigma$ credible regions.}  The stellar (\ajsiv{green}) and dark matter (\ajsiv{black}) components of the composite model are plotted separately. \ajsiv{The vertical dashed black lines mark the pixel size in the F160W band and the best fit Einstein radius.} {\bf Bottom:} Ratio of average convergence of the composite model to that of the power-law model as a function of radius.
    }
    \label{fig:glee_kappa_profile}
\end{figure}

The composite model fits the quasar positions to an rms of $\sim0\farcs01$, slightly worse than the power-law model.  We note that the dark matter component contributes a very small fraction of the mass (of order $\sim1\%$) relative to the stellar component, which has a large mass-to-light ratio.  While this may appear unusual, the stellar mass enclosed within the Einstein radius determined from stellar population synthesis (SPS) models fit to the imaging data assuming a Salpeter IMF is consistent with the total enclosed mass as constrained by the lensing.  In Fig.~\ref{fig:glee_kappa_profile}, we show the circularly averaged convergence of both the power-law and composite models.  The effective Einstein radii (at which \ajsiv{$\langle{\kappa}(< r)\rangle = 1$}) of the two models agree to within less than one UVIS pixel ($0\farcs04$), which corresponds to $\sim 2-3\%$.  At the Einstein radius, the composite model slope closely matches the slope of the power-law model. The magnitude of the external shear ($\gamma_{\mathrm{ext}}$) required for the power-law and composite models differs, resulting in a difference in the external convergence ($\kappa_{\mathrm{ext}}$) as determined by \citet{Buckley-Geer20}.

\begin{figure*}
        \centering
    \includegraphics[width=\textwidth]{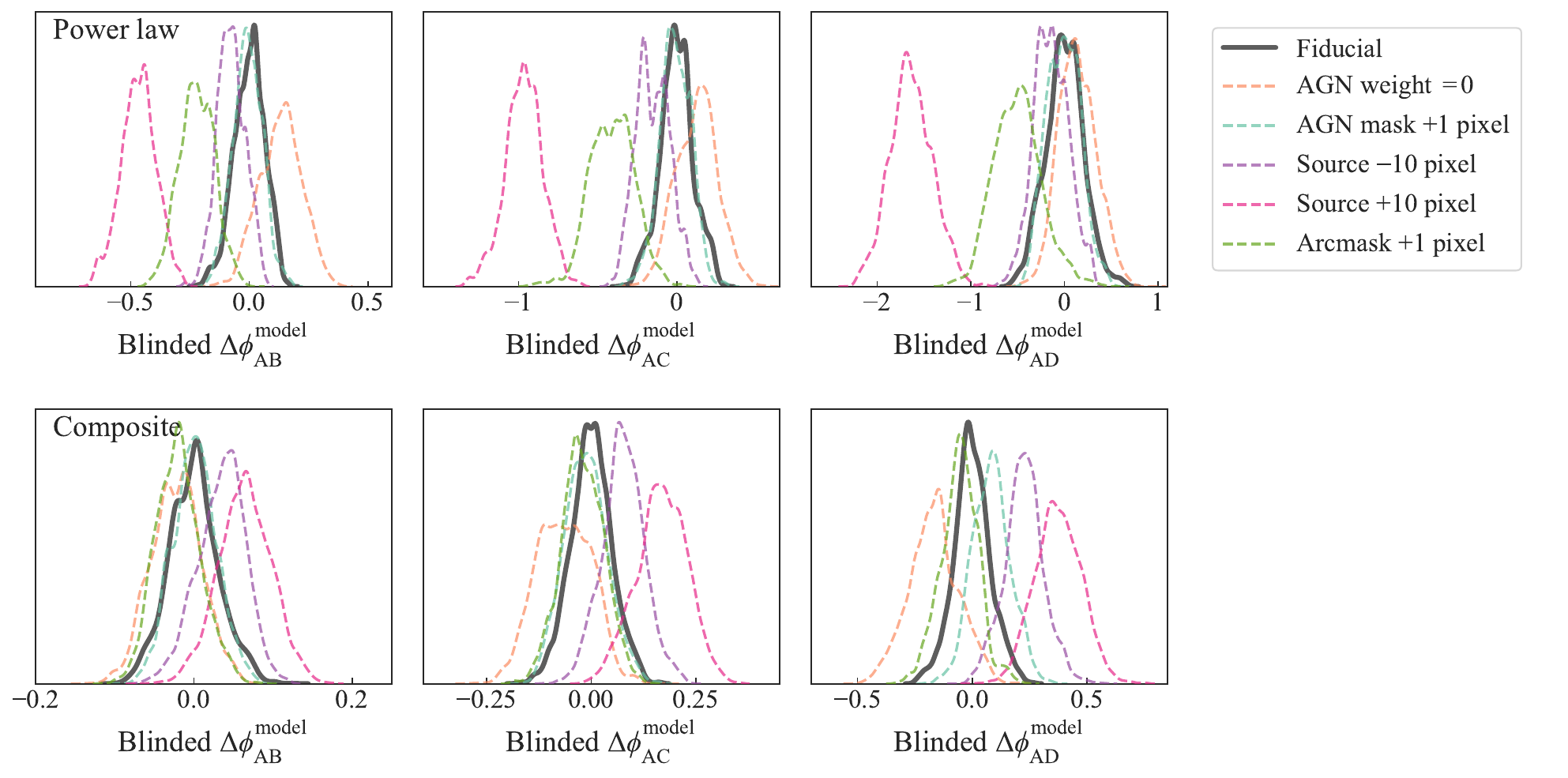}
    \caption{
    \ajsiii{Model-predicted distributions of Fermat potential differences (blinded)} for each of the \textsc{glee} models tested, with power-law models (top) and composite models (bottom). 
    }
    \label{fig:glee_td_indiv}
\end{figure*}

\reply{

The relative BIC weightings of each model are provided in Table~\ref{tab:glee_td_bic}}. The \ajsiii{blinded distributions of Fermat potential differences} are plotted individually for each model in Fig.~\ref{fig:glee_td_indiv}. The \ajsiii{un-blinded illustrations of} the BIC-weighted distributions are \ajsiii{provided later in Sect. \ref{sec:time_delay_comapare}}. 
Notably, the power-law and composite model have predicted time delays that are offset by $\sim 13\%$, indicating a difference in the two models.  Contributing to this difference is the larger $\kappa_{\mathrm{ext}}$ for the composite model.  As a result, the combined constraint has a larger uncertainty, reflecting this difference.  Without factoring in the different $\kappa_{\mathrm{ext}}$ distributions, the power-law and composite models would be offset by $\sim 8\%$.

\renewcommand*\arraystretch{1.5}
\begin{table}
\caption{BIC weighting for different lens models \ajsiv{from \textsc{glee}. The $\Delta$BIC values are calculated relative to the model with the lowest BIC value.} \label{tab:glee_td_bic}}
\begin{minipage}{\linewidth}
\begin{tabular}{lcc}
\hline
Model setting &
$\Delta$BIC &
BIC weight \\
\hline
\multicolumn{3}{c}{Power-law ellipsoid model} \\
\hline
Fiducial &
0 &
0.661
\\
AGN mask weight = 0 &
223 &
0.000
\\
AGN mask + 1 pix &
26 &
0.324
\\
$40\times40$ source &
84 &
0.015
\\
$60\times60$ source &
179 &
0.000
\\
Arcmask + 1 pix &
295 &
0.000
\\
\hline
\multicolumn{3}{c}{Composite model} \\
\hline
Fiducial &
34 &
0.218
\\
AGN mask weight = 0 &
252 &
0.000
\\
AGN mask + 1 pix &
45 &
0.132
\\
$40\times40$ source &
424 &
0.000
\\
$60\times60$ source &
0 &
0.650
\\
Arcmask + 1 pix &
137 &
0.000
\\
\hline
\end{tabular}
\\
\end{minipage}
\end{table}
\renewcommand*\arraystretch{1.0}

\section{Lenstronomy modelling} \label{sec:lenstronomy_modelling}

In this section we describe the \textsc{lenstronomy} model setups and modelling results. The software package \textsc{lenstronomy} \citep{Birrer18, Birrer21b} is a publicly available lens modelling software.\footnote{\faGithub\ \url{https://github.com/lenstronomy/lenstronomy}}. \ajsiv{In contrast} with \textsc{glee}, the software \textsc{lenstronomy} uses basis sets to reconstruct the flux distribution of the background source galaxy \citep{Birrer15}. In this section we describe the specific model settings for \textsc{lenstronomy} on top of the baseline models from Sect. \ref{sec:baseline_models}, then present our modelling results, and lastly combine the lens models with the measured stellar kinematics and the estimated external convergence.

\subsection{\textsc{lenstronomy} specific model settings}
We explain particular model settings related \ajs{to} the mass and light profiles of the deflector galaxy in Sect. \ref{sec:lenstronomy_deflector_settings}, the source light profiles in Sect. \ref{sec:lenstronomy_source_profiles}, and the image region for likelihood computation in Sect. \ref{sec:lenstronomy_mask_settings}. We summarize the set of all the lens models combining these different settings in Sect. \ref{sec:lenstronomy_settings_combination}.

\subsubsection{Mass and light profiles of the deflector galaxy} \label{sec:lenstronomy_deflector_settings}
We simultaneously model the HST images from all three bands. We join the centroids of the triple S\'ersic profiles across the three bands in the power-law model setup, and also the centroids of the triple Chameleon profiles in the composite model setup. We join the ellipticity parameters of the light profiles only between the two UVIS bands. We let the amplitudes $I_{\rm eff}$, effective radii $\theta_{\rm eff}$, and the S\'ersic indices $n_{\rm s}$ in the three bands \ajs{independently vary to allow} for a colour gradient. 

In the composite model setup, we adopt a Gaussian prior with mean $22\farcs6$ and standard deviation $3\farcs1$ for the NFW scale radius $r_{\rm s}$ based on the measurements of \citet{Gavazzi07} for a sample of SLACS survey lens systems \citep{Bolton06}. Since the velocity dispersion and the redshift of the central deflector of \lensname\ fall within the ranges spanned by the SLACS lenses, such a prior is appropriate \citep{Treu06}. Similar priors were also adopted in previous H0LiCOW and STRIDES analyses \citep[e.g.][]{Wong17, Rusu17, Shajib20}. \ajs{Although the measurement by \citet{Gavazzi07} are reported in the physical kpc unit, we use the same fiducial cosmology as \citet{Gavazzi07} to recover the scale in the observable angular unit.} We also impose a prior on the concentration parameter using the theoretical $M_{200}$--$c$ relation from \citet{Diemer19} with an intrinsic scatter of 0.11 dex.

\subsubsection{Source light profiles} \label{sec:lenstronomy_source_profiles}

We adopt a basis set of shapelets and one elliptical S\'ersic profile to describe the flux distribution of the quasar host galaxy. The S\'ersic profile describes the smooth component of the flux distribution of the host galaxy, and the shapelets account for the non-smooth features \citep{Refregier03, Birrer15}. The number of shapelets $n_{\rm shapelets}$ depends on the maximum polynomial order $n_{\rm max}$ as $n_{\rm shapelets} = (n_{\rm max}+1)(n_{\rm max} + 2)/2$, and the spatial extent of the shapelets is characterized with a scale size $\varsigma$. We model the quasar images as point sources on the image plane. \reply{We treat the positions of the quasar images as free parameters throughout the model optimization and MCMC procedures. The point source positions are constrained directly through the likelihood of the pixel-level flux values in the imaging data. The four image positions give six independent relative positional parameters. We chose the option within \textsc{lenstronomy} to solve the lens equation to constrain six parameters out of the set of the mass model parameters from these six independent relative positional parameters.\footnote{using the \texttt{`PROFILE\_SHEAR'} solver of \textsc{lenstronomy}} These six mass model parameters then have `one-to-one' correspondence with the sampled quasar image positions. Therefore, they are not treated as non-linear parameters anymore in the optimization and sampling procedures. For the power-law model, the six parameters chosen are the PEMD's centroid RA and Dec, axis ratio $q_{\rm m}$, position angle $\varphi_{\rm m}$, Einstein radius $\theta_{\rm E}$, and the external shear angle $\varphi_{\rm ext}$. For the composite model, the six parameters chosen are the NFW profile's centroid RA and Dec, axis ratio $q_{\rm NFW}$, position angle $\varphi_{\rm NFW}$, density normalization $\rho_{\rm s}$, and the external shear angle $\varphi_{\rm ext}$.}

We join the ellipticity parameters of the source S\'ersic profiles across the three bands. The centroids of all the light profiles are also joint across the three bands. \ajs{This centroid is set at the quasar position in the source plane that is} constrained through solving the lens equations for the four image positions. The effective radii $\theta_{\rm eff}$, the S\'ersic indices $n_{\rm s}$, the shapelet scale sizes $\varsigma$ for different bands are independent of each other.

We treat $n_{\rm max}$ as a hyper-parameter and fix it for a particular model optimization. A minimum number of shapelet components is necessary to describe the complex features in the lensed arcs; however, too many shapelet components will fit the noise in the imaging data. Thus, striking a balance between these two scenarios is necessary when choosing the number of shapelet components. We adopt three choices for $\{n_{\rm max}^{\rm IR}, n_{\rm max}^{\rm UVIS} \}$: $\{7, 11\},\ \{8, 12\},\ \{9, 13\}$. 

\subsubsection{HST image region for likelihood computation} \label{sec:lenstronomy_mask_settings}

We chose a circular aperture in each band encompassing the lensed arcs centred on the lens galaxy to compute the imaging likelihood. The radii of these apertures are hyper-parameters in the model. We take two sets of choices for $\{r_{\mathcal{L}}^{\rm IR},\ r_{\mathcal{L}}^{\rm UVIS} \}$: $\{ 2\farcs2,\ 3\farcs6 \},\ \{ 2\farcs3,\ 3\farcs7 \}$ with $r_{\mathcal{L}}$. Some nearby objects (stars or smaller galaxies) are masked out if they fall within the likelihood computation region (see Figs. \ref{fig:lenstronomy_powerlaw_model} or \ref{fig:lenstronomy_composite_model} for the shape and comparative size of the likelihood computation regions).

\subsubsection{Model choice combinations} \label{sec:lenstronomy_settings_combination}

Summarizing the above sections, we have the hyper-parameter choices \lreply{(i) for the lens galaxy mass profile: power-law and composite; (ii) for the source light $\{n_{\rm max}^{\rm IR}, n_{\rm max}^{\rm UVIS} \}$: $\{7, 11\},\ \{8, 12\}$, and $\{9, 13\}$; and (iii) for the likelihood computation region radii $\{r_{\mathcal{L}}^{\rm IR},\  r_{\mathcal{L}}^{\rm UVIS} \}$: $\{ 2\farcs2,\ 3\farcs6 \}$ and $\{ 2\farcs3,\ 3\farcs7 \}$}.
%
%
Taking a combination of these choices, we have 12 different model setups. We perform the optimization with the same models setups twice. These twin runs are different due to stochasticity in the PSF reconstruction and MCMC sampling procedures, and help us assess random errors. As a result, we have 24 different optimized models, on which we perform BMA. \ajs{The light profiles from the deflector, the lensed light profiles from the quasar host galaxy, and the point sources at the quasar image positions form a linear basis set for reconstructing the observed HST imaging. As a result, the amplitudes of these profiles are linear parameters, as they can be obtained through a linear inversion for a sampled set of non-linear parameters that describe all the mass and light profiles.} There are 206--281 linear parameters and 51--54 non-linear parameters in our models.

\subsection{Modelling workflow}

For each model setting, we reconstruct the PSF in each HST band. The reconstruction is initiated from a PSF estimate with a corresponding error map created from $\sim$4--6 bright stars within the HST image. The PSF reconstruction is carried out in multiple iterations with model optimization having fixed PSFs interlaced in between the PSF reconstruction iterations \ajs{(see \citet{Birrer19b} for details, and also \citet{Chen16} for a similar algorithm}. There is an offset between the recorded coordinates between IR and UVIS images. After each iteration of PSF reconstruction, we re-align the coordinate system of the IR image with that of the UVIS images using the quasar positions \citep{Shajib19}. Thus, we have a block of three operations constructing one unit of PSF reconstruction iteration:
(i) IR-band image re-alignment,
(ii) lens model optimization,
and (iii) PSF reconstruction.

We optimized the lens model using \reply{the} particle swarm optimization (PSO) method \citep{Kennedy95}, which is implemented in \textsc{lenstronomy}. After the PSF reconstruction procedure, we performed MCMC sampling of the model posterior using \textsc{emcee} \citep{Foreman-Mackey13}, which is an affine-invariant ensemble sampler \citep{Goodman10}. We chose the number of walkers to be eight times the number of sampled parameters. We run the chain for 10000 steps. We check for convergence of the chain by manually inspecting that the median and standard deviations of the parameters within the walkers at each step has reach equilibrium for at least 1000 steps. We take the walker distribution from the last 1000 steps of the chain to be the model posterior.

\begin{figure*}
        \centering
    \includegraphics[width=1.14\textwidth]{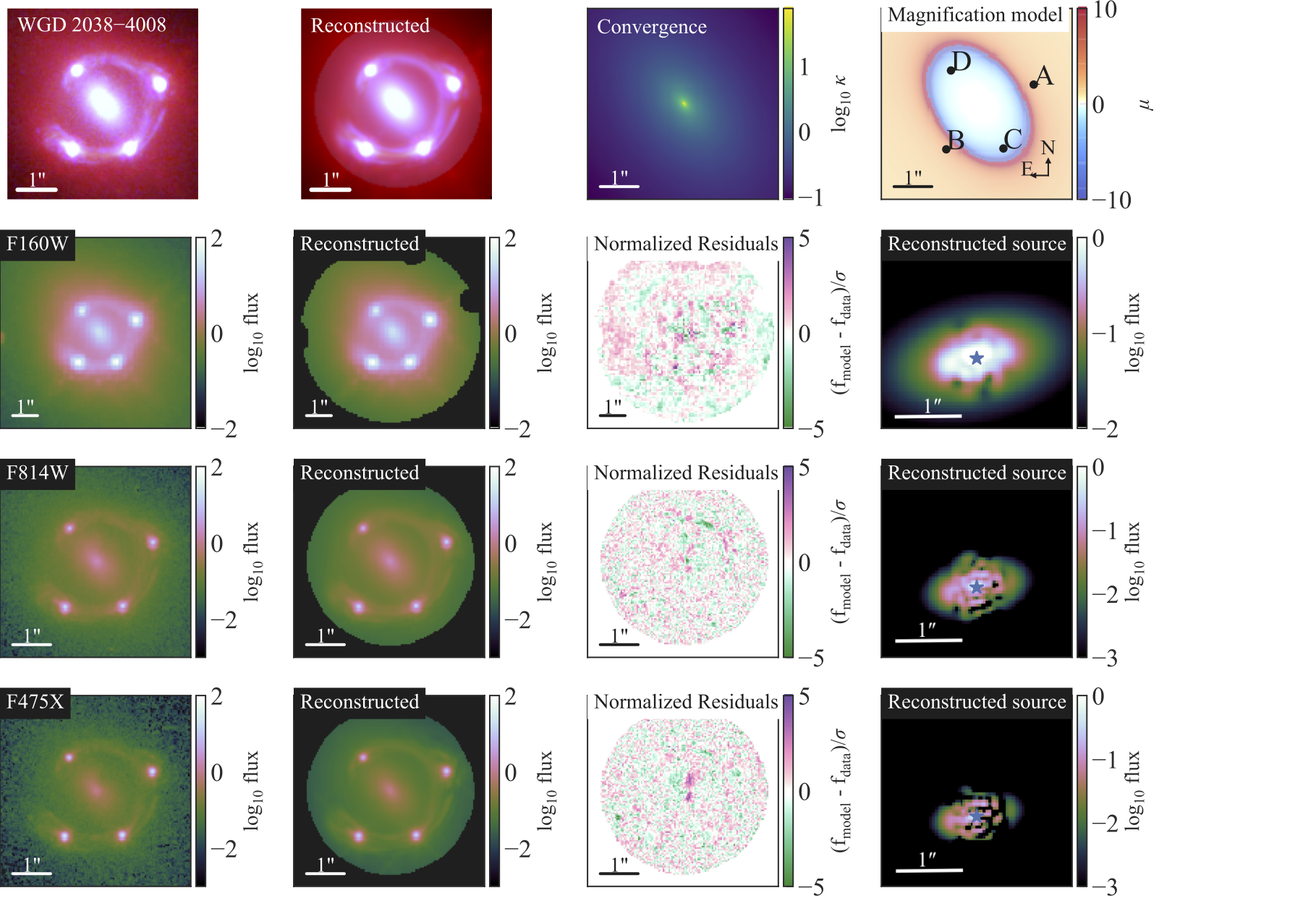}
    \caption{
    Most likely \textsc{lenstronomy} lens model and reconstructed image of \lensname\ using the power-law model. The top row shows,  from left to right, the observed RGB image, the reconstructed RGB image, the convergence profile, and the magnification model. The next three rows show, from left to right, the observed image, the reconstructed image, the residual, and the reconstructed source for each of the HST filters. The three filters are F160W (second row), F814W (third row), and F475X (fourth row). All the scale bars in each panel correspond to 1$\arcsec$. \reply{The star symbol in the reconstructed source panels marks the position of the quasar host galaxy's centroid.}
    }
    \label{fig:lenstronomy_powerlaw_model}
\end{figure*}

\begin{figure*}
        \centering
    \includegraphics[width=1.14\textwidth]{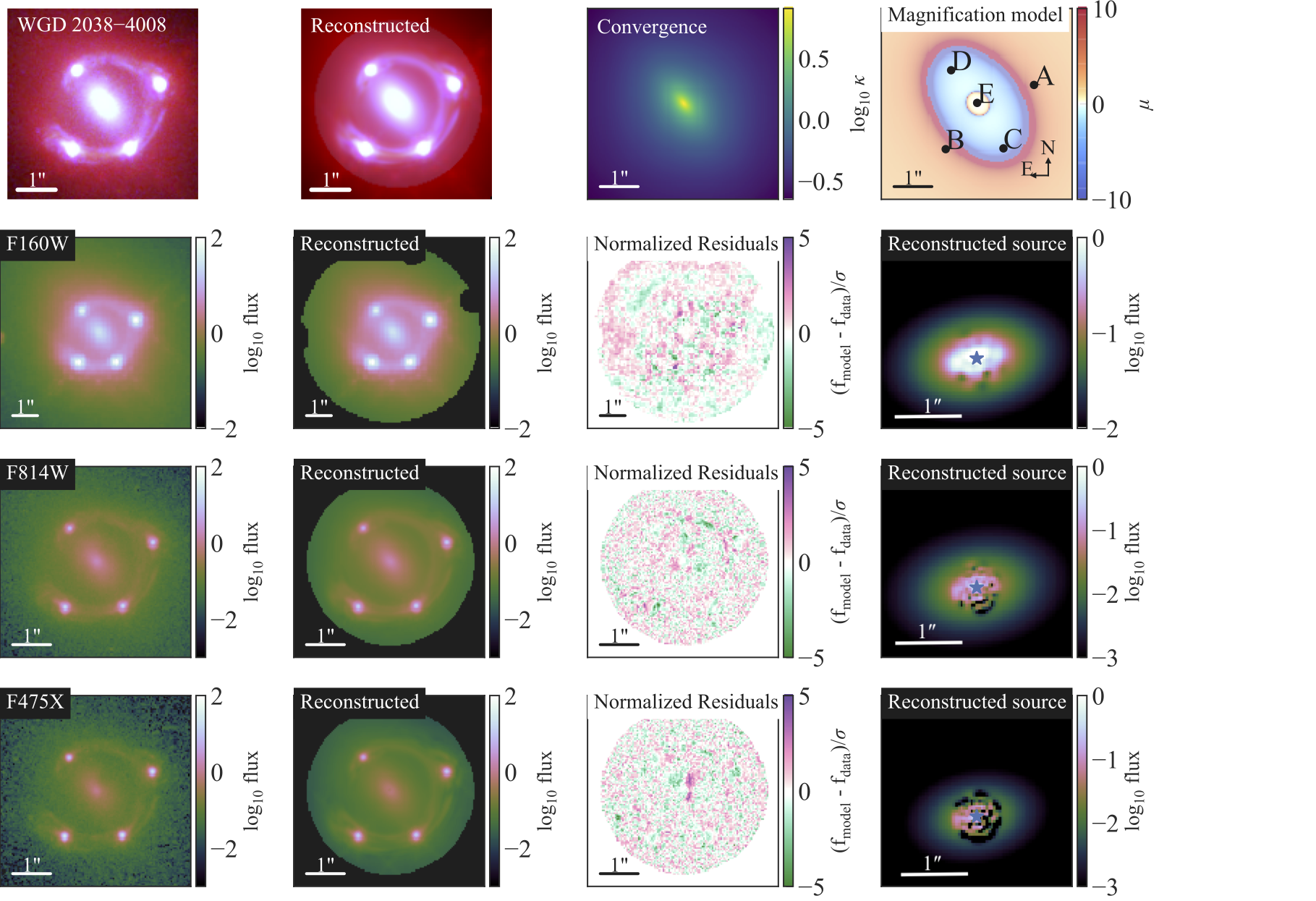}
    \caption{
    Most likely \textsc{lenstronomy} lens model and reconstructed image of \lensname\ using the composite model. The top row shows, from left to right, the observed RGB image, reconstructed RGB image, the convergence profile, and the magnification model. The next three rows show, from left to right, the observed image, the reconstructed image, the residual, and the reconstructed source for each of the HST filters. The three filters are F160W (second row), F814W (third row), and F475X (fourth row). All the scale bars in each panel correspond to 1$\arcsec$. \reply{The star symbol in the reconstructed source panels marks the position of the quasar host galaxy's centroid.} In the magnification model, a central image is predicted due to a central core in the triple Chameleon light profile. However, this central image is highly de-magnified, with magnification \reply{0.019$\pm$0.02}, and thus its presence cannot be ascertained in our imaging data. 
    }
    \label{fig:lenstronomy_composite_model}
\end{figure*}

We illustrate the best-fit model from the model setup with the \ajsiv{lowest} BIC value among the power-law and composite model families in Figs. \ref{fig:lenstronomy_powerlaw_model} and \ref{fig:lenstronomy_composite_model}, respectively.

\subsection{Bayesian model averaging} \label{sec:lenstronomy_bma}

We have 24 models that make up our set of models $\{ S, M \}$, with each lens model family from $M \equiv \{\textrm{power law}, \textrm{composite}\}$ has  12 different hyper-parameter settings in $S$. We approximate the integral on the right-hand side of Eq. \ref{eq:imaging_evidence} as a discrete summation over $S$ as
\begin{equation} \label{eq:model_averaging_summation}
\begin{split}
        &\int p(\xi \mid O_{\rm img}, M, S) \ p(O_{\rm img} \mid M, S) \ p(S) \ \rmd S \\ 
        &\approx \sum_{n} \Delta S_n\ p(S_n) \ p(\xi \mid O_{\rm img}, M, S_n) \ p(O_{\rm img} \mid M, S_n).
\end{split}
\end{equation}
Here, $\Delta S_n$ can be interpreted as the model space volume that represents the model $S_n$. Although the models $\{S_n\}$ differ from each other by discrete steps, an appropriately chosen expression for $\Delta S_{n}$ can account for sparse sampling from the model space, as we cannot adopt a sufficiently large number of models that are densely populated in the model space due to computational limitation. We use \ajsiv{the} BIC score  of a model as a proxy for the model evidence $p(O_{\rm img} \mid M, S_n)$. Thus, $\exp(- \Delta {\rm BIC }/2)$ acts as the evidence ratio and provides the relative weight between two models. The $\Delta S_n$ term is effectively an additional weighting on top of this BIC weighting \citep{Birrer19b, Shajib20}. We take $p(S_n) = 1$ and therefore need to effectively implement the weighted sum of $p(\xi \mid O_{\rm img}, M, S_n)$ in the right-hand side of Eq. \ref{eq:model_averaging_summation} through sampling.

We tabulate the BIC values of the models in Table \ref{tab:lenstronomy_bic_values}. \ajs{The BIC values are computed from the maximum sampled likelihood in each MCMC chain.} We estimate the sparsity of models $\{S_n \}$ by taking the variance ${\sigma^{\rm model}_{\Delta {\rm BIC}}}^2$ of $\Delta {\rm BIC}$ between `neighbouring' models that differ with each other by one step in only one setting \citep{Shajib20}. We furthermore accounted for the numeric uncertainty in estimation of $\Delta {\rm BIC}$ by taking the variance ${\sigma^{\rm numeric}_{\Delta {\rm BIC}}}^2$ of $\Delta {\rm BIC}$ between identical models that we have optimized twice. These twin runs produce slightly different posteriors -- and thus BIC values -- due to stochasticity in the PSF reconstruction, PSO, and MCMC sampling steps, similar to what was done in \citet{Birrer19b}. Thus, our total variance in $\Delta {\rm BIC}$ is
\begin{equation}
        \sigma_{\Delta {\rm BIC}}^2 \equiv {\sigma^{\rm model}_{\Delta {\rm BIC}}}^2 + {\sigma^{\rm numeric}_{\Delta {\rm BIC}}}^2.
\end{equation}
We compute that ${\sigma^{\rm model}_{\Delta {\rm BIC}}} = 304$ and ${\sigma^{\rm numeric}_{\Delta {\rm BIC}}} = 69$. To implement the $\Delta S_{\rm n}$  weighting through sampling, we first follow \cite{Birrer19b} to obtain the absolute weight $W_{n, \text{abs}}$ of the $n$\textsuperscript{th} model by convolving the $\Delta$BIC with the evidence ratio function $f(x)$ as
\begin{equation} \label{eq:abs_model_weight}
        W_{n, \text{abs}} = \frac{1}{\sqrt{2 \pi} \sigma_{\Delta {\rm BIC}}} \int_{-\infty}^{\infty} f(x) \exp \left[-\frac{({\rm BIC}_n - x)^2}{2 \sigma_{\Delta {\rm BIC}}^2} \right] \mathrm{d} x,
\end{equation}
where the evidence ratio function $f(x)$ is defined using the BIC difference as
\begin{equation}
        f(x) \equiv 
     \begin{cases}
       1 &\quad x < {\rm BIC}_{\rm min}, \\\
       \exp\left( {\rm BIC}_{\rm min} - x \right) &\quad x \geq {\rm BIC}_{\rm min}. \
     \end{cases}
\end{equation}
Then, we obtain the relative weight $W_{n, \text{rel}}$ by normalizing the absolute weights by the maximum absolute weight as
\begin{equation}
        W_{n, \text{rel}} = \frac{W_{n, \text{abs}}}{\max \left(\left\{ W_{n, \text{abs} } \right\} \right)}.
\end{equation}
Finally, we combine the individual model posteriors following Eq. \ref{eq:model_averaging_summation} as
\begin{equation}
\begin{split}
        &\sum_{n} \Delta S_n\ p(S_n) \ p(O_{\rm img} \mid M, S_n) \ p(\xi \mid O_{\rm img}, M, S_n) \\
        & \qquad \qquad \propto \sum_{n} W_{n, {\rm rel}}\ p(\xi \mid O_{\rm img}, M, S).
\end{split}
\end{equation}

In Sect. \ref{sec:lenstronomy_lensing_results}, we compare the two mass model families after performing the above model averaging procedure within each model family.

\begin{table*}
\caption{\label{tab:lenstronomy_bic_values}
        BIC values for different \textsc{lenstronomy} model setups. The difference $\Delta$BIC is calculated only within the particular mass profile family -- power law or composite. The model setups are ordered from \reply{lower to higher} BIC values within each mass profile family. \reply{The $\Delta$BIC values are calculated relative to the model setup with the lowest BIC value.} The relative weights for each model are obtained from $\Delta$BIC adjusted for sparse sampling from the model space as described in Sect. \ref{sec:lenstronomy_bma}.
        }
\begin{tabular}{cccrc}
\hline
Source light $n_{\rm max}$ & Likelihood computation region size & Run number & $\Delta$BIC & BIC weight \\
\hline
\multicolumn{5}{c}{Power-law ellipsoid model} \\
\hline
$\{9, 13\}$ & $\{ 2.2,3.6 \}$ & 2 & 0 & 1.00 \\
$\{9, 13\}$ & $\{ 2.2,3.6 \}$ & 1 & 43 & 0.95 \\
$\{8, 12\}$ & $\{ 2.2,3.6 \}$ & 2 & 209 & 0.76 \\
$\{8, 12\}$ & $\{ 2.2,3.6 \}$ & 1 & 235 & 0.73 \\
$\{7, 11\}$ & $\{ 2.2,3.6 \}$ & 2 & 606 & 0.37 \\
$\{7, 11\}$ & $\{ 2.2,3.6 \}$ & 1 & 726 & 0.29 \\
$\{9, 13\}$ & $\{ 2.3,3.7 \}$ & 2 & 2129 & 0.00 \\
$\{8, 12\}$ & $\{ 2.3,3.7 \}$ & 2 & 2350 & 0.00 \\
$\{9, 13\}$ & $\{ 2.3,3.7 \}$ & 1 & 2366 & 0.00 \\
$\{7, 11\}$ & $\{ 2.3,3.7 \}$ & 2 & 2778 & 0.00 \\
$\{8, 12\}$ & $\{ 2.3,3.7 \}$ & 1 & 2786 & 0.00 \\
$\{7, 11\}$ & $\{ 2.3,3.7 \}$ & 1 & 2793 & 0.00 \\
\hline
\multicolumn{5}{c}{Composite model} \\
\hline
$\{9, 13\}$ & $\{ 2.2,3.6 \}$ & 1 & 0 & 1.00 \\
$\{9, 13\}$ & $\{ 2.2,3.6 \}$ & 2 & 10 & 0.99 \\
$\{8, 12\}$ & $\{ 2.2,3.6 \}$ & 1 & 318 & 0.64 \\
$\{8, 12\}$ & $\{ 2.2,3.6 \}$ & 2 & 449 & 0.51 \\
$\{7, 11\}$ & $\{ 2.2,3.6 \}$ & 1 & 604 & 0.37 \\
$\{7, 11\}$ & $\{ 2.2,3.6 \}$ & 2 & 675 & 0.32 \\
$\{9, 13\}$ & $\{ 2.3,3.7 \}$ & 2 & 2009 & 0.00 \\
$\{9, 13\}$ & $\{ 2.3,3.7 \}$ & 1 & 2191 & 0.00 \\
$\{8, 12\}$ & $\{ 2.3,3.7 \}$ & 2 & 2373 & 0.00 \\
$\{8, 12\}$ & $\{ 2.3,3.7 \}$ & 1 & 2378 & 0.00 \\
$\{7, 11\}$ & $\{ 2.3,3.7 \}$ & 2 & 2807 & 0.00 \\
$\{7, 11\}$ & $\{ 2.3,3.7 \}$ & 1 & 2912 & 0.00 \\
\hline
\end{tabular}
\end{table*}

\subsection{Lensing-only constraints on the Fermat potential difference} \label{sec:lenstronomy_lensing_results}

\reply{We constrain the image positions with uncertainty 0\farcs002 for the power-law model, and with uncertainty 0\farcs004 for the composite model. Given the longest predicted time-delay for this system, these precisions are well below the astrometric requirement of $\sim$0\farcs02 uncertainty so that the astrometric uncertainty is subdominant to achieve $\leq$5\% uncertainty in $H_0$  from this single system \citep{Birrer19}.}

\begin{figure*}
        \includegraphics[width=\textwidth]{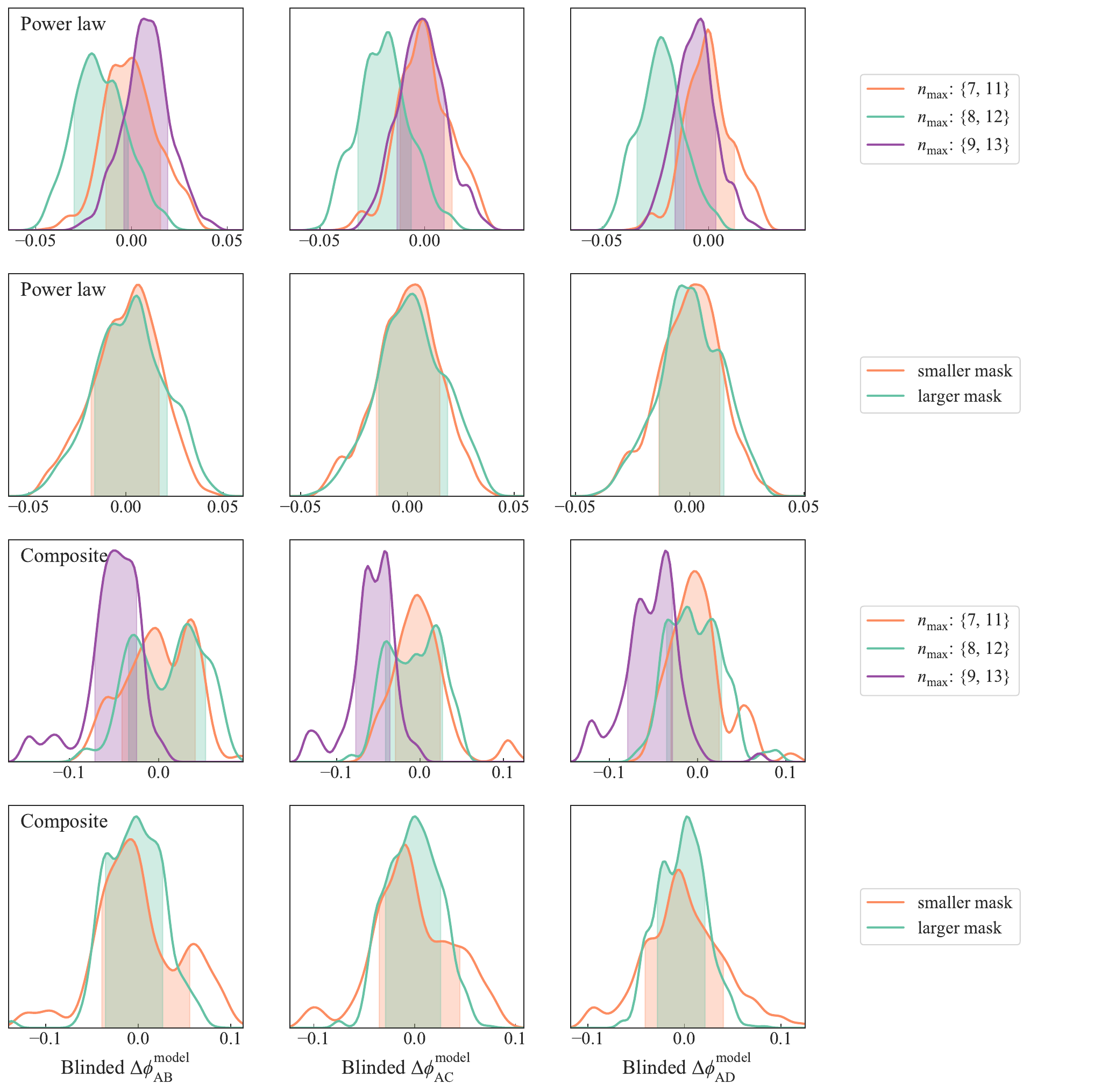}
        \caption{\label{fig:model_setting_compare}
        Comparison between the Fermat potential difference posteriors from different \textsc{lenstronomy} model settings. The posterior for a particular setting is obtained by averaging over models that differ in the other model settings, but within a particular model family using the procedure described in Sect. \ref{sec:lenstronomy_bma}. The top two rows correspond to the power-law mass model families, and the bottom two rows correspond to the composite mass model families. The shaded regions illustrate the 68\% credible regions. The posteriors are blinded by $\Delta \phi^{\rm blinded} \equiv \Delta \phi/\overline{\Delta \phi_{\rm ref}} - 1$, where the `ref' subscript refers to one of the compared models. The potential differences are consistent within 1$\sigma$ between our adopted choices of model settings.
        }
\end{figure*}

In Fig. \ref{fig:model_setting_compare} we compare between the model settings -- namely, the choices for $n_{\rm max}$ and the size of the image likelihood computation region. All the posteriors within a particular lens model family (i.e. power law or composite) are statistically consistent with each other within $1\sigma$.

\begin{figure*}
        \includegraphics[width=\textwidth]{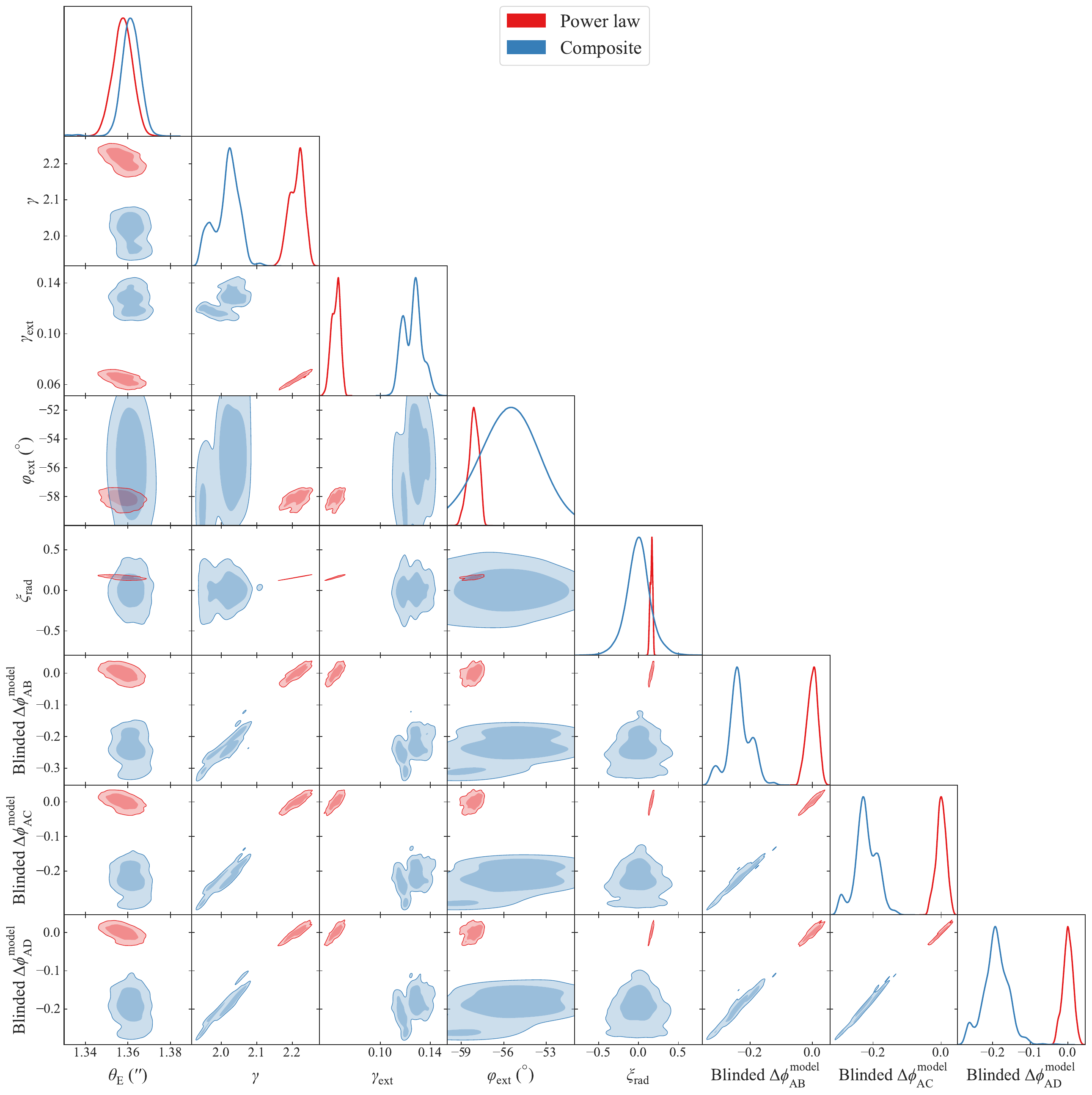}
        \caption{\label{fig:lenstronomy_powerlaw_vs_composite}
        Comparison of the \textsc{lenstronomy} lens model parameters and Fermat potential difference between the power-law (red) and composite (blue) mass models. The posteriors are obtained by averaging over all the model settings following the procedure described in Sect. \ref{sec:lenstronomy_bma}. \ajs{The parameter $\gamma$ for the composite model is computed from circularly averaging the quantity $2 - \left[\mathrm{d} \log \alpha(r)/\mathrm{d} \log r  \right]_{r = \theta_E}$.} Some parameters are blinded as $p^{\rm blinded} \equiv p / \overline{p_{\rm pl}} - 1$ for $p \in \{\gamma$, $\Delta \phi\}$, where the subscript `pl' refers to the power-law model posteriors. The composite model-predicted $\gamma$ is approximately 8\% smaller than that from the power law. Consequently, the Fermat potential differences are smaller by approximately 16\% for the composite model.
        }
\end{figure*}

We compare the combined model posteriors between the power-law and composite models in Fig. \ref{fig:lenstronomy_powerlaw_vs_composite}. The Fermat potential differences deviate by 16--21\% between the power-law and composite model setups. However, the MST-invariant quantity $\xi_{\rm rad} \propto \theta_{\rm E} \alpha^{\prime\prime}_{\rm E} / (1 - \kappa_{\rm E})$ is consistent between the two model setups (see Eq. 42 of \citealt{Birrer21} for the full definition of $\xi_{\rm rad}$, and also \citealt{Kochanek20}). Thus, the difference in the Fermat potential from the two model setups can be interpreted as a manifestation of the internal MSD. We combine the stellar kinematics and estimated external convergence with the lens models to mitigate the internal MSD in Sect. \ref{sec:lenstornomy_kinematics}.

\ajs{However, we first check for potential unphysical properties in our best fit composite models as the source of the large difference in the Fermat potential in Sects. \ref{sec:lenstronomy_halo_properties} and \ref{sec:lenstronomy_halo_prior_systematic_test}.} \ajsii{These checks were performed prior to un-blinding the models.}

\subsubsection{Halo properties in the composite model} \label{sec:lenstronomy_halo_properties}
\begin{figure}
        \includegraphics[width=0.5\textwidth]{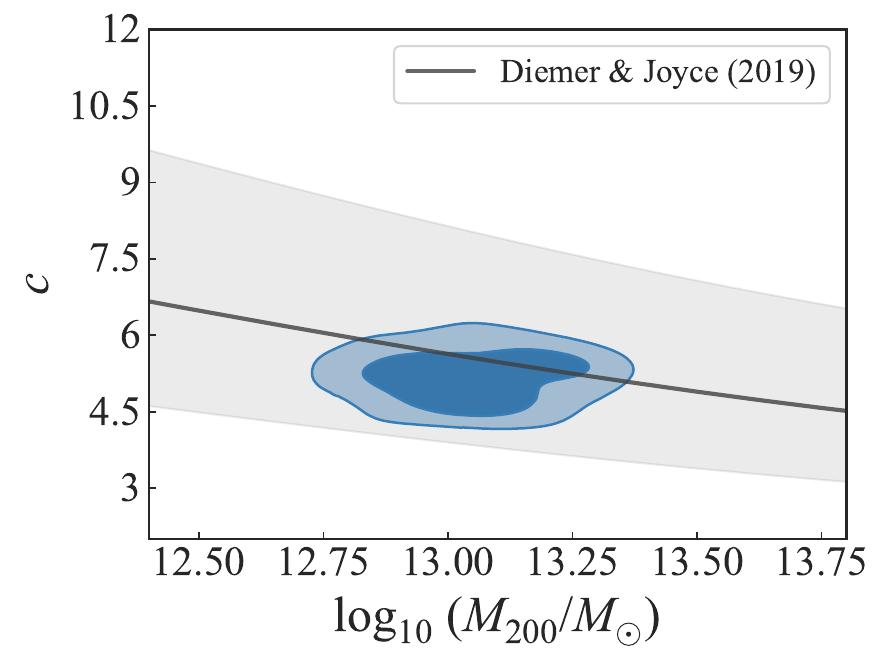}
        \caption{\label{fig:lenstronomy_mc_relation}
        Distribution of $M_{\rm 200}$ and $c_{\rm 200}$ parameters for the NFW halos in \textsc{lenstronomy} composite model (blue shaded region). This distribution is averaged over all the model settings within the composite model family following the procedure described in Sect. \ref{sec:lenstronomy_bma}. The 2 contours correspond to the 68\% and 95\% credible regions, respectively. The black solid line traces the theoretical prediction of the $M_{200}$--$c_{200}$ relation at $z_{\rm d} = 0.230$ from \citet{Diemer19} with the grey shaded region corresponding to the 68\% confidence interval. We adopt this $M_{200}$--$c$ relation as a prior in our analysis in addition to a $M_{200}$ prior based on \citet{Sonnenfeld18c}.
        }
\end{figure}

\begin{figure}
        \includegraphics[width=0.5\textwidth]{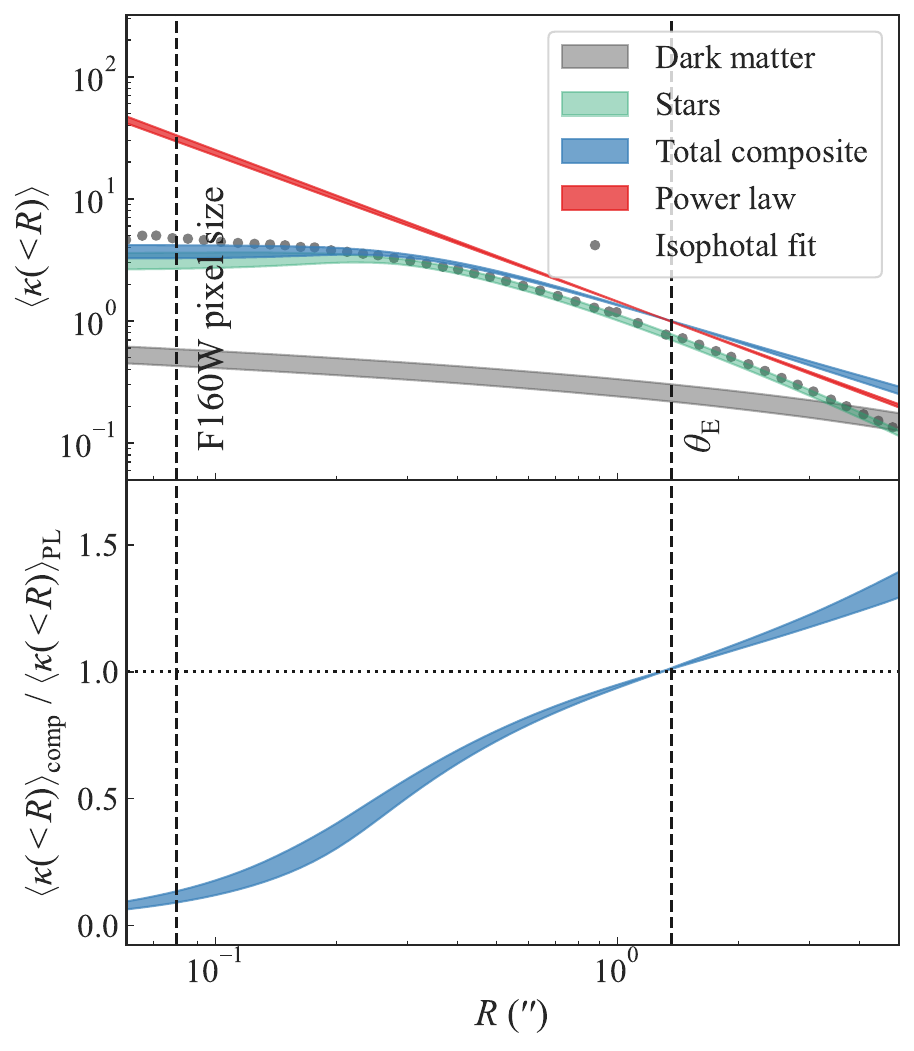}
        \caption{\label{fig:lenstronomy_radial_convergence_profile}
        \ajsiv{\lreply{Radial mass profiles of the central deflector constrained by the \textsc{lenstronomy} models.} \textbf{Top:} Circularly averaged convergence $\langle \kappa(<R) \rangle$ as a function of radius} from \textsc{lenstronomy} \ajsiv{power-law (red) and composite (blue) models}. \ajsiv{The stellar (green) and dark matter (grey) distributions in the composite model are also individually illustrated.} The shaded regions encompass the 16th and 84th percentile of the sampled profiles for the corresponding case. 
        The grey points illustrate the surface brightness profile fitted with isophotes as described in Sect. \ref{sec:lenstronomy_deflector_settings}. 
        The amplitude of the isophotal fit is normalized to match with the triple Chameleon profile (\ajsiv{green} shaded region) at $\theta_{\rm E}$ for the purpose of this illustration. The vertical black dashed lines mark the pixel size in the F160W band and the best fit Einstein radius. \ajsiv{\textbf{Bottom:} Ratio of the circularly averaged convergence profiles between the composite and the power-law models.} At the Einstein radius the convergence slope deviates by 16--21\% between the two model setups.
        }
\end{figure}

\begin{figure}
        \includegraphics[width=\columnwidth]{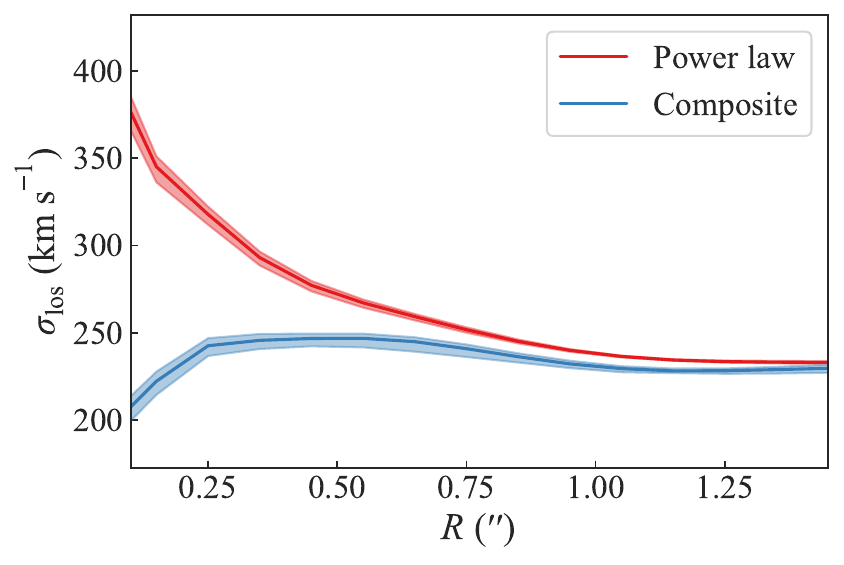}
        \caption{ \label{fig:lenstronomy_velocity_dispersion_profile}
        \ajsii{Line-of-sight velocity dispersion profile (circularized) corresponding to the lens model posteriors in the power-law (red) and composite (blue) models from \textsc{lenstronomy}. Isotropic stellar orbits are assumed in computing these velocity dispersion profiles. The solid lines correspond to the median and the shaded regions encompass the 16th and 84th percentiles. The composite profile predicts a decrease in the velocity dispersion towards the centre due to the flattened core -- which is not observed in local massive ellipticals \citep[e.g.][]{Cappellari16, Ene19} -- thus pointing to the atypicality of the posterior mass profile in the composite model.}
        }
\end{figure}

\lreply{Fig.} \ref{fig:lenstronomy_mc_relation} illustrates the $M_{200}$--$c$ relation posterior for our system in comparison with the adopted prior; \ajs{the median of the concentration posterior is consistent with the concentration prior within 1$\sigma$.} Our combined posterior from the composite model setup provides the total halo mass $\log_{10} (M_{\rm 200}/M_{\odot}) = 13.04_{-0.13}^{+0.14}$ 
and the total stellar mass is $\log_{10} (M_{\star}/M_{\odot}) = 11.87_{-0.03}^{+0.01}$. 
\ajs{The total stellar mass is obtained by doubling the enclosed mass within the half-light radius of 3\farcs2 corresponding to the F160W band.}
The projected dark matter fraction within the Einstein radius is $0.22^{+0.06}_{-0.02}$. \ajs{The total baryon-to-dark-matter fraction is $0.07_{-0.02}^{+0.03}$, which is consistent with the upper limit set by the cosmic baryonic fraction 0.19 \citep{PlanckCollaboration18}.}
In Fig. \ref{fig:lenstronomy_radial_convergence_profile} we plot the azimuthally averaged convergence profiles for the power-law and composite models to  illustrate the difference in the convergence slope at the Einstein radius $\theta_{\rm E}$. The inner region ($\lesssim$0\farcs2) of the triple Chameleon profile is flat unlike the singular centre in the power-law model. The flat or cored convergence profile at the centre gives rise to an inner critical curve in the image plane (Fig. \ref{fig:lenstronomy_composite_model}). 
This core in the centre of the stellar mass distribution follows from the stellar flux distribution, as the radial flux profile from isophotal fitting also shows a stellar core. \ajs{We fit a core S\'ersic profile -- that is defined by Eq. (2) in \citet{Dullo19} -- to the azimuthally averaged light profile from our isophotal fitting to obtain the stellar core radius. We obtain $0.780\pm0.004$ kpc assuming a fiducial flat \lcdm\ cosmology with $H_0 = $ 70 km s$^{-1}$ Mpc$^{-1}$ and $\Omega_{\rm m} = 0.3$. This stellar core radius is consistent with the core radii measured in local elliptical galaxies \citep[0.64--2.73 kpc;][]{Bonfini16, Dullo19}.} \ajsii{In Fig. \ref{fig:lenstronomy_velocity_dispersion_profile}, we illustrate the velocity dispersion profiles predicted by the lens model posteriors assuming isotropic orbit and a flat \lcdm\ cosmology with $H_0 = 70$ km s$^{-1}$ Mpc$^{-1}$. The composite-model-predicted velocity dispersion profile decreases towards the centre by $\sim$20\% due to the flattened mass profile. Such a large decrease in the velocity dispersion has not been observed in local massive ellipticals \citep[e.g.][]{Cappellari16, Ene19}. As a result, the composite lens model posterior \ajsiv{suggests an} inconsistency with kinematic observations of the local ellipticals.} Interestingly, an inner critical curve in our composite model predicts a central image. The magnifications of the 5 images are A: $-1.6_{-0.2}^{+0.1}$, B: $4.0_{-0.3}^{+0.2}$, C: $-3.6_{-0.4}^{+0.2}$, D: $4.8\pm0.3$, and central: $0.019\pm0.002$. \ajs{The predicted appearance} of the central image is invariant under the MST. However, the demagnified and potentially dust-extincted central image is indistinguishable in the present imaging data. Thus, we are unable to distinguish the two profile families on the basis of the presence or absence of the central image.

\subsubsection{Test of potential systematics from modelling choices and priors} \label{sec:lenstronomy_halo_prior_systematic_test}

\begin{figure*}
        \includegraphics[width=\textwidth]{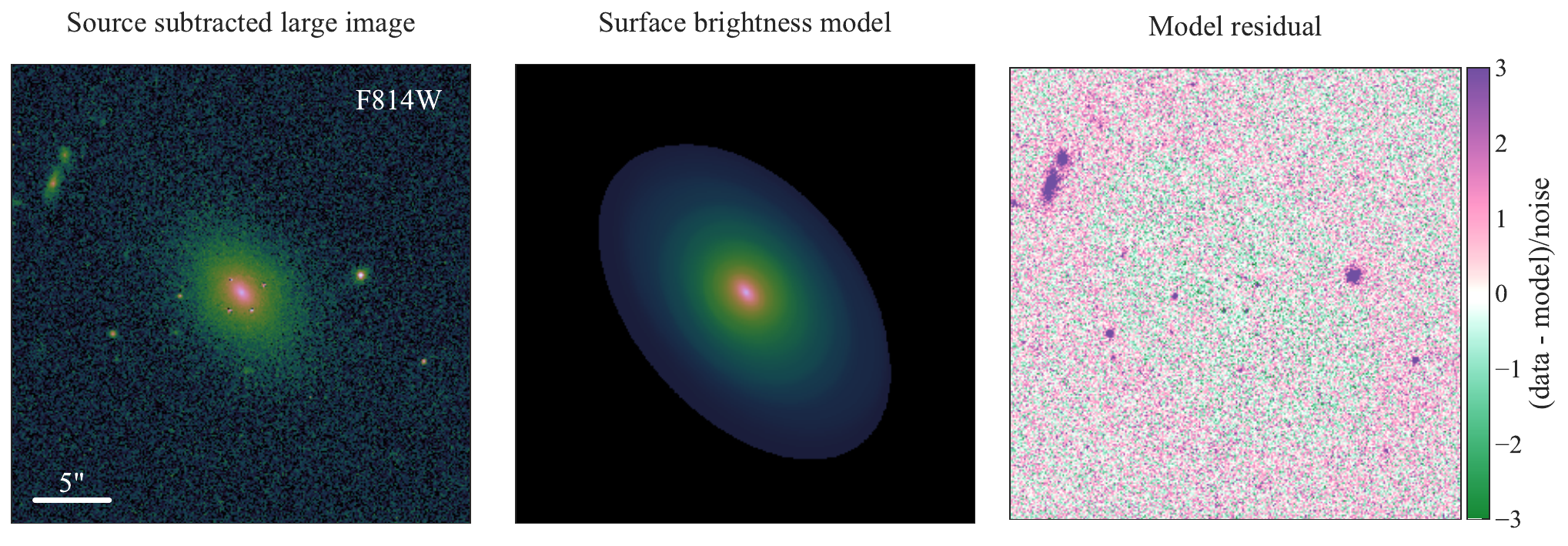}
        \caption{\label{fig:large_cutout_light_model}
        Fitting the lens galaxy light profile from a much larger cutout than that used for lens modelling. \textbf{Left:} Large cutout around the deflector galaxy in the \ajs{F814W} band, with the lensed arcs and quasar images subtracted using the best-fit of a power-law lens model. The galaxy extends much further beyond the Einstein radius with $\theta_{\rm eff}/\theta_{\rm E} \approx 1.7$. \textbf{Middle:} Model for the surface brightness profile of the deflector constructed using elliptical isophotes. \textbf{Right:} Model residual showing that the model has captured the overall light distribution despite numerous overlapping objects. The model from middle panel was further approximated with an elliptical multi-Gaussian expansion \citep[MGE;][]{Cappellari02} to allow deprojection along the LOS for kinematic modelling.
        }
\end{figure*}

We checked if our composite models are robust against potential biases from our particular model choices, for example the likelihood computation region and the prior on the NFW halo scale radius.  
We first optimized a lens model with the power-law mass profile and the triple Chameleon profile for the light instead of the triple S\'ersic profile. We then took these best fit parameters for the triple Chameleon profile and fixed them in the test composite setup. We let the overall scaling of the baryonic mass and light distributions be free, thus effectively allowing for a free mass-to-light ratio ($M/L$). We adopted a halo mass prior for the NFW profile dependent on the stellar mass given by
\begin{equation} \label{eq:lenstronomy_halo_mass_prior}
        \begin{aligned}
        &p(\log_{10} M_{200} \mid \log_{10} M_{\star}^{\rm Chab}) \\\
        &\qquad \equiv \mathcal{N} \left(\mu_{\rm h} + \beta_{\rm h} \left[\log_{10} M_{\star}^{\rm Chab} - 11.3\right],\ \sigma_{\rm h} \right),
        \end{aligned}
\end{equation}
where $M_{\star}^{\rm Chab}$ is the total stellar mass based on the SPS method assuming a Chabrier IMF \citep{Sonnenfeld18c}. Here, the parameters are $\mu_{\rm h} = 13.11\pm0.04$, $\beta_{\rm h} = 1.43 \pm 0.15$, and $\sigma_{\rm h} = 0.23 \pm 0.04$.
We measure the stellar mass $M_{\star}^{\rm Chab}$ from the total fluxes in the three HST bands. We fit the surface brightness profile of the deflector separately in three bands from large cutouts that fully contains the light distribution of the deflector (see Fig. \ref{fig:large_cutout_light_model}). First, we subtract lensed arcs and the quasar images from these cutouts using our best fit power-law model. Then, we fit elliptical isophotes using the \textsc{Photutils} software \citep{Bradley20}. The method \texttt{photutils.isophote.Ellipse.fit\_image()} allows a convenient way to ignore the overlapping objects (i.e. stars and galaxies) by sigma-clipping pixels along an isophote. Thus, we do not need to mask out these overlapping objects. We reconstructed the surface brightness profile of the deflector based on the fitted isophotes, which effectively interpolates through the pixels that are contaminated by overlapping objects. We obtain the total flux in each HST band from the reconstructed surface brightness profile using the fitted isophotes. We used \textsc{pygalexev}\footnote{\faGithub\ \url{https://github.com/astrosonnen/pygalaxev}} to obtain $M_{\star}^{\rm Chab}$, which is a \textsc{Python} wrapper for \textsc{galaxev} \citep{Bruzual03}. By adopting the \lreply{Basel Stellar Library \citep[BaSeL;][]{Lejeune98}}, exponentially decaying stellar formation history, and free metallicity, we obtain $\log_{10} M_{\star}^{\rm Chab} = 11.57^{+0.16}_{-0.13}$. Thus, from Eq. \ref{eq:lenstronomy_halo_mass_prior}, our Gaussian prior on the halo mass $\log_{10} M_{200}$ has mean 13.5 and standard deviation 0.3.
We additionally adopt a prior for the total stellar mass $\log_{10} M_{\star}$. We take an ad hoc prior that is uniform between 11.51 and 11.88 and drops off like a Gaussian function outside these limits. The range between 11.51 and 11.88 accounts for the unknown IMF and spans the range between light (e.g. Chabrier) and heavy (e.g. Salpeter) IMFs that differ by $\sim$0.25 dex in the stellar mass. The exact form of this prior is
\begin{equation}
        p(\log_{10} M_{\star}^{/\odot}) =
     \begin{cases}
       A \exp \left[-\frac{(\log_{10} M_{\star}^{/\odot} - 11.51)^2}{2 \times 0.13^2}\right], \quad \log_{10} M_{\star}^{/\odot} < 11.51, \\\
       A , \quad 11.51 \leq \log_{10} M_{\star}^{/\odot} \leq 11.88, \\\
       A \exp \left[-\frac{(\log_{10} M_{\star}^{/\odot} - 11.88)^2}{2 \times 0.16^2}\right] , \quad \log_{10} M_{\star}^{/\odot} > 11.88,
     \end{cases}
\end{equation}
where $M_{\star}^{/\odot} \equiv M_{\star}/M_{\odot}$, $A$ is the amplitude that normalizes the probability distribution to have $\int p(\log_{10} M_{\star}^{/\odot}) \ \rmd (\log_{10} M_{\star}^{/\odot}) = 1$. The actual value of $A$ is not required for sampling in the MCMC method. \ajs{We also allow an additional uncertainty of $\pm$0.06 dex in $M_{\star}$ to allow 15\% uncertainty on the assumed $H_0$ in the SPS-based stellar mass estimation.}

We furthermore mask out the central region in the deflector galaxy in \ajsiv{the} test composite model setup so that the optimization does not incentivize the presence of a central image to make up for residuals in the deflector light galaxy model that cannot be fully accounted by the triple Chameleon light profile.

\begin{figure*}
        \includegraphics[width=\textwidth]{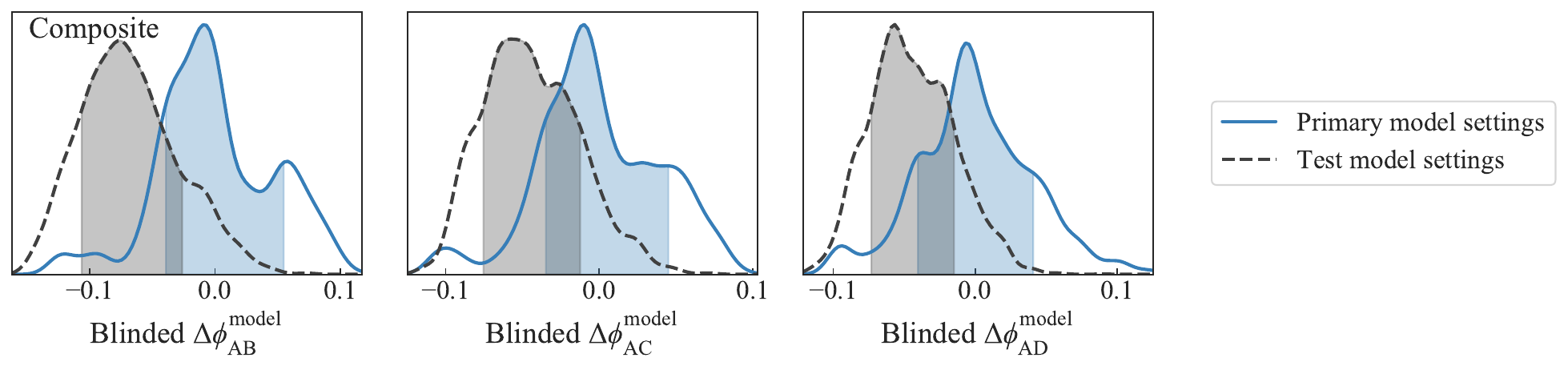}
        \caption{\label{fig:lenstronomy_composite_model_check}
        Comparison of the Fermat potential posteriors between two different model settings for the \textsc{lenstronomy} composite model. The blue solid distributions correspond to our primary model settings. In the test model setup (black dashed distributions), we fix the triple Chameleon profile for the stellar mass and light distributions to best fit values from a separately optimized model with the power-law mass profile. We also mask the central region of the deflector galaxy in the test model setup. Additionally, the prior on the halo mass profile is different. Whereas we adopt a prior on the NFW scale radius $r_{\rm s}$ in the primary setup, we adopt a combination of priors on $M_{200}$ and $M_{\star}$ in the test setup. Despite multiple differences in the model settings, the Fermat potential differences are consistent within 1$\sigma$ with each other.
        }
\end{figure*}

We perform the 12 different model setups also for this test composite model and combine the posteriors based on their BIC values. We compare our primary composite model with the test composite model in Fig. \ref{fig:lenstronomy_composite_model_check}. Although the Fermat potential differences are consistent between these two model setups \ajs{within $1\sigma$, the ones from the test setup are smaller by 4--7\% than the primary setup. For the test setup the stellar mass is $\log_{10} (M_{\star}/M_{\odot}) = 11.84_{-0.01}^{+0.02}$, the halo mass is $\log_{10} (M_{200}/M_{\odot}) = 13.3_{-0.3}^{+0.1}$, the total baryonic fraction is $0.04_{-0.01}^{+0.03}$, and the dark matter fraction within the Einstein radius is $0.28_{-0.03}^{+0.02}$. The total halo mass increases in the test setting over the one obtained from our primary setting due to the adopted halo mass prior. As a result, the increase in the halo normalization leads to a decrease in the Fermat potential, or equivalently the shallowing of the convergence profile. This impact of the prior on the halo profile normalization is the same as observed by \citet{Shajib21} for the SLACS lenses.} 
\ajs{As adopted priors can systematically shift Fermat potential differences, more physically motivated priors are not sufficient to explain all the differences between the composite and the power-law models.}

\ajs{\ajsii{The mass difference of 3.8$\times10^{10} M_{\odot}$ within 0\farcs2  (assuming flat \lcdm\ cosmology with $H_0=70$ km s$^{-1}$ Mpc$^{-1}$) between the power-law and composite models \ajsiv{could} be explained by an ultra-massive black hole \citep[e.g.][]{Mehrgan19, Dullo19}.} Another potential solution that would push the Fermat potential differences from the composite models towards the ones from the power-law models is to have stellar mass-to-light ratio gradient \ajsii{($\eta \sim 0.27$ with $M_{\star}/L \propto R^{-\eta}$)} to steepen up the total mass density profile. Although \citet{Shajib21} find no evidence for a significant mass-to-light ratio {on average} for SLACS lenses at the similar redshift ($\langle z \rangle \sim 0.2$) as \lensname\ ($z_{\rm d} = 0.23$), there can still be individual cases with steep mass-to-light ratio gradients. \ajsii{Moreover, the value $\eta \sim 0.27$ will be consistent with the constraints $\langle \eta \rangle = 0.24 \pm 0.04$ reported by \citet{Sonnenfeld18c}.} Adopting a mass-to-light ratio gradient for the luminous component or including a central black hole in the composite lens model is beyond the scope of this study. At this point, however, there is no {a priori} reason to modify the composite model to push the Fermat potential differences towards the ones from the power-law models. We use the kinematics data to bridge the discrepancy between the two models or to be the decider between them next in Sect. \ref{sec:lenstornomy_kinematics}. As the composite models are related to the true underlying mass distribution through an approximate MST, we rely on the kinematics data to appropriately adjust the Fermat potential differences along the MSD towards the true values.}



\subsection{Combining with stellar kinematics and external convergence}\label{sec:lenstornomy_kinematics}

\begin{figure}
        \includegraphics[width=0.5\textwidth]{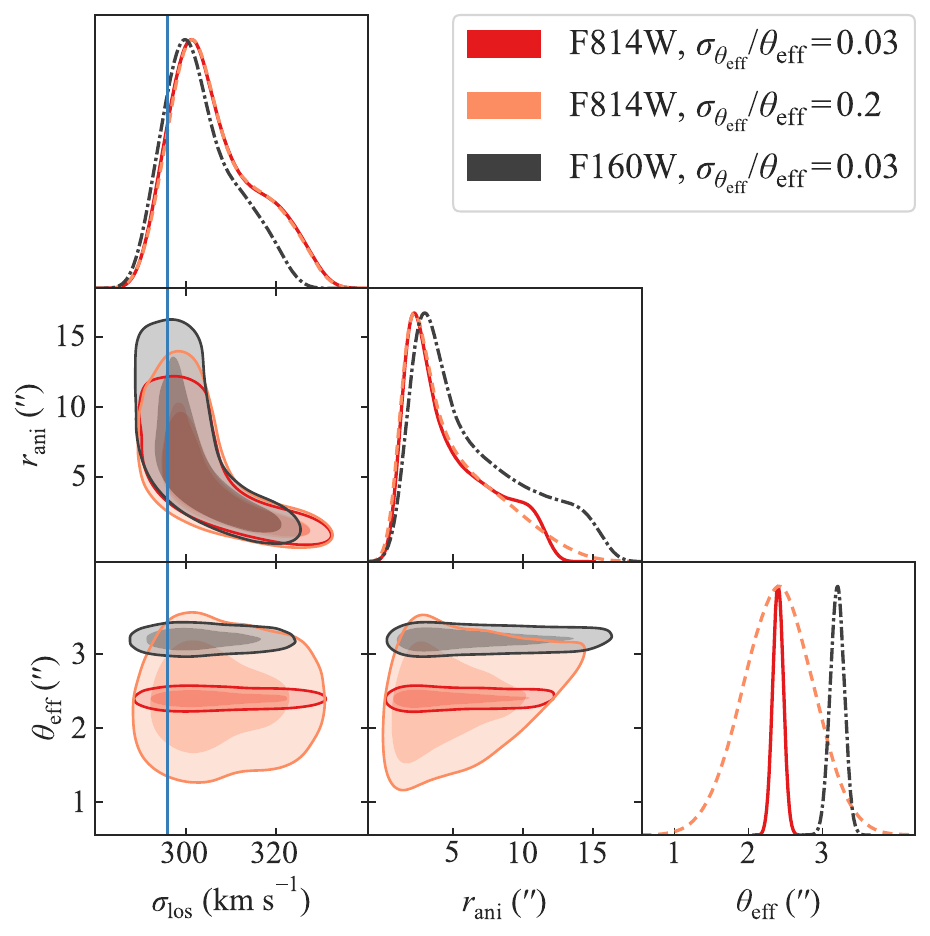}
        \caption{ \label{fig:kinematics_systematic_compare}
        \ajs{Checking for systematics in the kinematic computation. Our primary settings for kinematic computation adopts the F814W light profile with 3\% uncertainty in $\theta_{\rm eff}$ (red). The orange contours are for the case with 20\% uncertainty in $\theta_{\rm eff}$, and the black contours are for the case with F160W light adopted in the kinematic computation. \reply{The vertical blue line marks the measured LOS velocity dispersion, which has an uncertainty of 19 km s$^{-1}$.} The difference in the computed velocity dispersion between these cases is negligible, and it impacts the Fermat potential differences by $\lesssim0.1$\%.}
        }
\end{figure}

In this subsection we perform dynamical modelling of the deflector based on our lens models using Eq. \ref{eq:jeans_solution}. We need to estimate the luminosity density $l(r)$ to use in this equation by deprojecting the surface brightness profile of the deflector. The surface brightness of \lensname\ extends far beyond the size of our likelihood computation region (Fig. \ref{fig:large_cutout_light_model}). Thus, deprojecting the light profile fit from our lens models may potentially produce early truncation in the 3D luminosity distribution along the LOS. Therefore, we use the reconstructed surface brightness profile from the fitted isophotes in the \ajs{F814W} band from Sect. \ref{sec:lenstronomy_deflector_settings} (see Fig. \ref{fig:large_cutout_light_model}). \ajs{We chose the light profile from the F814W band, because the velocity dispersion was measured in the optical.}
From this reconstructed surface brightness profile, we numerically find the circular aperture that contains half of the total light as $\theta_{\rm eff} = 2\farcs4$. The aperture size in this numeric computation can only grow by a size of a pixel, which is 0\farcs08. We adopt a 3\% Gaussian uncertainty for $\theta_{\rm eff}$, \ajs{which combines one pixel size as a systematic uncertainty with a typical 2\% uncertainty for $\theta_{\rm eff}$ from fitting surface brightness profiles with continuous parameters from \citet{Shajib21}}. \reply{We check that adopting 20\% uncertainty for $\theta_{\rm eff}$ or using the $\theta_{\rm eff}$ measured from the F160W band does not significantly impact ($\lesssim0.1$\%) the resultant Fermat potential difference (Fig. \ref{fig:kinematics_systematic_compare}).} We approximate the reconstructed surface brightness profile with an elliptical 2D multi-Gaussian expansion (MGE) series \citep{Cappellari02}. The MGE approximation allows for a straightforward deprojection into a 3D light profile, which we use as $l(r)$ in Eq. \ref{eq:jeans_solution}. Since we only solve the Jeans equation in the spherical case, we adopt the spherical equivalent of the elliptical Gaussian components by taking the Gaussian scales along the intermediate axes. We also apply a self-consistent 2\% uncertainty to the MGE scale parameters by letting them vary with $\theta_{\rm eff}$. We adopt \ajs{a uniform prior on $\log a_{\rm ani}$}, where $a_{\rm ani}$ is the anisotropy scaling factor defined by $r_{\rm ani} \equiv a_{\rm ani}\theta_{\rm eff}$. \citet{Birrer16, Birrer20} find that \ajs{the uniform prior for $\log a_{\rm ani}$} is a less informative choice than the uniform prior \ajs{on $a_{\rm ani}$}.

We performed a test for systematics in our velocity dispersion modelling. We adopted two test cases: (i) where the $\theta_{\rm eff}$ uncertainty is taken as 20\%, and (ii) where the light profile from the F160W band is used in the kinematic computation.

\begin{figure}
        \includegraphics[width=\columnwidth]{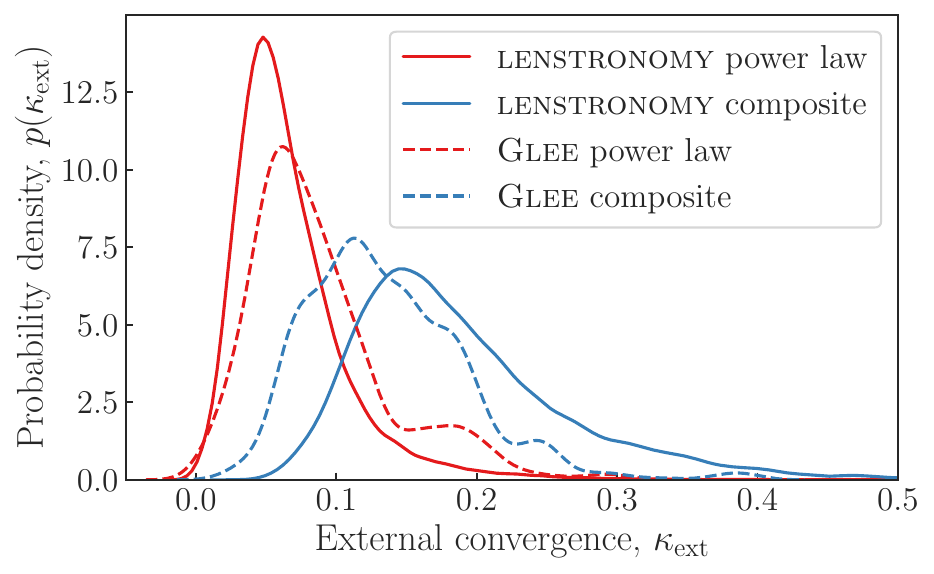}
        \caption{\label{fig:lenstronomy_kappa_ext}
        External convergence distribution from \citet{Buckley-Geer20} with additional weighting applied based on the predicted external shears for the power-law \ajsiv{(red lines) and composite (blue lines) models from \textsc{lenstronomy} (solid lines) and \textsc{glee} (dashed lines). \ajsv{Each illustrated \textsc{glee} distribution is a BIC-weighted combination of multiple $\kappa_{\rm ext}$ distributions corresponding to the external shear constraint from individual lens model setups. In contrast, each illustrated \textsc{lenstronomy} distribution is a single $\kappa_{\rm ext}$ distribution corresponding to the combined (through BIC weighting) external shear value from all the model setups within a model family (i.e. power law or composite). The $\kappa_{\rm ext}$ distributions used by one team were not revealed to the other team before the un-blinding to maintain independence.}}
        }
\end{figure}

For the external convergence $\kappa_{\rm ext}$, we impose a selection criterion on the $P(\kappa_{\rm ext})$ estimated in \citet{Buckley-Geer20} by requiring that the selected LOSs also correspond to the \ajsv{combined (through BIC weighting)} external shear value \ajsv{from our lens models within a lens model family (i.e. power law or composite)}. In Fig. \ref{fig:lenstronomy_kappa_ext} we illustrate the two $\kappa_{\rm ext}$ distributions consistent with the external shear values for the power-law and composite mass profiles.\footnote{\ajsv{While \citet{Buckley-Geer20} apply a joint constraint of number counts inside the 45$\arcsec$ and 120$\arcsec$ apertures, this would lead to too few LOSs selected from the Millennium simulation once the large shear value constraint of the composite model obtained by the \textsc{lenstronomy} team is imposed. To contain enough LOSs for a robust distribution, the \textsc{lenstronomy} team therefore removed the 45$\arcsec$ aperture constraint. For consistency, this was done for both the power-law and composite mass models. However, the number counts from the 45$\arcsec$ aperture were still retained by the \textsc{glee} team, whose composite model shear value is smaller. This difference between the external convergence distributions used by the two teams was revealed to each other only after the un-blinding. While this creates an inconsistency between the two teams, \citet[][]{Rusu20} show that for large shear values, the $\kappa_{\rm ext}$ distribution is dominated by the shear constraint, and therefore the imposition of the 45$\arcsec$ aperture or the lack thereof is expected to have a negligible impact.}}

We combine the stellar kinematics and external convergence information with the lens model posterior in two different ways: (i) with fixed $\lambda_{\rm int} = 1$ (Sect. \ref{sec:lenstronomy_lambda_int_fixed}), and (ii) with free $\lambda_{\rm int}$ constrained by the stellar kinematics (Sect. \ref{sec:lenstronomy_lambda_int_free}).

\subsubsection{The case with $\lambda_{\rm int} = 1$} \label{sec:lenstronomy_lambda_int_fixed}

\begin{figure*}
        \includegraphics[width=\textwidth]{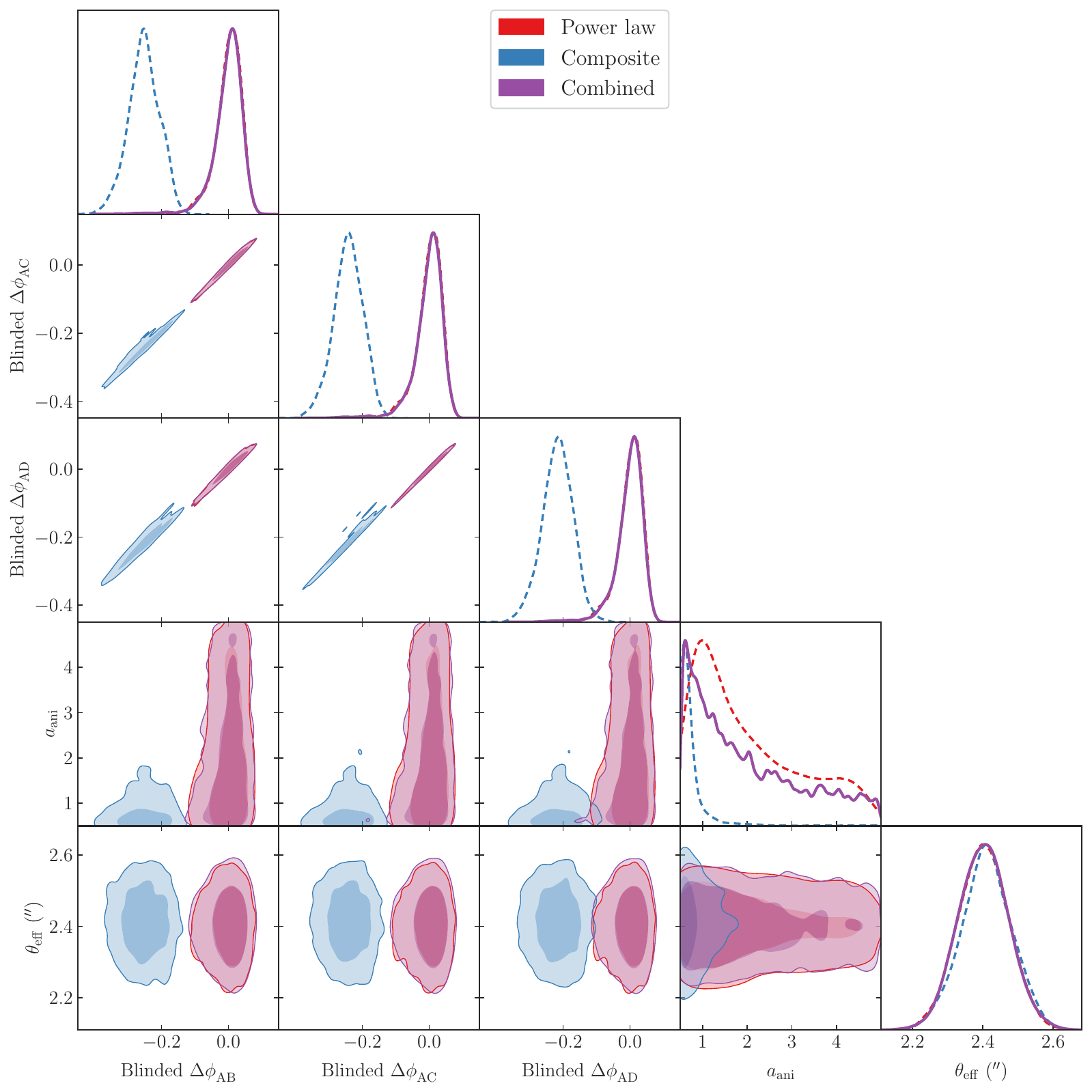}
        \caption{\label{fig:lenstronomy_mst_uncorrected_fermat_potential}
        Fermat potential differences $\Delta \phi = \Delta \phi^{\rm model} (1 - \kappa_{\rm ext})$ from \textsc{lenstronomy} for the power-law (red) and composite (blue) model with the kinematics information folded into these individual model posteriors. The folding in of the kinematics information is performed through importance sampling from the posterior weighted by the kinematics likelihood (Eq. \ref{eq:lenstronomy_kinematic_likelihood}). Next, the two model posteriors are joined together with equal weights and then the kinematics information is folded in. The combined posterior (purple) mostly resemble the power-law posterior, as the composite posterior is heavily down-weighted by the kinematics likelihood.}
\end{figure*}

Assuming $\lambda_{\rm int} = 1$, the model-predicted velocity dispersion can be written as
\begin{equation}
        \sigma_{\rm ap}^2 = (1 - \kappa_{\rm ext}) \frac{\Ds}{\Dds} c^2 J(\xi_{\rm lens}, \xi_{\rm light}, \beta_{\rm ani}).
\end{equation}
This assumption when combining the stellar kinematics information with the lens model posteriors is the same as done in earlier TDCOSMO analyses prior to \citet[][TDCOSMO-IV]{Birrer20}, for example in \citet{Suyu13}, \citet{Wong17}, and \citet{Rusu20}. We assume a flat $\Lambda$CDM cosmology with $\Omega_{\rm m} = 0.3$ to compute the fiducial distance ratio $\Ds/\Dds$. To combine the stellar kinematic information, we consider the kinematics likelihood function
\begin{equation} \label{eq:lenstronomy_kinematic_likelihood}
        \log \mathcal{L}_{\rm kin} = - \frac{\left(\sigma_{\rm model} - \sigma_{\rm measured}\right)^2}{2 \sigma_{\sigma_{\rm measured}}^2} - \frac{1}{2} \log \left(2 \uppi \sigma_{\sigma_{\rm measured}}^2\right).
\end{equation}
We first combine the lens model posteriors from the power-law and composite models with equal weights, and then importance sample from this combined posterior with weight $\mathcal{L}_{\rm kin}$ \citep{Lewis02}. We note that each of the power-law and composite models are already averaged over the various adopted model settings within each mass model family following our BMA procedure from Sect. \ref{sec:lenstronomy_bma}. We illustrate the Fermat potential differences from each of the power-law and composite models in Fig. \ref{fig:lenstronomy_mst_uncorrected_fermat_potential}. \ajs{}

After combining the kinematics information with $\lambda_{\rm int}$, the combined posterior for the Fermat potential differences end up mostly similar to the power-law posterior, as the kinematic likelihood heavily down-weights the posterior from the composite model. \ajs{Although the composite model was designed with physical motivations to mimic a real galaxy structure, the kinematics data heavily disfavours the composite lens model posterior for $\lambda_{\rm int} = 1$. Furthermore, applying more physical priors to resolve this inconsistency rather makes the kinematics data disfavour the composite model more, which suggest that our composite model is not adequate in describing the true galaxy mass distribution. Further generalization in the composite model (e.g. mass-to-light ratio gradient and generalized NFW halo) may thus be necessary for a composite model to be simultaneously consistent with the lensing data, the kinematics data, and the cosmological expectations for galaxies, e.g. the $M$--$c$ relation, baryonic fraction.} The uncertainties on the combined Fermat potential differences, and thus on the predicted time delays, are approximately 4\%, which is comparable with those from the previous TDCOSMO analyses under the same assumption of $\lambda_{\rm int} = 1$ \citep{Wong20}.

\subsubsection{The case with free $\lambda_{\rm int}$} \label{sec:lenstronomy_lambda_int_free}

Now, we treat $\lambda_{\rm int}$ as a free parameter and constrain it using the stellar kinematics by re-expressing Eq. \ref{eq:vdisp_ap} as
\begin{equation} \label{eq:lenstronomy_lambda_int}
        \lambda_{\rm int} = \frac{\sigma^2_{\rm ap}}{(1 - \kappa_{\rm ext}) (\Ds/\Dds) c^2 J(\xi_{\rm lens}, \xi_{\rm light}, \beta_{\rm ani})}.
\end{equation}
Such constraining of $\lambda_{\rm int}$ from stellar kinematics is the same approach as \citet[][TDCOSMO-IV]{Birrer20}, albeit these authors achieved a tighter constraint on $\lambda_{\rm int}$ from a joint sample of seven time-delay lens systems and 33 non-time-delay lenses through a hierarchical Bayesian analysis.

We obtain the $D_{\rm s}/D_{\rm ds}$ distribution to use in Eq. \ref{eq:lenstronomy_lambda_int} from the relative distance constrained by the Pantheon SN sample \citep{Scolnic18}. We approximate the luminosity distance up to the Pantheon supernovae using a fourth-order Taylor expansion. The coefficients in the Taylor expansion allow increasing complexity by including the deceleration parameter $q_0$, the jerk parameter $j_0$, the snap parameter $s_0$, and the curvature density parameter $\Omega_{k}$. We compute the model evidence using nested sampling for different size of the parameter set \citep{Skilling04}. We select the model that goes up to the jerk parameter $j_0$ based on its highest model evidence.  We use the relation $D_{\rm A} = D_{\rm L}/(1+z)^2$ to convert the luminosity distance to angular diameter distance and transform the 2D posterior distribution of ($q_0$, $j_0$) to obtain the $D_{\rm s}/D_{\rm ds}$ distribution.

\begin{figure*}
        \includegraphics[width=\textwidth]{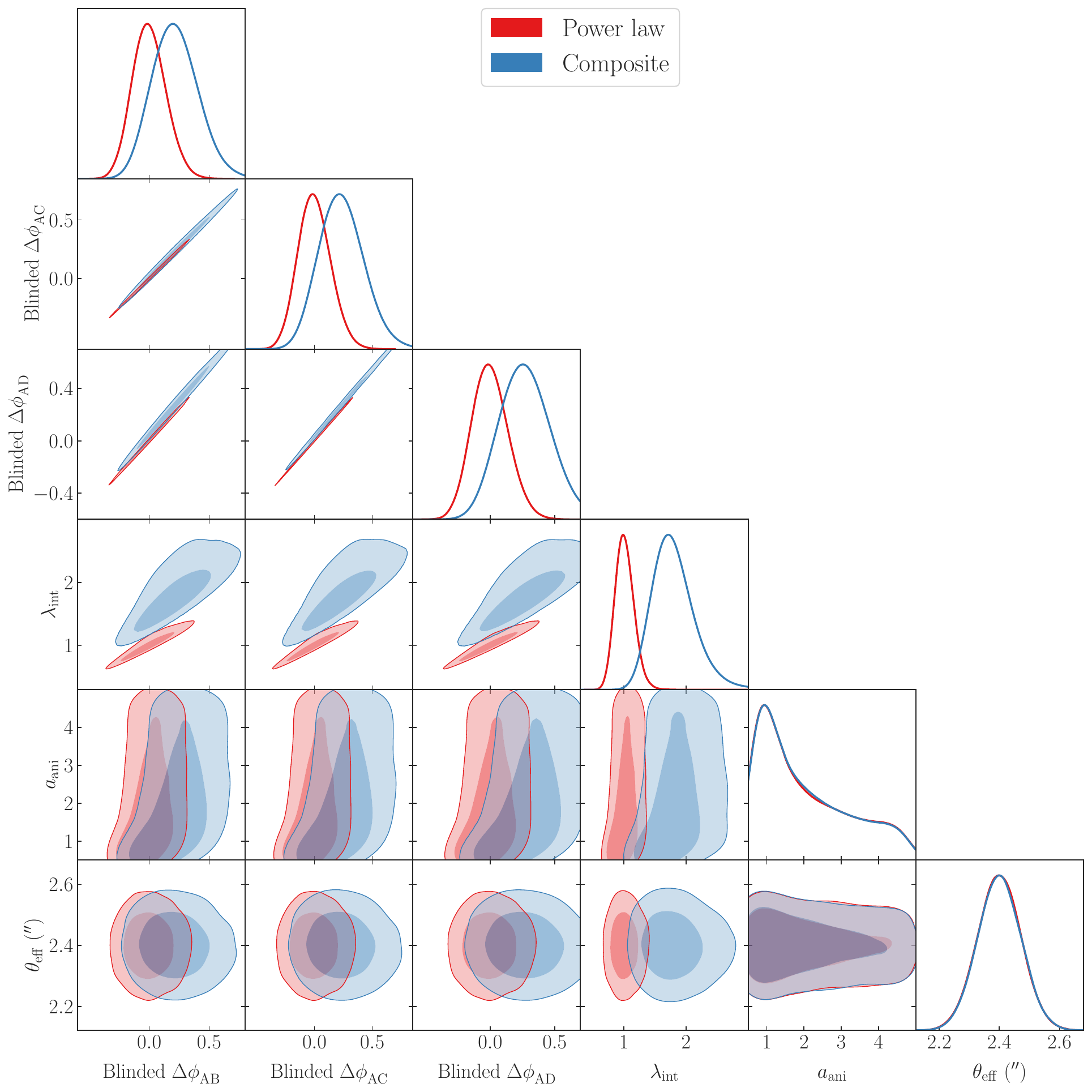}
        \caption{\label{fig:lenstronomy_mst_corrected_fermat_potential}
        MST-adjusted Fermat potential differences $\Delta \phi = \Delta \phi^{\rm model} \lambda_{\rm int} (1 - \kappa_{\rm ext})$ from \textsc{lenstronomy} for the power-law (red) and composite (blue) model families. The internal MST parameter $\lambda_{\rm int}$ is estimated by combining the lens models with the measured velocity dispersion and the external convergence estimate. After adjusting for the MST, the power-law and composite model predictions for the Fermat potential differences become consistent with each other within 1$\sigma$.}
\end{figure*}

In Fig. \ref{fig:lenstronomy_mst_corrected_fermat_potential} we compare the MST-corrected Fermat potential differences between the power-law and the composite models. We find $\lambda_{\rm int}^{\rm pl} = 1.02 \pm 0.15$ for the power-law model and $\lambda_{\rm int}^{\rm comp} = 1.79 \pm 0.53$ for the composite model. \ajs{This large median value of $\lambda_{\rm int}^{\rm comp}$ falls in the excluded region in \citet[][TDCOSMO-IV]{Birrer20} that is based on physical arguments on the mass density distribution. However, the mass profile adopted by \citet{Birrer20} is a cored power-law profile, which can be interpreted as the presence of a cored component in the NFW profile with its radius being much larger than the Einstein radius. \citet{Shajib21} demonstrate that deviations from a power-law profile can be explained by shifting the normalization of the NFW profile without any core component. \ajsii{Whereas the MST considered by \citet[][TDCOSMO-IV]{Birrer20} allows redistribution of matter only within the dark component, the composite model considered here allows redistribution of matter between dark and luminous components.} Thus, large deviations of $\lambda_{\rm int}$ from 1 is still physically plausible, so the large $\lambda_{\rm int}$ produced by our composite model is not in tension with the exclusion range set by \citet[][TDCOSMO-IV]{Birrer20}.
}

The predicted Fermat potential difference between the power-law and composite models are consistent within $1\sigma$ after adjusting for the internal MST and the external convergence. Thus, we demonstrate that the two model families we adopted are linked through an approximate MST. Therefore, to predict the time delays or to measure \ho\ through constraining $\lambda_{\rm int}$ from stellar kinematics, the choice of mass model family is largely irrelevant as the same result can be obtained with any of the conventional model families. In this case, the uncertainty on the Fermat potential difference is dominated by the velocity dispersion uncertainty, which is at 6.4\%. As $\lambda_{\rm int} \propto \sigma_{\rm ap}^2$, the uncertainty on the $\lambda_{\rm int}$ is thus expected to be twice the uncertainty of the velocity dispersion. Our obtained uncertainties of $\lambda_{\rm int}^{\rm pl}$ and $\lambda_{\rm int}^{\rm comp}$ are consistent with this expectation. The uncertainties on the predicted time delays from the final combined posterior with free $\lambda_{\rm int}$ is 19--22\%.

\subsection{\ajsii{Discussion on \textsc{lenstronomy} models}}

\ajsii{The predicted Fermat potential differences from \textsc{lenstronomy} are discrepant between the power-law and composite models, with the predictions from the composite model being 16--21\% lower than those from the power-law model. Both model families fit the imaging data almost equally well, with the composite model providing a slightly higher likelihood value (Table \ref{tab:lenstronomy_bic_values}). The discrepancy between the two model families are caused by the NFW profile normalization in the composite model, which makes the logarithmic slope of the density profile shallower at the Einstein radius (Fig. \ref{fig:lenstronomy_radial_convergence_profile}). We check for any potential unphysical properties in the mass profile posterior for the composite model (see Sect. \ref{sec:lenstronomy_halo_properties}). The size of the central core observed in stellar mass profile is consistent with previous observations \citep[e.g.][]{Bonfini16, Dullo19}. The halo properties (e.g. the $M$--$c$ relation, the total baryonic fraction, and the dark matter fraction within the Einstein radius) are consistent with previous observations of galaxy properties and cosmology. However, the predicted velocity dispersion profile (Fig. \ref{fig:lenstronomy_velocity_dispersion_profile}) from the composite mass model shows a decrease towards the centre, which has not been observed in local massive elliptical galaxies \citep[e.g.][]{Cappellari16, Ene19}, thus pointing to a potential inconsistency in the composite profile. We tested with additional physically motivated priors for the halo mass profile; however, that \ajsiv{only} amplified the discrepancy further. Alternatively, the discrepancy between the power-law and composite mass profiles can be reconciled by including an ultra-massive black hole ($M_{\rm BH} \sim 3.8 \times 10^{10}\ M_{\odot}$), or by incorporating a stellar mass-to-light ratio gradient with exponent $\eta \sim 0.27$, or a combination of both. These values are plausible based on previous observations \citep[e.g.][]{Sonnenfeld18c, Mehrgan19, Dullo19}.}

\ajsii{Furthermore, the predicted central velocity dispersion from the composite model is inconsistent with the observed one. As a result, when no internal MSD is assumed (i.e. $\lambda_{\rm int} = 1$), the kinematics likelihood largely excludes the posterior from the composite model in the final combined posterior. As a result, the final combined posteriors from \textsc{lenstronomy} is almost entirely contributed by the power-law model.
}

\section{Comparison of the two \ajs{software programs}} \label{sec:software_comparison}

The lens model posteriors from both teams were un-blinded on October 22, 2021, and no further modification to the lens models was performed afterwards. We only performed tests to investigate the differences or the lack thereof between the two modelling teams\suyu{;} 
the final time-delay predictions are kept frozen at the values during un-blinding. Table \ref{tab:model_params_comparison} compares the model parameters, derived quantities, \ajsiv{predicted time delays} between \textsc{glee} and \textsc{lenstronomy}. We compare the un-blinded time-delay predictions from the combination of lensing, kinematics, and LOS analyses in Sect. \ref{sec:time_delay_comapare}. Then, we compare the lens model parameters and  Fermat potential differences from lens modelling only in Sect. \ref{sec:model_param_compare}. We compare the pixelized PSF reconstructions between the software programs in Sect. \ref{sec:psf_comparison} and the computational requirements in Sect. \ref{sec:computational_requirement}. Finally, we discuss our findings in Sect. \ref{sec:discussion}.

\renewcommand*\arraystretch{1.5}
\begin{table*}
\caption{Comparison of \textsc{glee} and \textsc{lenstronomy} model parameters and derived quantities.
\label{tab:model_params_comparison}
}
\begin{minipage}{\linewidth}
\begin{tabular}{l|c|c}
\hline
Parameter &
\textsc{glee} constraints &
\textsc{lenstronomy} constraints
\\
\hline
\multicolumn{3}{c}{Power-law ellipsoid model}
\\
\hline
$\theta_{\mathrm{E}}~(\arcsec)$\footnote{Spherical-equivalent Einstein radius} &
$1.379_{-0.001}^{+0.001}$ &
$1.380_{-0.001}^{+0.001}$
\\
$q_{\rm m}$ &
$0.610_{-0.005}^{+0.005}$ &
$0.643_{-0.005}^{+0.005}$ 
\\
$\varphi_{\rm m}$ ($^{\circ}$) &
$36.8_{-0.3}^{+0.3}$ &
$37.1_{- 0.2}^{ + 0.2}$
\\
$\gamma$ &
$2.30_{-0.01}^{+0.01}$ &
$2.22_{-0.03}^{+0.02}$
\\
$\gamma_{\rm ext}$ &
$0.078_{-0.002}^{+0.001}$ &
$0.065_{-0.004}^{+0.003}$
\\
$\varphi_{\rm ext}$ ($^{\circ}$) &
$-57.0_{-0.3}^{+0.3}$ &
$-58.1_{- 0.4}^{+ 0.3}$
\\
\hline
\multicolumn{3}{c}{Composite model}
\\
\hline
Stellar $M/L$ ($M_{\odot}/L_{\odot}$)\footnote{$M/L$ for rest-frame $V$ band. The given uncertainties are statistical uncertainties only. The stellar mass is calculated assuming $H_{0} = 70~\mathrm{km~s^{-1}~Mpc^{-1}}$, $\Omega_{\mathrm{m}} = 0.3$, and $\Omega_{\Lambda} = 0.7$, but changes in the cosmology affect the $M/L$ by a negligible amount.} &
$6.3_{-0.1}^{+0.1}$ &
$2.30_{-0.20}^{+0.06}$
\\
NFW $\kappa_{0,\mathrm{h}}$ &
$0.015_{-0.008}^{+0.003}$ &
$0.16_{-0.01}^{+0.04}$
\\
NFW $r_{\mathrm{s}}~(\arcsec)$ &
$19.3_{-1.2}^{+1.2}$ &
$22.8_{-3.5}^{+2.6}$
\\
NFW $q_{\rm m}$ &
$0.85_{-0.01}^{+0.01}$ &
$0.76_{-0.04}^{+0.07}$
\\
NFW $\varphi_{\rm m}$ ($^{\circ}$) &
$24.8_{-1.3}^{+1.7}$  & 
$-54.2^{+2.3}_{-3.4}$
\\
$\gamma_{\mathrm{ext}}$ &
$0.101_{-0.001}^{+0.002}$ &
$0.128_{-0.008}^{+0.005}$
\\
$\varphi_{\rm ext}$ ($^{\circ}$) &
$-57.3_{-0.2}^{+0.2}$ &
$-55.3^{+0.9}_{-1.4}$
\\
\hline
\multicolumn{3}{c}{\ajsiv{Predicted time delays from power-law and composite models combined}\footnote{\ajsiv{Assuming a flat \lcdm\ cosmology with $H_{0} = 70~\mathrm{km~s^{-1}~Mpc^{-1}}$, $\Omega_{\mathrm{m}} = 0.3$, and $\Omega_{\Lambda} = 0.7$.}}} \\
\hline
$\Delta t_{\rm AB}$ (d) & $-4.4_{-0.5}^{+0.4}$ & $-5.0_{-0.2}^{+0.2}$ \\
$\Delta t_{\rm AC}$ (d) & $-9.4_{-0.8}^{+0.7}$ & $-10.0^{+0.4}_{-0.3}$ \\
$\Delta t_{\rm AD}$ (d) & $-23.0_{-2.4}^{+1.8}$ & $-24.2_{-0.7}^{+1.0}$ \\
\hline
\end{tabular}
\\
{\footnotesize Reported values are medians, with errors corresponding to the 16th and 84th percentiles.}
\\
{\footnotesize Angles are measured east of north.}
\end{minipage}
\end{table*}
\renewcommand*\arraystretch{1.0}

\subsection{Predicted time delays} \label{sec:time_delay_comapare}

\begin{figure*}
        \includegraphics[width=\textwidth]{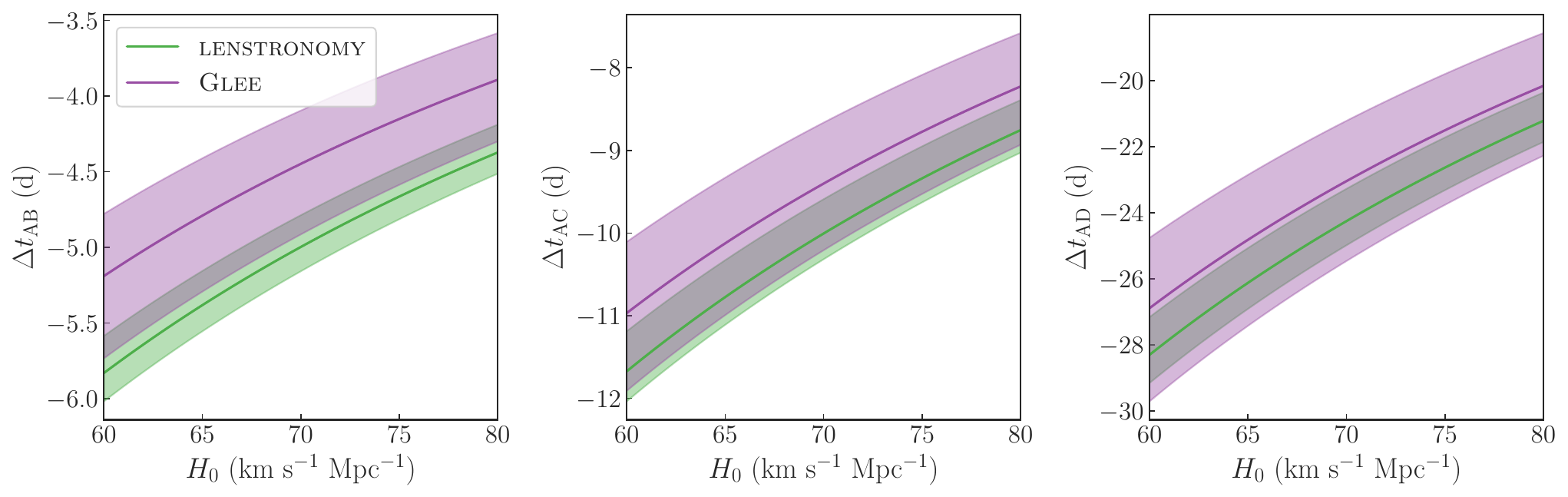}
        \caption{\label{fig:time_delay_prediction_comparison}
        Comparison between the two modelling teams for the predicted time delays \ajsiv{(un-blinded)} as a function of \ho\ for the three image pairs\suyu{, in flat $\Lambda$CDM cosmology with $\Omega_{\mathrm{m}} = 0.3$}. Each posterior is the final combined posterior from the two mass-model setups: power-law and composite, including external convergence and stellar kinematics, and assuming $\lambda_{\rm int} = 1$.
        }
\end{figure*}

\begin{figure*}
        \includegraphics[width=\textwidth]{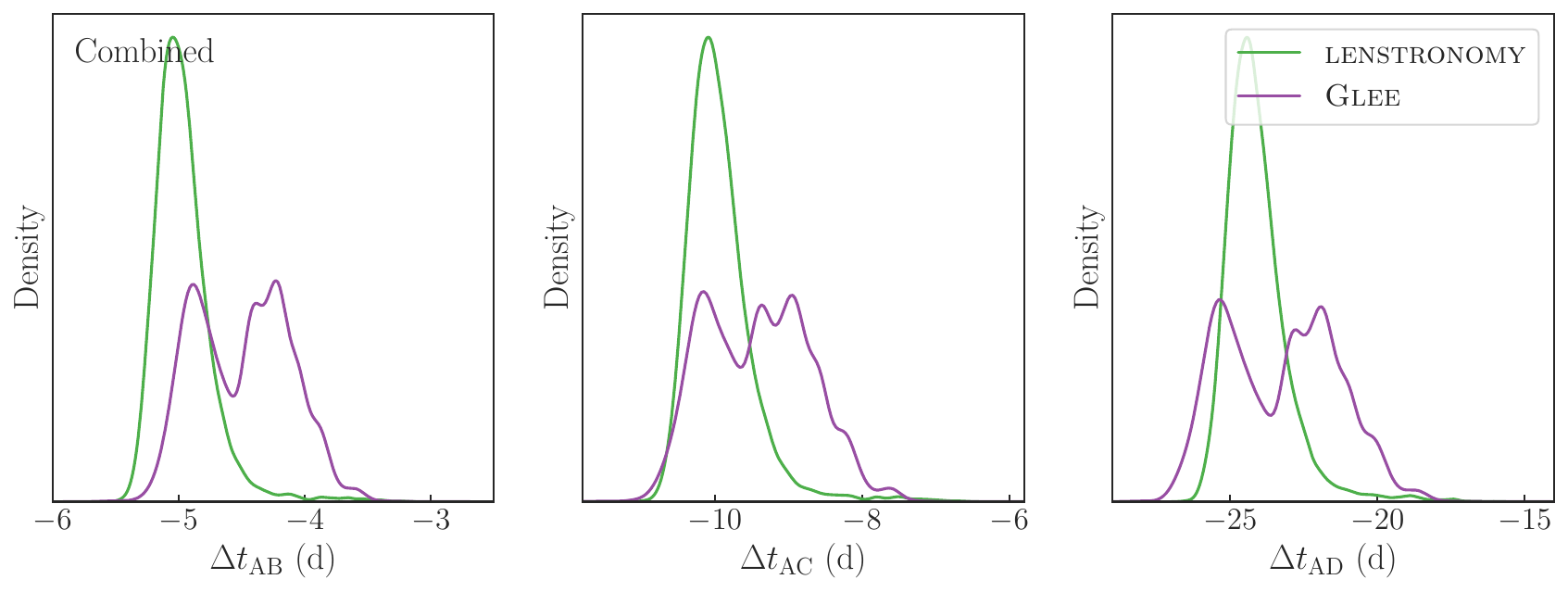} \
        \includegraphics[width=\textwidth]{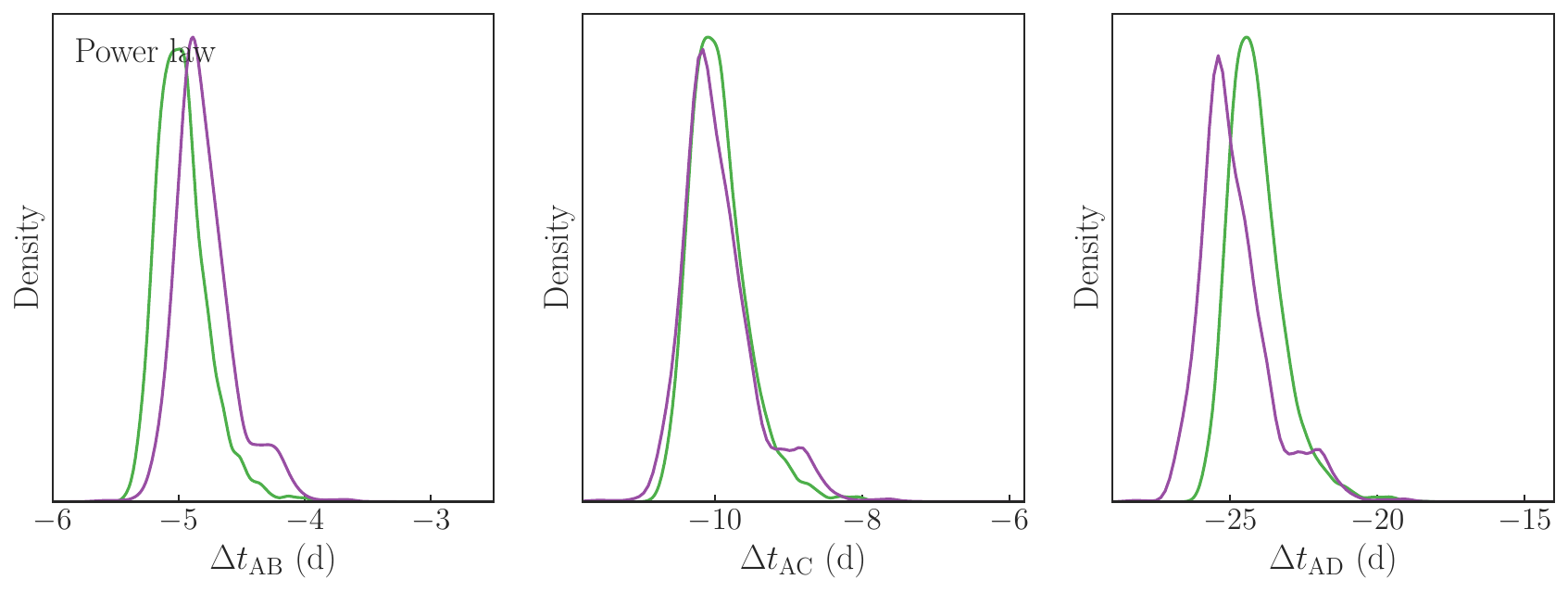} \\
        \includegraphics[width=\textwidth]{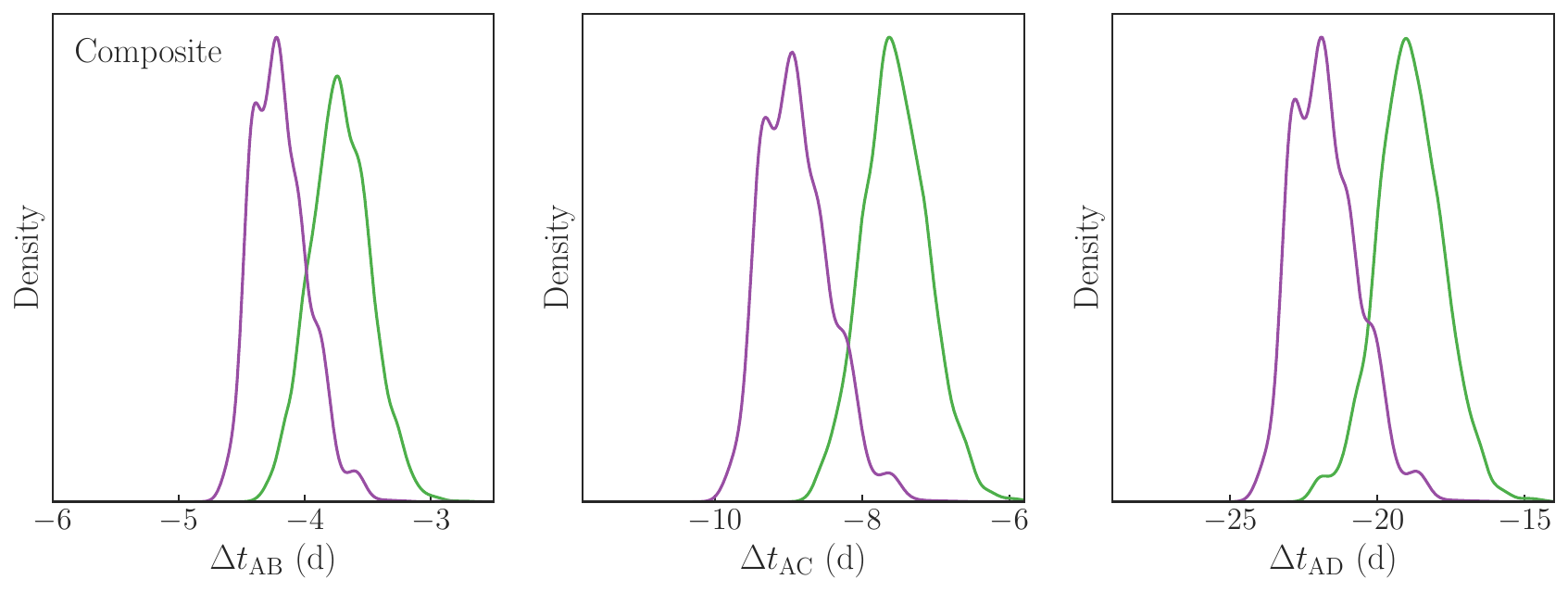} 
        \caption{\label{fig:time_delay_compare_breakdown}
        Comparison of predicted time delays between the \textsc{lenstronomy} (green) and \textsc{glee} (purple) teams. The three rows show the comparison for combined (top), power-law-only (middle), and composite-only (bottom) cases. \reply{The combined posteriors are consistent within $<1.2\sigma$, the power-law  model's posteriors are consistent within $<0.6\sigma$, and the composite model's posteriors are consistent within $<2.1\sigma$}. In percentage, the maximum deviation for the power-law model is between AB image pair by 3.4\%, and for the composite model is between AC image pair by 15\%. The \textsc{glee} team combined the power-law and composite model posteriors with equal weighting after applying the kinematics weighting within each model family, whereas the \textsc{lenstronomy} team combined the two model families with kinematics weighting, leading to the final combined posterior being dominated by the power-law model. 
        }
\end{figure*}

We illustrate the final predicted time delay from both teams as a function of \ho\ in Fig. \ref{fig:time_delay_prediction_comparison}\kw{, assuming a flat $\Lambda$CDM cosmology with $\Omega_{\mathrm{m}} = 0.3$}. The predictions for all image pairs are consistent with each other within $\sim1\sigma$. We further compare the time delay predictions for combined, power-law-only, and composite-only cases in Fig. \ref{fig:time_delay_compare_breakdown}. The combined time-delay posteriors differ the largest for the AB image pair by 11\% (1.2$\sigma$). We note that the \textsc{glee} team applied kinematics weighting to the lens model posteriors only within the mass families and then combined the mass families with equal weighting, whereas the \textsc{lenstronomy} team weighted the mass families by kinematics. As a result, the combined posteriors from the \textsc{glee} team receives equal contribution from both model families giving rise to bi-modalities, where the combined posterior from the \textsc{lenstronomy} team is almost entirely contributed by the power-law model posterior. The power-law-model-predicted time delays agree better between the two teams, with the largest difference appearing for AB image pair by 3.4\% (0.6$\sigma$). The composite model predictions are more discrepant, with the largest difference appearing for AC image pair by 15\% (2.1$\sigma$).

\subsection{Model parameters and Fermat potential differences} \label{sec:model_param_compare}

\reply{The astrometric uncertainty on the constrained AGN image positions are consistent between the two teams, as both teams constrain the image positions with uncertainty 0\farcs004 at maximum. This astrometric precision satisfies the requirement for cosmographic measurements \citep{Birrer19}.}

\begin{figure*}
        \includegraphics[width=\textwidth]{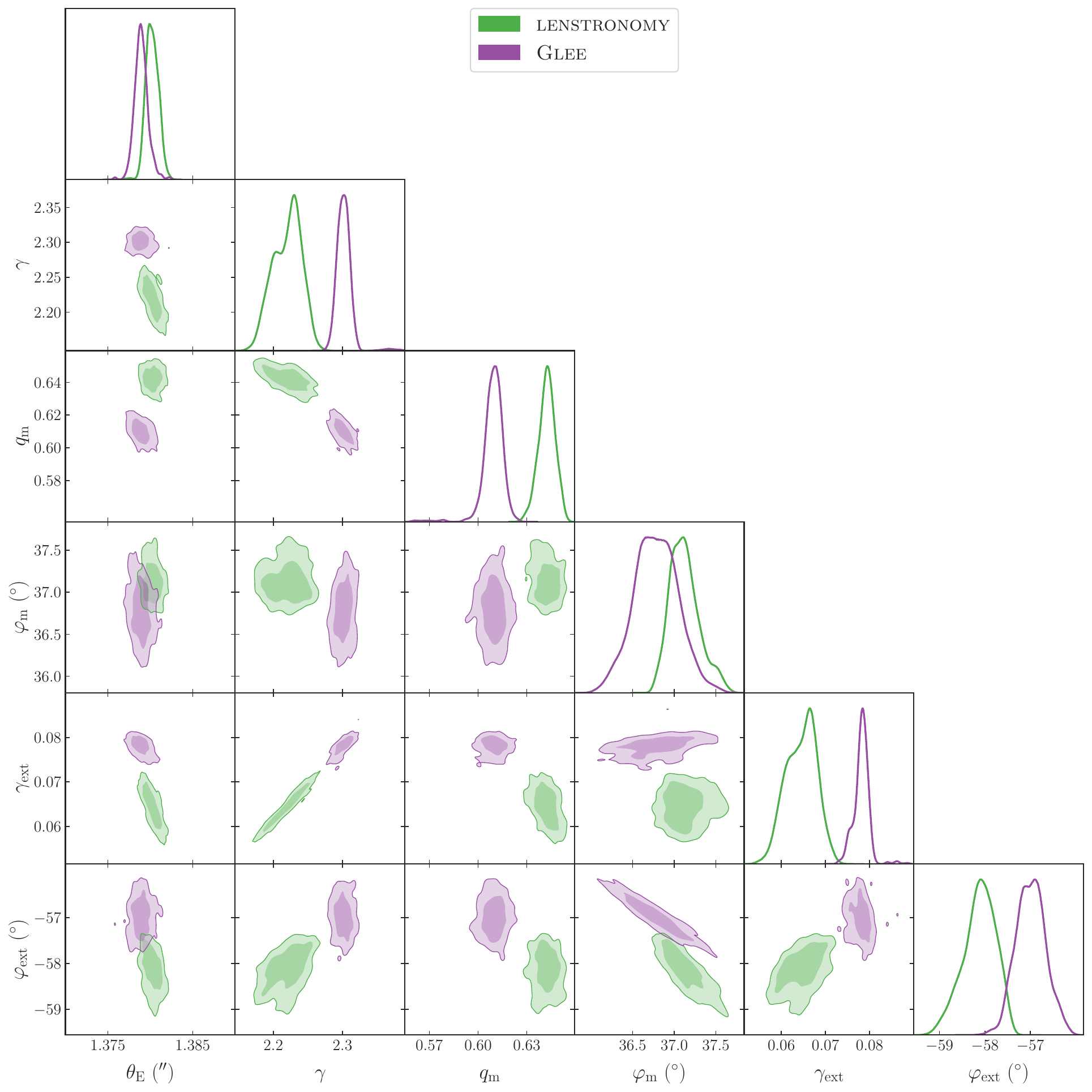}
        \caption{\label{fig:pl_params_compare}
        Comparison of lens model parameter differences for the power-law model between the \textsc{lenstronomy} (green) and the \textsc{glee} (purple) teams.
        }
\end{figure*}

In Fig. \ref{fig:pl_params_compare} we compare the lens model parameter posteriors and the predicted Fermat potential posteriors from the power-law lens models from both teams. The power-law exponent constrained by the \textsc{lenstronomy} model is $\gamma = 2.21\pm 0.02$, whereas that constrained by the \textsc{glee} model is $\gamma = 2.30 \pm 0.01$, which is a discrepancy at 4$\sigma$. The external shear magnitude constrained by the \textsc{lenstronomy} model is $\gamma_{\rm ext} = 0.065 \pm 0.004$, and that by the \textsc{glee} model is $\gamma_{\rm ext} = 0.078 \pm 0.001$, which is at a $3.4\sigma$ discrepancy. We identify a degeneracy between $\gamma$ and $\gamma_{\rm ext}$ internal to both models, and the discrepancy between the posteriors from the two models lie along this degeneracy (Fig. \ref{fig:pl_params_compare}). \reply{The external shear magnitudes are typical for quadruply lensed quasar systems \citep[e.g. see][]{Schmidt22}. We investigated for deviation from the simple ellipticity description in our mass models as a potential source of the external shear. We find that boxy-ness or discy-ness in the luminous component is negligible within the Einstein radius (i.e. $\sqrt{a_4^2 + b_4^2} \lesssim 0.005$), which implies that allowing boxy-ness or discy-ness in the description of the mass profile is not required \citep[for definitions of $a_4$ and $b_4$, and their impact on $H_0$ measurement, see][]{VandeVyvere22}. As a result, we attribute the LOS galaxies around the central deflector to be the main source of the external shear, with additional potential contribution from the mild isophotal twist beyond the Einstein radius in the central deflector \citep{VandeVyvere22b}.} 

\begin{figure*}
        \includegraphics[width=\textwidth]{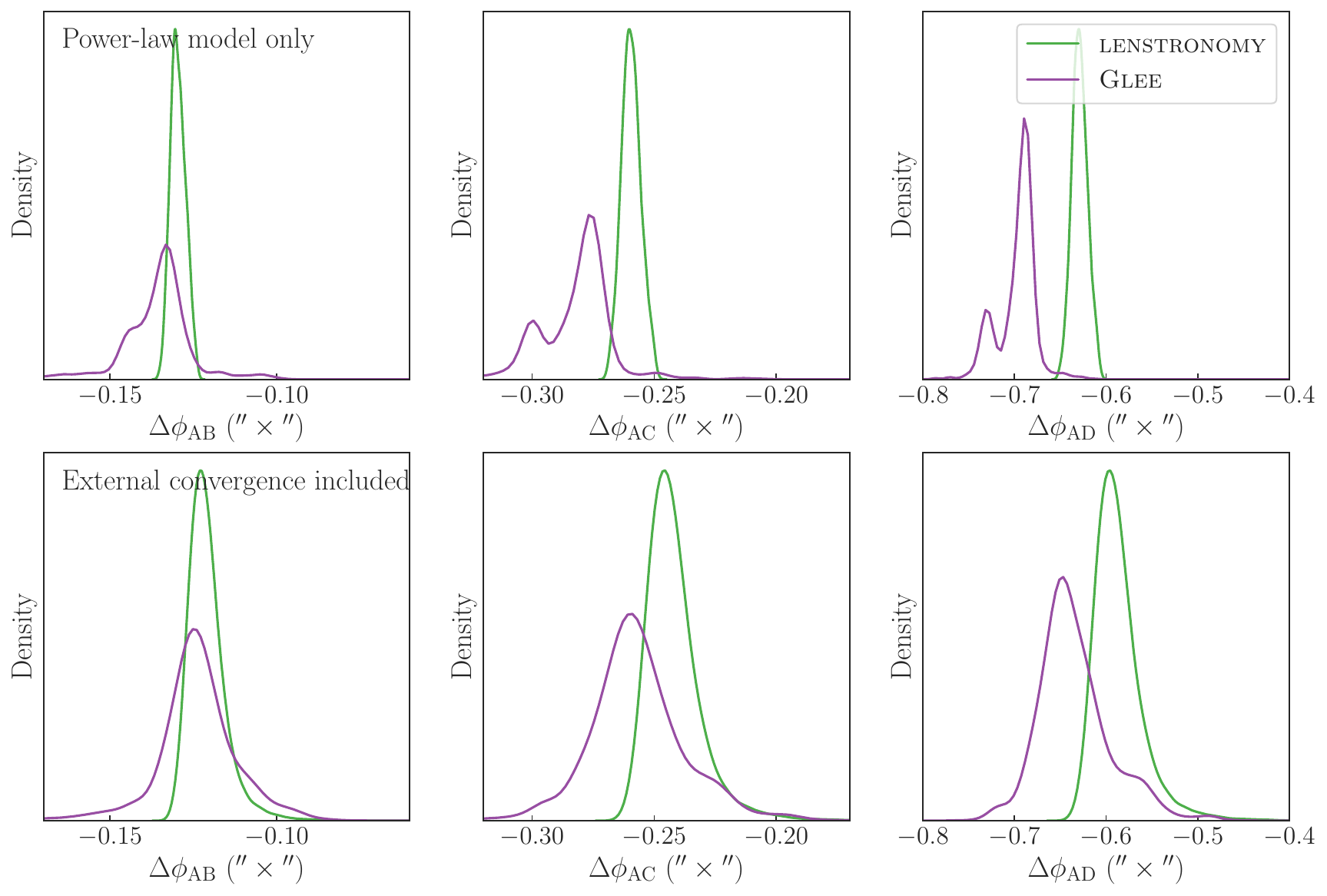}
        \caption{\label{fig:pl_potential_compare_w_kappa_ext}
        Comparison of Fermat potential differences for the power-law model between the \textsc{lenstronomy} (green) and the \textsc{glee} (purple) teams, without including the external convergence (top row), and with including the external convergence (bottom row). 
        }
\end{figure*}

Next in Fig. \ref{fig:pl_potential_compare_w_kappa_ext}, we compare the Fermat potential differences only from the power-law models of both teams without adjusting for the external convergence \suyu{(top row)} and with \suyu{adjustment}
for the external convergence \suyu{(bottom row)}. The Fermat potential differences from the lens model are discrepant, for example by $5.5\sigma$ (8.9\%) for the AD image pair that has the longest predicted time delay. However, after combining the corresponding external convergence -- based on selection cuts 
\suyu{using} 
the best fit external shear from each model -- the Fermat potential differences all become consistent within $1\sigma$, for example by 0.26$\sigma$ (1\%) for the AD image pair. Interestingly, the positive correlation between $\gamma$ and $\gamma_{\rm ext}$ allows the estimated $\kappa_{\rm ext}$ selected on $\gamma_{\rm ext}$ to bring the time-delay posteriors closer, as higher $\gamma_{\rm ext}$ selects higher $\kappa_{\rm ext}$. However, the strength of the positive correlation between $\gamma$ and $\gamma_{\rm ext}$ depends on the particular morphology of the quad lenses \citep{Shajib19}. Thus, we cannot conclude if this effect -- that the $\gamma_{\rm ext}$-selected $\kappa_{\rm ext}$ brings the time delay posteriors more into agreement -- applies to all lensing systems and \kw{is} not \kw{just} a particular occurrence for the lens system \lensname. A detailed analysis of a larger sample of lenses is required to reach a conclusion on this matter, and it is left for future work.


%

\subsection{Reconstructed PSFs} \label{sec:psf_comparison}

\begin{figure}
        \includegraphics[width=0.5\textwidth]{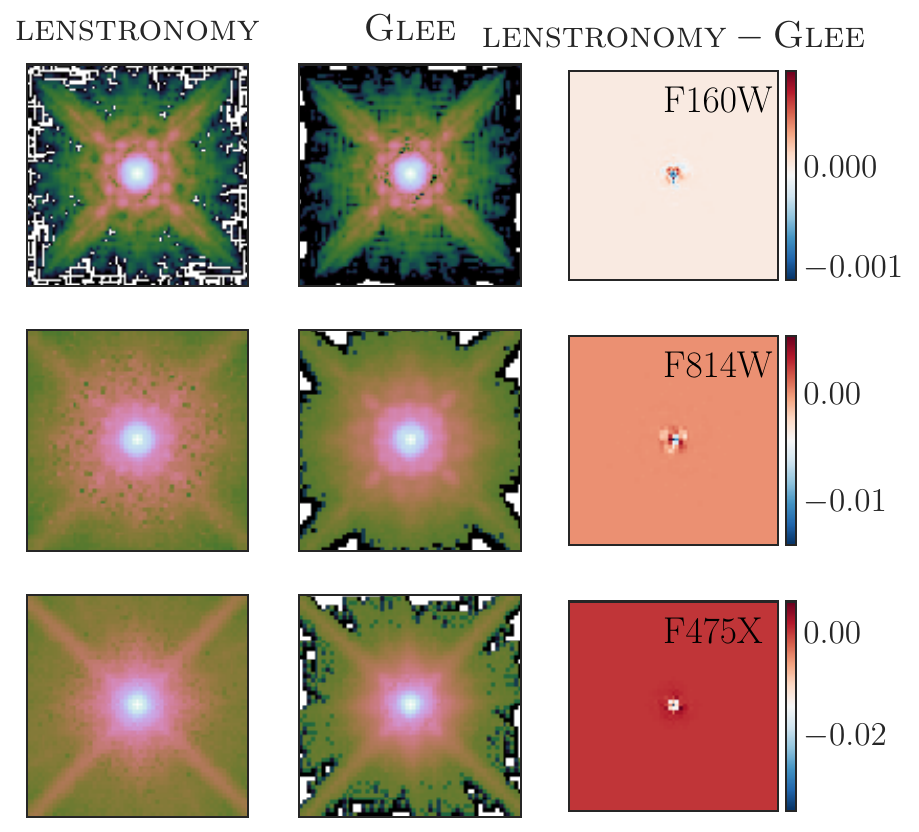}
        \caption{\label{fig:psf_compare}
        Comparison of the reconstructed pixelized PSFs by the \textsc{lenstronomy} (first column) and the \textsc{glee} (second column) modelling procedures. The third column illustrates the difference between the \textsc{lenstronomy} and \textsc{glee} PSFs. The three rows correspond to the F160W, F814W, and F475X filters from top to bottom. The F160W PSF is supersampled with a supersampling factor of 3. We note that the illustrated supersampled \textsc{lenstronomy} PSF was not used in the pre-un-blinding models and was reconstructed after the un-blinding to perform further tests. The original reconstructed PSF in \textsc{lenstronomy} in F160W was not supersampled. The \textsc{glee} PSF FWHMs are smaller by $\sim$2--7\% than the \textsc{lenstronomy} ones.
        }
\end{figure}

The reconstructed pixelized PSFs by the \textsc{glee} modelling procedure have smaller FWHM by $\sim$2--7\% than the ones from \textsc{lenstronomy}: 1.7\% in F160W, 3.9\% in F814W, and 6.7\% in F475X (Fig. \ref{fig:psf_compare}). Furthermore, the F160W PSF in \textsc{glee} is supersampled with a supersampling factor of 3, whereas the F160W PSF in \textsc{lenstronomy} has the same pixel resolution (0\farcs08) as the drizzled image. The PSFs for the UVIS filters have the same pixel resolution as the drizzled image (0\farcs04) for both \textsc{glee} and \textsc{lenstronomy}.

\begin{figure}
        \includegraphics[width=0.5\textwidth]{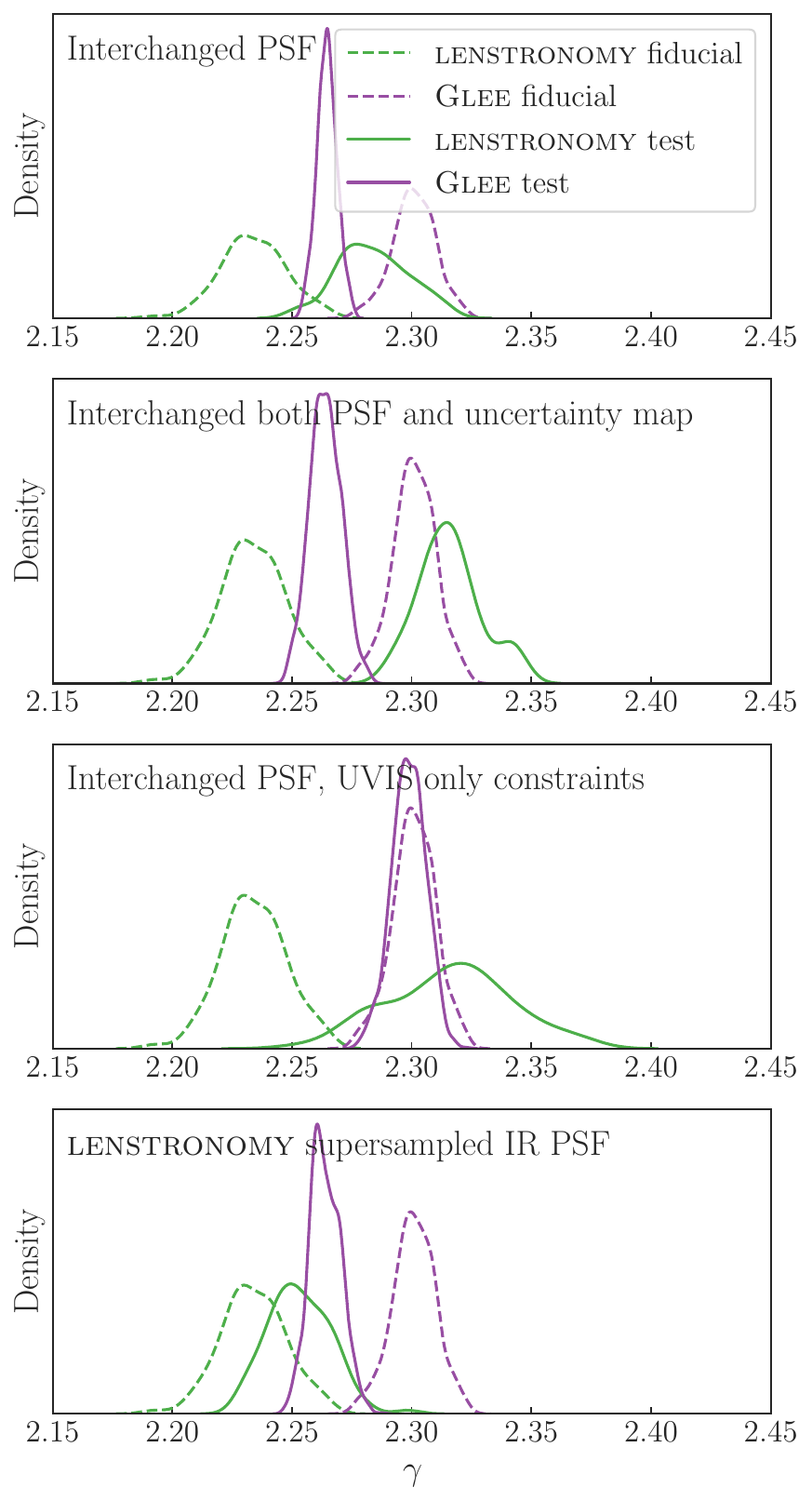}
        \caption{\label{fig:psf_test_compare}
        Deviations in the logarithmic slope $\gamma$ of the power-law model with different PSF settings. In all panels, the dashed distributions show the fiducial constraints -- \textsc{lenstronomy} in green and \textsc{glee} in purple. The \textsc{lenstronomy} fiducial constraint is from the highest BIC value power-law model (Table \ref{tab:lenstronomy_bic_values}), and the \textsc{glee} fiducial constraint is from the `power-law fiducial' setup (Table \ref{tab:glee_td_bic}). \textbf{First panel:} When only the reconstructed PSFs are interchanged to optimize the models, the constraint from one software program moves towards the other's fiducial constraint to be consistent within $1.2-1.6\sigma$. We note that \textsc{glee} model artificially creates supersampled version of the \textsc{lenstronomy} PSF through interpolation to perform the tests here. \textbf{Second panel:} When both the PSFs and weighted uncertainty maps are interchanged, 
        the resultant $\gamma$ becomes slightly more consistent between the software programs within $1.0-1.5\sigma$. Therefore, we conclude that the \kw{differences in the} PSF itself significantly contributes to the discrepancy in the power-law $\gamma$ constraint and not the particular method of \ajsiii{weighting the uncertainty map to account for PSF uncertainty}. 
         \textbf{Third panel:} When models from both of the software programs are optimized only with UVIS data and interchanged PSFs, the \textsc{glee} constraint does not shift significantly; however, the \textsc{lenstronomy} constraint shifts significantly towards the \textsc{glee} constraints. This result indicates that the difference between the \textsc{glee} and \textsc{lenstronomy} fiducial models are largely created by the difference in the IR PSF. \textbf{Fourth panel:} When both software programs use a supersampled PSF with a supersampling factor of 3 reconstructed by \textsc{lenstronomy}, the resultant $\gamma$ constraint agrees very well, within $0.6 \sigma$.
        }
\end{figure}

We test how the differences in the reconstructed PSFs contribute to the differences in the logarithmic slope parameter for the power-law model between the two software programs. The test results are illustrated in Fig. \ref{fig:psf_test_compare}. In these tests, we change the adopted PSFs and optimize a fiducial lens model from each software program. For \textsc{lenstronomy}, the fiducial model is the power-law model with the highest BIC score (Table \ref{tab:lenstronomy_bic_values}). For \textsc{glee}, the fiducial model is the `power-law fiducial model'(Table \ref{tab:glee_td_bic}). We first interchanged all the PSFs between the software programs. Due to numerical requirements, the \textsc{glee} team artificially supersample the \textsc{lenstronomy}-reconstructed PSF through interpolation for this and subsequent tests. With the PSFs interchanged, the power-law slope parameter $\gamma$ constrained by one software program shifts towards the fiducial constraint from other software program. These shifts bring the $\gamma$ distributions within $\sim$1.2--1.6$\sigma$ consistency between the two software programs given the same PSFs, although still leaving some unexplained deviations.

We further interchange \ajsiii{the weighted uncertainty maps} in addition to the reconstructed PSFs between the software programs. \ajsiii{Originally, the \textsc{glee} team creates a weighted uncertainty map by boosting the noise levels around the quasar positions (see Sect. \ref{sec:glee_modelling}), whereas the \textsc{lenstronomy} team adds the PSF uncertainty map of the initial PSF estimate in quadrature with the data uncertainty map at the positions of the quasars.}  In this test, the resultant $\gamma$ distributions become slightly more consistent within $\sim$1.0--1.5$\sigma$ compared to the previous test. Therefore, the particular choice or method to estimate the \ajsiii{weighted uncertainty maps} does not significantly contribute to the deviation in the power-law $\gamma$ parameter distribution from the two software programs. 

In the next test, we constrained the lens models from UVIS data only with interchanged PSFs. In this test, the \textsc{glee} constraint on $\gamma$ remained stable; however, the \textsc{lenstronomy} constraint shifted towards the \textsc{glee} constraint to be consistent within $\sim0.5\sigma$. As a result, we conclude that the discrepancy in the $\gamma$ distribution between \textsc{lenstronomy} and \textsc{glee} is dominated by the difference in the IR PSF. 

For the last test, we add a new feature in \textsc{lenstronomy} to reconstruct the IR PSF with a supersampled resolution and reconstruct the PSF with a supersampling factor of 3 following the \textsc{glee} team. If this supersampled IR PSF from \textsc{lenstronomy} is used by both software programs, then the resultant $\gamma$ distribution becomes consistent within $\sim0.6\sigma$. However, the $\gamma$ distribution of the \textsc{glee} model with its own supersampled PSF still differs by $2.6\sigma$ from that of the \textsc{lenstronomy} model with its own supersampled PSF. It is not possible to evaluate which reconstructed PSF is more accurate {a priori}. Therefore, it is recommended to marginalize over multiple PSF reconstructions to account for the stochasticity within one particular reconstruction algorithm and as well as different reconstruction algorithms. \suyu{In addition, supersampled PSFs are recommended especially for the IR band with large pixels; the subsampling factor can be set to the minimal value to produce stable results while keeping computational time low. }


\subsection{Requirements for computational resources} \label{sec:computational_requirement}

The entire \textsc{lenstronomy} modelling procedure required $\sim$~$4 \times 10^5$ CPU hours including initial modelling trials, running full MCMC chains of the adopted models, post-processing of posteriors, and post-un-blinding tests. The estimated usage for \textsc{glee} models are $\mathcal{O}(10^5)$ CPU hours only to run the final models and MCMC chains to produce the final un-blinded result. However, the total usage of CPU hours can be $\mathcal{O}(10^6)$ CPU hours including modelling trials, robustness tests, and reruns of chains to recover lost progress due to numerical issues. 

\subsection{Discussion} \label{sec:discussion}

{The final un-blinded time-delay predictions agree within $<1.2\sigma$ between the two modelling teams. As a result, the inferred Hubble constants from the two teams based on the observed time delays will be consistent within $<1.2\sigma$.} However, the predictions from the composite models only are less in agreement between the two teams. Interestingly, the composite-model predictions deviate from the power-law ones towards the same direction for both teams. We were already aware prior to the un-blinding that the composite model is \ajsiii{atypical with respect to previous observations}, for example the velocity dispersion profile significantly decreases towards the centre for the one from the \textsc{lenstronomy} (Fig. \ref{fig:lenstronomy_velocity_dispersion_profile}), \ajsiii{the one from the \textsc{glee} team has a very low inner dark matter fraction.} 
Although such discrepancies between composite and power-law model predictions have been observed in the previously analysed systems \citep[e.g.][]{Suyu14}, this system \lensname\ demonstrates the largest discrepancy to date out of the systems analysed by H0LiCOW/TDCOSMO. However, unlike the previously analysed systems, the Einstein radius of this system encompasses only the very central region of the very extended lens galaxy, and thus the imaging observables probe a different region of the elliptical galaxy: at $ \sim \theta_{\rm eff}/3$ instead of $\sim \theta_{\rm eff}$. This discrepancy  in the composite model illustrates that this model is not an adequate description for the mass distribution at the central region of {all} elliptical galaxies. Rather an appropriate combination of a mass-to-light ratio gradient and a supermassive black hole can be necessary to sufficiently describe the mass distribution at the scales considered here. In such cases when the different mass model families fit the imaging observables equally well, but lead to different predictions in the Fermat potential and kinematics, the observed kinematics is crucial to act as the differentiator between the mass model families through appropriate weighting of the kinematics likelihood. In the future, spatially resolved velocity dispersion from integral field spectra\lreply{, for example from the Multi Unit Spectroscopic Explorer (MUSE) on the VLT,} will be able to constrain such an improved composite model with additional degrees of freedom allowed. 

For the composite model setup, the modelling teams had more freedom in choosing particular priors and model settings, allowing for discrepancies between the results from the two teams. However, the power-law models are specified with less room for independent choices to be made by  the modelling teams. Therefore, we compare the results from the power-law models between the two teams to identify systematic differences at the level of the software packages. 

In particular, we focus on the power-law logarithmic slope parameter $\gamma$, as this parameter is the most sensitive lens model parameter to the predicted time delays and thus the inferred Hubble constant. The $\gamma$ distributions between the modelling teams are discrepant at $4\sigma$. We identify the difference in the reconstructed PSFs, especially in the IR band, to be the dominant source of this discrepancy. Given the same supersampled PSF in the IR band and non-supersampled PSFs in the UVIS bands, both modelling softwares produce $\gamma$ constraints with differences below 0.5\%. The $\gamma$ distributions from the two modelling software programs are not expected to be identical due to differences in the numeric implementation, for example the likelihood computation region and the source reconstruction method. Thus, we can conclude that the systematic differences in the model fitting part of the software programs are below 0.5\% in $\gamma$ corresponding to \ajsiii{$\sim$1\%} on H$_0$. However, the PSF reconstruction method can lead to systematic differences as large as $\sim4$\%. This difference is reconciled in the time-delay predictions from the two teams after combining the lens model posteriors with the kinematics data and the estimated external convergence for the particular system \lensname. However, it is inconclusive if such differences can be similarly reconciled for all lens systems in general, and we leave this investigation with a larger lens sample for a future study. If such differences cannot be reconciled for other lens systems and these differences do not average out when a sample of lenses is considered, then these differences would be non-negligible in the long run when a large sample of lens systems are combined to infer the Hubble constant. As it is currently not possible to evaluate the appropriateness of one reconstructed PSF over the other, we recommend to marginalize over different realizations of the PSF reconstruction and also over different reconstruction algorithms to avoid any potential bias. Furthermore, a supersampled PSF in the IR band is also recommended, as the drizzled pixel scale of 0\farcs08 in the IR does not optimally sample the PSF.

\section{Summary and conclusion} \label{sec:conclusion}

In this study two teams independently modelled the lens system \lensname\ from three-band HST imaging using two different software programs: \textsc{lenstronomy} and \textsc{glee}. Two families of models were specified as baseline models before the modelling was performed -- one family with a power-law ellipsoidal mass distribution and the other family with a two-component mass distribution that accounts for the dark and baryonic components separately. The baseline settings were pre-specified to allow a fair comparison between the modelling results. Individual teams were allowed to improve upon the baseline settings as they deemed appropriate, for example the choice of priors and the numerical settings specific to the software program being used. The two modelling procedures were carried out blindly with regards to the other team. The models were un-blinded on October 22, 2021, after an internal review \ajsiv{by scientists from the TDCOSMO collaboration not directly involved with either modelling team}, and no further modifications to the lens models were performed. The predicted Fermat potential differences from both teams were combined with the observed kinematics data and estimated external convergence to predict the time delays between the three image pairs. A future study will infer the Hubble constant by comparing the predicted time delays with the observed ones from ongoing monitoring campaigns. We investigated the observed systematic differences between the model outputs from the two teams and identify that \kw{differences in} the reconstructed PSF are the dominant source of systematic differences. The main results of this study are as follows:
\begin{itemize}
        \item The final predicted time delays from \textsc{lenstronomy} are: $\Delta t_{ \rm AB } = -5.0^{ +0.2 }_{ -0.2 }$ d,
$\Delta t_{ \rm AC } = -10.0^{ +0.4 }_{ -0.3 }$ d, and
$\Delta t_{ \rm AD } = -24.2^{ +1.0 }_{ -0.7 }$ d; and the ones from \textsc{glee} are: $\Delta t_{ \rm AB } = -4.4^{ +0.4 }_{ -0.5}$ d,
$\Delta t_{ \rm AC } = -9.4^{ +0.7 }_{ -0.8 }$ d, and
$\Delta t_{ \rm AD } = -23.0^{ +1.8 }_{ -2.4 }$ d. These values assume a flat \lcdm\ cosmology with $H_0 = 70$ km s$^{-1}$ Mpc$^{-1}$ and $\Omega_{\rm m} = 0.3$. The negative value of $\Delta t_{\rm AB}$, for example, signifies that image B lags image A. \ajsiii{This system is currently being monitored under the COSMOGRAIL programme to measure the time delays \citep{Eigenbrod05}}. \ajsiv{Once the time delays are measured, the mutual agreement between the predicted values will result in a mutually consistent inference of the Hubble constant.} 
        
        \item The logarithmic slope, $\gamma$, and \ajsiv{the external shear, $\gamma_{\rm ext}$, of the power-law model deviate} by 4$\sigma$ (3.9\%) between the \textsc{lenstronomy} and \textsc{glee} models. This discrepancy is predominantly created by the difference in the reconstructed pixelized PSFs. When the same PSF is used by both modelling programs, then the resultant $\gamma$ \ajsiv{and $\gamma_{\rm ext}$} distributions agree within $\sim$0.6$\sigma$ ($\sim$0.5\%), which is compatible with the $\sim$1\% precision goal in the Hubble constant measurement from a large sample of $\sim$40 quad lenses \citep[e.g.][]{Shajib18, Birrer21c}. 
        The particular method \ajsiii{of weighting the uncertainty map to account for the PSF uncertainty} is non-dominant in this discrepancy. 
        
        \item Our composite model posteriors are not generally in good agreement with the one from the power-law model, and the discrepancy is more prominent for \textsc{lenstronomy} models due to the adoption of more stringent physical priors on the halo mass. This inconsistency points to the inadequacy of our composite model in describing the mass distribution \suyu{for this particular lens system, \lensname;} 
        \lensname\ is atypical compared to previously analysed TDCOSMO lenses in that the \ajsiii{$\theta_{\rm E}/\theta_{\rm eff}$} 
        is relatively small and thus the imaging information only probes the inner region of the deflector galaxy where the combination of NFW and mass-follows-light profiles is not an adequate model. \ajsiv{We stress that for all the seven systems previously analysed by the TDCOSMO collaboration, the power-law and composite models were in excellent agreement.} For such discrepancies between the power-law and composite models, additional data, such as the stellar kinematics, should be used to select the better model. \ajsiv{The two models predict significantly different velocity dispersion profiles and will therefore be easily separable by spatially resolved kinematics.}
\end{itemize}

In the context of the recent debate around the Hubble constant, it is paramount to thoroughly investigate for potential systematic biases in the measurement methods \citep[e.g.][]{Freedman21, Riess22}. This study performs one such crucial systematic check for time-delay cosmography to investigate the robustness of lens modelling software programs. By keeping the modelling systematics under control, future large samples of lensed quasars and SNe will robustly measure the Hubble constant to $\lesssim1$\% precision \citep[e.g.][]{Jee16, Birrer21c, Birrer22}.

\begin{acknowledgements}
We thank Christopher~D.~Fassnacht, Veronica Motta, and Abigail Lee for useful comments and discussions that improved this manuscript.
Support for this work was provided by NASA through the NASA Hubble Fellowship grant
HST-HF2-51492 awarded to AJS by the Space Telescope Science Institute, which is operated by the Association of Universities for Research in Astronomy, Inc., for NASA, under contract NAS5-26555.
AJS, SB, and TT were supported by the National Aeronautics and Space Administration (NASA) through the Space Telescope Science Institute (STScI) grant HST-GO-15320. AJS was also supported by the Dissertation Year Fellowship from the University of California, Los Angeles (UCLA) graduate division. This research was supported by the U.S. Department of Energy (DOE) Office of Science Distinguished Scientist Fellow Program. 
This work was supported by World Premier International Research Center Initiative (WPI Initiative), MEXT, Japan. This work was supported by JSPS KAKENHI Grant Number JP20K14511.
SHS thanks the Max Planck Society for support through the Max Planck Research Group.
This research is supported in part by the Excellence Cluster ORIGINS which is funded by the Deutsche Forschungsgemeinschaft (DFG, German Research Foundation) under Germany's Excellence Strategy -- EXC-2094 -- 390783311.
TT acknowledges support by the Packard Foundation through a Packard Research fellowship, by the National Science Foundation through NSF grants AST-1836016, AST-1906976, and by the Gordon and Betty Moore Foundation through grant 8548.
AA’s research is funded by Villum Experiment Grant \textit{Cosmic Beacons} (project number 36225).
This work was supported by the Swiss National Science Foundation (SNSF) and by the European Research Council (ERC) under the European Union’s Horizon 2020 research and innovation programme (grant agreement No 787886). 
\\
This work used computational and storage services associated with the Hoffman2 Shared Cluster provided by UCLA Institute for Digital Research and Education’s Research Technology Group. Additionally, this work was completed in part with resources provided by the University of Chicago’s Research Computing Center.
\\
AJS \suyu{and SHS} acknowledge the hospitality of the Aspen Center of Physics (ACP) and the Munich Institute for Astro- and Particle Physics (MIAPP) of the Excellence Cluster "Universe", where part of this research was completed. ACP is supported by National Science Foundation grant PHY-1607611.
\\
This research made use of \textsc{glee} \citep{Suyu10b, Suyu12c}, \textsc{lenstronomy} \citep{Birrer15, Birrer18, Birrer21b}, \textsc{fastell} \citep{Barkana99}, \textsc{numpy} \citep{Oliphant15}, \textsc{scipy} \citep{Jones01}, \textsc{Astropy} \citep{AstropyCollaboration13, AstropyCollaboration18}, \textsc{Photutils} -- an \textsc{Astropy} package for detection and photometry of astronomical sources \citep{Bradley20},  \textsc{jupyter} \citep{Kluyver16}, \textsc{matplotlib} \citep{Hunter07}, \textsc{seaborn} \citep{Waskom14}, \textsc{sextractor} \citep{Bertin96}, \textsc{emcee} \citep{Foreman-Mackey13}, \textsc{colossus} \citep{Diemer18}, \textsc{getdist} (\url{https://github.com/cmbant/getdist}), \textsc{pygalaxev} \citep{Bruzual03}, and \textsc{dynesty} \citep{Skilling04, Speagle19}.

\end{acknowledgements}

\bibliographystyle{aa}
\bibliography{ajshajib}

\end{document}